\documentclass[structabstract]{aa}
\usepackage{graphicx}
\usepackage{txfonts}
\usepackage{aalongtable}
\usepackage{color}

\begin{document}

\title{Abundances of neutron-capture elements in stars of the galactic disk substructures
\thanks{Based on spectra collected with the ELODIE spectrograph at the
1.93-m telescope of the Observatoire de Haute Provence (France).}
\thanks{Tables \ref{t_abund} and \ref{t_abund2} are only available in electronic form at the CDS
via the anonymous ftp to cdsarc.u-strasbg.fr (130.79.128.5)
or via http://cdsweb.u-strasbg.fr/cgi-bin/qcat?J/A+A/(vol)/(page)}
}

\author{T.V. Mishenina \inst{1,2} \and
M. Pignatari\inst{3}\and
S.A. Korotin \inst{1} \and
C. Soubiran \inst{2} \and
C. Charbonnel \inst{4,5} \and
F.-K. Thielemann\inst{3}\and
T.I. Gorbaneva\inst{1} \and
N.Yu. Basak\inst{1}
}

\institute{
Astronomical Observatory, Odessa National University,
 and Isaac Newton Institute of Chile, Odessa Branch,
T.G.Shevchenko Park, Odessa 65014 Ukraine, \\email:{tamar@deneb1.odessa.ua }
\and
Universit\'e de Bordeaux 1 - CNRS - Laboratoire d'Astrophysique de Bordeaux,
UMR 5804, BP 89, 33271 Floirac Cedex, France, \\
email:{Caroline.Soubiran@obs.u-bordeaux1.fr}
\and
Department of Physics, University of Basel, Klingelbergstrabe 82, 4056 Basel,
Switzerland \\email:{marco.pignatari@unibas.ch}
\and
Geneva Observatory, University of Geneva, 1290 Versoix, Switzerland
\and
IRAP, UMR 5277 CNRS and Universit\'e de Toulouse, 31400 Toulouse, France}

\date{}
\titlerunning{Abundances of neutron-capture elements in stars of the galactic disk substructures}
\authorrunning{Mishenina et al.}

\abstract
{}
{  The aim of this work is to present and discuss the observations of
the iron peak (Fe, Ni) and neutron-capture
element (Y, Zr, Ba, La, Ce, Nd, Sm, and Eu) abundances for 276 FGK dwarfs,
located in the galactic disk with metallicity -1 $<$ [Fe/H] $<$ +0.3.}
{Atmospheric parameters and chemical composition of the studied stars were
determined from an high resolution, high signal-to-noise echelle spectra
obtained with the echelle spectrograph ELODIE at the
Observatoire de Haute-Provence (France).
Effective temperatures were estimated by the line depth ratio method and from
the $\rm H_{\alpha}$ line-wing fitting. Surface gravities ($\log~g$) were
determined by parallaxes and the ionization balance of iron.
Abundance determinations were carried out using the LTE approach,
taking  the hyperfine structure for Eu into account, and the abundance of Ba was
computed under the NLTE approximation.}
{We are able to assign most of the stars in our sample to the substructures of
the Galaxy thick disk, thin disk, or Hercules stream according to their
kinematics.
The classification of 27 stars is uncertain.
For most of the stars in the sample, the abundances of neutron-capture
elements have not been measured earlier.
For all of them, we provide the chemical composition
and  discuss the contribution from different nucleosynthesis processes.
}
{

The [Ni/Fe] ratio shows a flat value close to the solar one
for the whole metallicity range, with a small scatter, pointing to a
nearly solar Ni/Fe ratio for the ejecta of both core-collapse SN and SNIa.
The increase in the [Ni/Fe] for metallicity higher than solar
is confirmed, and it is due to the metallicity dependence of $^{56}$Ni ejecta
from SNIa.
Under large uncertainty in the age determination of observed stars, we verified
that there is  a large dispersion in the AMR in the thin disk,
and no clear trend as in the thick disk. That may be one of the main
reasons for the dispersion, observed for the $s$-process elements in the thin disk
(e.g., Ba and La), whereas much narrower dispersion can be seen for $r$-process
elements (e.g., Eu).
Within the current uncertainties, we do not see a clear
decreasing trend of [Ba/Fe] or [La/Fe] with metallicity
in the thin disk, except maybe for
super-solar metallicities.
We cannot confirm an increase in the
mentioned ratios with decreasing stellar age.
}

\keywords{
Nucleosynthesis --
Stars: abundances --
Stars: late-type --
Galaxy: evolution
}

\maketitle

\section{Introduction}

The chemical abundances that we observe today in the solar system and in stars
provide fundamental constraints in our understanding of the stellar evolution and
nucleosynthesis, of the galactic formation and chemical evolution, and of the near-field
cosmology observations.
In particular, despite their low abundance, elements heavier than iron have been
observed over a large sample of stars, spreading over Gyrs in age and over
orders of magnitude in metal content, in our Galaxy, and in more distant objects
(e.g., Sneden et al. \cite{sneden08}; Tolstoy et al. \cite{tolstoy09} and reference
therein). Most of their abundances are produced by neutron capture processes:
the slow neutron capture (or the $s$-process) and the rapid neutron capture process
(or the $r$-process) (Burbidge et al. \cite{bur57}; Cameron \cite{cameron:57}).

The first phenomenological studies introduced three different components for the
$s$-process (e.g., K\"{a}ppeler et al.  \cite{kap89}): the weak $s$-process
component, producing most of the $s$-species between Fe and Sr;
the main $s$-process component for abundances between Sr and Pb; and the strong
$s$-process component, responsible for the production of 50 \% of the solar
\element[][208]{Pb}.
Full nucleosynthesis simulations based on realistic stellar models mostly
confirm this scenario. The bulk of the weak $s$-process component is made in
massive stars, triggered by the activation of the
$^{22}$Ne($\alpha$,n)$^{25}$Mg reaction in the convective He-burning core and in
the following convective C-burning shell (e.g., Rauscher et al.
\cite{rauscher:02}; The et al. \cite{the:07}; Pignatari et al.
\cite{pignatari:10}).
The main and strong $s$-process components are produced in the AGB stars at the solar-like
and low metallicity, respectively (e.g., Gallino et al. \cite{gallino:98};
Bisterzo et al. \cite{bisterzo:11}).
Most of the neutrons are provided by the $^{13}$C($\alpha$,n)$^{16}$O reaction
in the radiative $^{13}$C-pocket formed right after the third dredged-up  event
Straniero et al. \cite{st03}, with a relevant contribution from the partial activation of the
$^{22}$Ne($\alpha$,n)$^{25}$Mg in the convective thermal pulse
(Gallino et al. \cite{gallino:98}).
In particular, the Galactic chemical evolution (GCE) computations confirmed that different
generations of the AGB stars have to be taken into account to properly study the
$s$-process distribution of the solar system (Travaglio et al. \cite{tr04};
Serminato et al. \cite{se09}).

The origin of heavy $r$-process elements remains uncertain.
At least three sources have been proposed, namely:
1) the neutrino-induced winds from the core-collapse supernovae (e.g., Woosley et
al. \cite{wo94});
2) the enriched neutron-rich matter from merging neutron stars
(e.g., Freiburghaus et al. \cite{fr99}), and/or neutron-star/black
hole mergers (Surman et al. \cite{su08});
3) polar jets from rotating MHD core-collapse supernova
(Nishimura et al. \cite{nishimura:06}).
For a more detailed description of different $r$-process scenarios, we refer
to Thielemann et al. (\cite{thielemann:11}). For the recent results related to the
$r$-process, we refer to Winteler et al. (\cite{winteler:12}) and
Korobkin et al. (\cite{korobkin:12}).

The surface abundances of the FGK dwarf stars do not show any noticeable change due
to the stellar evolution, reflecting their pristine chemical composition.
Stars in the range of metallicity -1 $<$ [Fe/H] $<$ +0.3 dex were born
in the interstellar medium, which had been enriched by several generations of
stars.
Those stars do not have homogeneous kinematics, and  were possibly formed in
different galactic subsystems or were captured from outside of the Galaxy
(Feltzing et al. \cite{fe09}; Marsakov \& Borkova \cite{ma05};
Klochkova et al. \cite{kl11}). The analysis of their main features may
be a powerful tool for tracing the formation of the galactic substructure and the
galactic chemical enrichment.

According to Gilmore \& Reid (\cite{gi83}),
the stellar distribution from the galactic plane towards
the southen galactic pole is described by two exponentials with different height
and density, introducing the concept of the $thick$ $disk$.
In the past decades, it has been shown that the stars of the thick disk and
the $thin$ $disk$ have different kinematics, ages, and the abundances of
$\alpha$-elements.
The behavior of the neutron-capture elements relative to metallicity
and the study of the $s$- and $r$-process contributions
%to the enrichment of their main elements
for those two substructures
were presented by different authors
(Prochaska et al. \cite{pr00}; Mashonkina \& Gehren \cite{ma00}, \cite{ma01};
Mashonkina et al. \cite{ma04}; Alende Prieto et al. \cite{ar04};
Brewer \& Carney \cite{br06};  Bensby et al. \cite{be05};
Reddy et al. \cite{re06}; Nissen \& Shuster \cite{ni08}; Felting et al.
\cite{fe09}; etc.).

The galactic disk also includes stellar clusters and  groups of
stars with their peculiar motion. Among them  is
the Hercules stream, first investigated in detail by Eggen (\cite{eg58}),
and then by Fux (\cite{fux01}), Famaey et al. (\cite{fa05}).  According to Famaey et al.(\cite{fa05})
the Hercules stream has dynamical origin and can be made of stars with very different birth locations and ages.
Thus, no coherent chemical trend is expected for them, and on the contrary, a large dispersion
of their properties should be observed. Kinematically,  the Hercules stream is somewhere between
the thin disk and the thick disk and complicates their separation. When performing the deconvolution of
the thin and the thick disks on kinematical criteria, it is important to consider that group of stars
to obtain pure samples.
In the papers by Soubiran \& Girard (\cite{so05}), Bensby et al. (\cite{be07}), and 
Pakhomov (\cite{pakh11})  the chemical composition and kinematics
of the stars was studied. The work by Soubiran \& Girard (\cite{so05})
 shows that those stars are chemically closer to the thin disk, but
according to Bensby et al. (\cite{be07}) and Pakhomov (\cite{pakh11}),
the stars represent a mixture of stars of the thick and thin disks.

The accuracy of determining of the element abundances, parameters of the thick
and thin disks, stellar ages, as well as the criteria for the star's assignments to
different substructures, play  important roles in interpreting the observational
data.
One of the reliable ways of investigating the formation and evolution
of different structures of the disk is to study  the chemical
composition of the stars belonging to those structures in detail.
Thus, theoretical stellar abundance yields can be used to investigate how the
Galactic substructures and the Galaxy as a whole have evolved up to the present
stage.

The goal of the present work is to provide and analyze the abundance signature
 of Ni and the neutron-capture elements for 276 dwarfs
in the solar neighborhood in the metallicity range -1 $<$ [Fe/H] $<$ 0.3.
The paper is structured as follows. The observations, processing, and selection
of stars are described in Section 2.
The  atmospheric parameters, the abundance determinations for
Y, Zr, Ba, La, Ce, Nd, Sm, and Eu, and the error analysis are
presented in Sections 3, 4, and 5 , respectively.
In Section 6, the final results are given and discussed.
Conclusions are drawn in Section 7.

\section{Observations,  processing, and selection of stars}
\label{sec: obs_proc_sel}
This study is a part of a wider project, in which the
metallicity distribution and behavior of  some elements  in the local thin disk
are investigated
for a complete sample of the G and K dwarfs and giants in the solar neighborhood
(Mishenina et al. \cite{mi04}, \cite{mi06}, \cite{mi08}).  The transition between the thin and
thick disks in kinematics and abundance trends
(Mishenina et al. \cite{mi04}), the vertical distribution of the Galactic disk, the measurement
of its surface mass density  (Bienaym\'{e} et al. \cite{bien06}), its age metallicity relation (AMR)
and  age velocity relation   (AVR)
(Soubiran et al. \cite{soub08}), as well as the construction
of the chemical evolution model (Nikityuk \& Mishenina \cite{nikmish06})  -
all of those were studied on the basis of the collection of the above -
mentioned data. In the present paper, we have considered
the neutron-capture elements
for the entire set of dwarfs in our example.
Following the approach in which the kinematical and chemical
information is combined, the G and K dwarfs within 25 pc
from the Sun were selected from the
Hipparcos catalog: for this study we analyzed
the spectra of 276 stars (F-G-K V)
with metallicities in the range -1 $<$ [Fe/H]$ <$ +0.3.
The spectra were obtained in the wavelength region
${\lambda~4400-6800\,\AA}$ and with the signal-to-noise ratios ($S/N$) of about 100-350,
using  the 1.93 m telescope at the Observatoire de
Haute-Provence (OHP, France) equipped  with the echelle-spectrograph ELODIE
(Baranne et al. \cite{ba96}), which provides the resolving  power of R = 42000.

The complex preprocessing of the images is available online, and it
allows the spectroscopic data  to be obtained in digital form with the
radial velocity $ V_{r} $ (Katz et al. \cite{ka98}) immediately after the
exposure. The spectra have been treated to correct the blaze efficiency and
cosmic and telluric lines following Katz et al. (\cite{ka98}).
The subsequent processing of
the studied spectra (including the continuous spectrum level set up, the
development of the dispersion curve, the measurement of equivalent widths,
etc.) was performed
by us with the DECH20 software package (Galazutdinov \cite{ga92}). The
equivalent widths (EWs) of the spectral lines were measured by the Gaussian profile
fitting.

Specific features were used to select the stars that belong to the thin and thick
disks or to other galactic substructures. Those include the spatial
distribution and local density of stars, space velocity, metallicity, and age.
Since the velocity distribution is well studied for those substructures, we
applied the kinematic approach for separating  the stars, determining the
probability that each star is a member of the thin or thick disks or of the
Hercules stream, based on its spatial velocity, kinematic parameters of the
disks, and the stream, as well as the percentage of the stars of the studied
sample in each disk, and in the Hercules stream. The probability of each star
belonging to the thin or thick disks
or to the Hercules stream was computed using the (U, V, W) velocities
by the method of Soubiran \& Girard (\cite{so05}) with parallaxes  and
the proper motion from van Leeuwen (\cite{le07}).
In our  sample (276 stars), 21 stars belong to the thick disk, 212 to the
thin disk, 16   to the Hercules stream, and 27 are unclassified.
The  probabilities stars are classified
by their belonging to the thick and thin disks or to the Hercules
stream are presented in Table \ref{t_abund}.

\section{Atmospheric parameters}

The atmospheric parameters for the target stars were determined in our previous
studies (Mishenina \& Kovtyukh \cite{mi01}; Mishenina et al. \cite{mi04};
Mishenina et al. \cite{mi08}).
The effective temperatures T$_{\rm eff}$ were estimated by calibrating
the ratio of the central depths of the lines with different potentials of the
lower level developed by Kovthykh et al. (\cite{ko04}).
For metal-poor stars the effective temperatures were determined from the
$\rm H_{\alpha}$ line-wing fitting (Mishenina \& Kovthyukh \cite{mi01}).
The surface gravities log g were computed by two methods.
For the stars with $\rm T_{eff}$ higher than 5000 K by the iron ionization
balance and the parallax ,  the parallax was the only method
used for the cooler stars. The microturbulent velocity V$_{\rm t}$ was derived considering that the
iron abundance log A(Fe) obtained from the given \ion{Fe}{I} line is not
correlated with the  EW of that line.

The adopted value of the metallicity [Fe/H] was calculated using the iron
abundance obtained from the \ion{Fe}{I} lines. As is known, the lines of neutral
iron are influenced by deviations from the local thermodynamic equilibrium (LTE),
and  therefore, that affects the iron abundances,  which are determined from
those lines. However, in the temperature and metallicity ranges of our
target stars, the NLTE corrections do not exceed 0.1 dex
(Mashonkina et al. \cite{ma11}).

\begin{figure}
\resizebox  {8.4cm}{3.7cm}
{\includegraphics{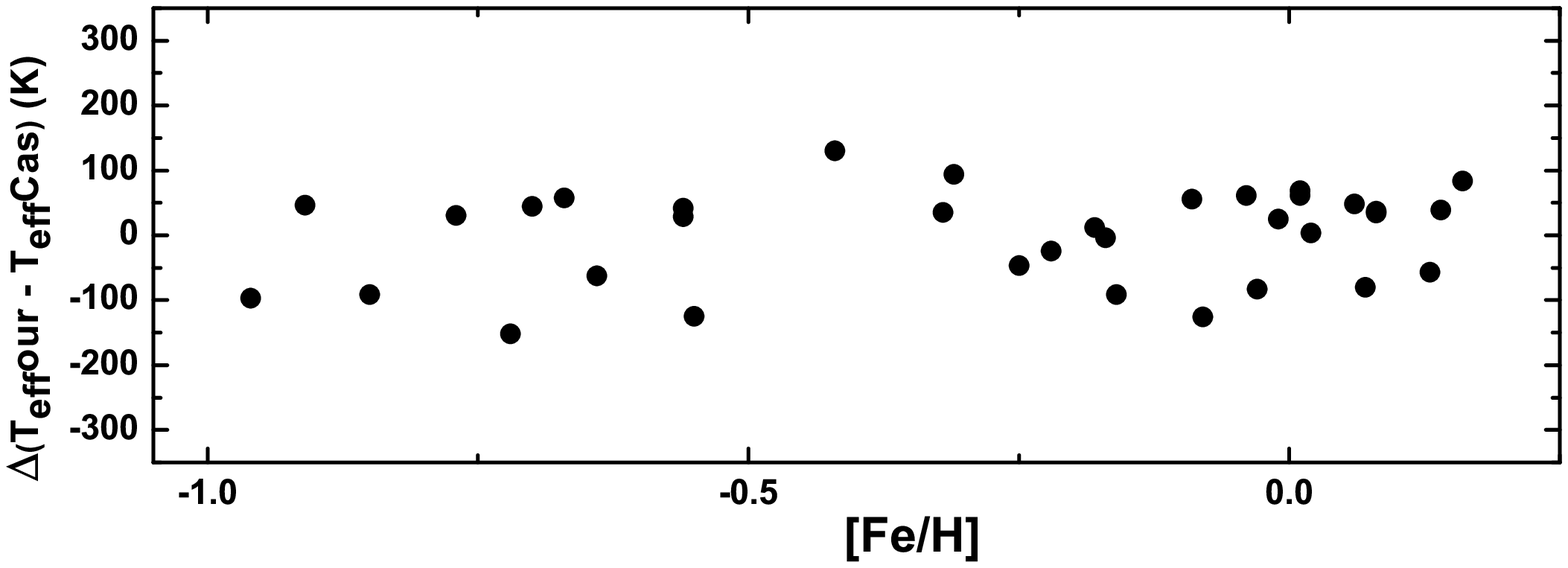}}
\resizebox  {8.4cm}{3.7cm}
{\includegraphics{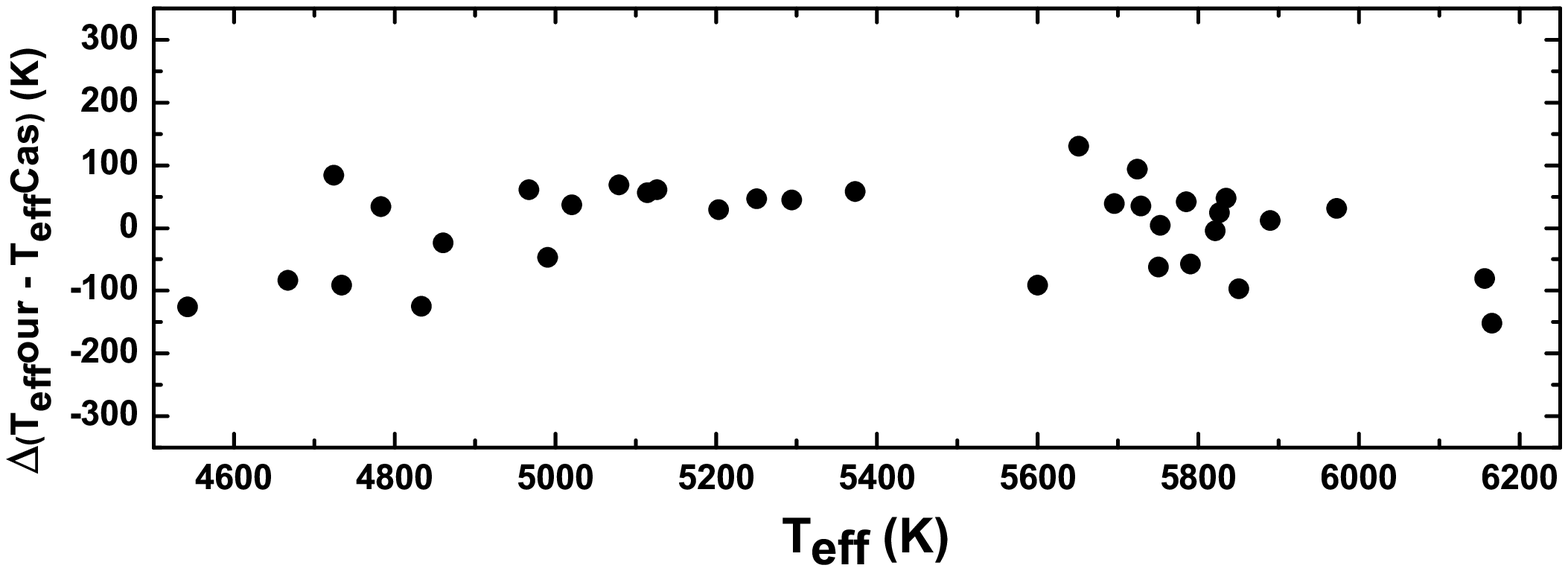}}
\caption[]{
Difference between the effective temperatures obtained in this work and those
reported  in Casagrande et al. (\cite{cas10}) for 33 stars in common
($\Delta T_{eff}$ vs. $T_{eff}$ and $\Delta T_{eff}$ vs. [Fe/H])
}
\label {dT_Cas}
\end{figure}

\begin{figure}
\resizebox  {8.4cm}{3.7cm}
{\includegraphics{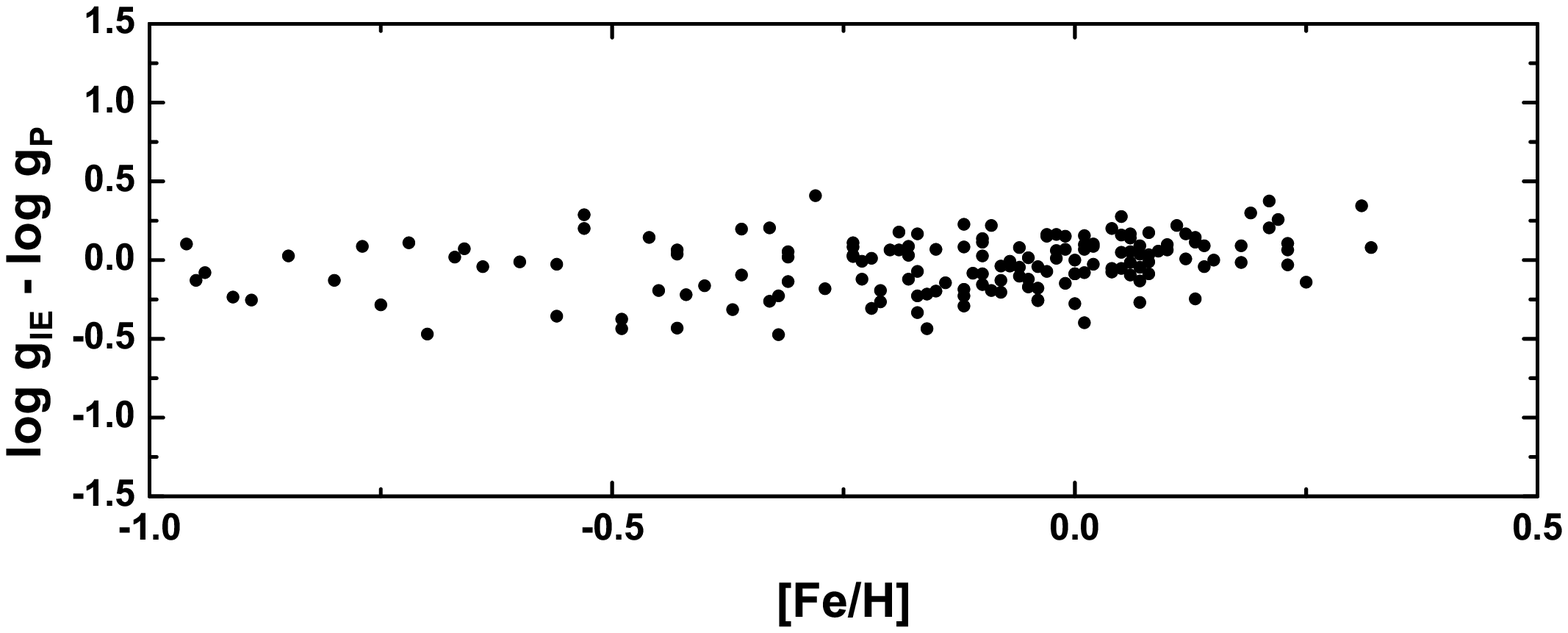}}
\resizebox  {8.4cm}{3.7cm}
{\includegraphics{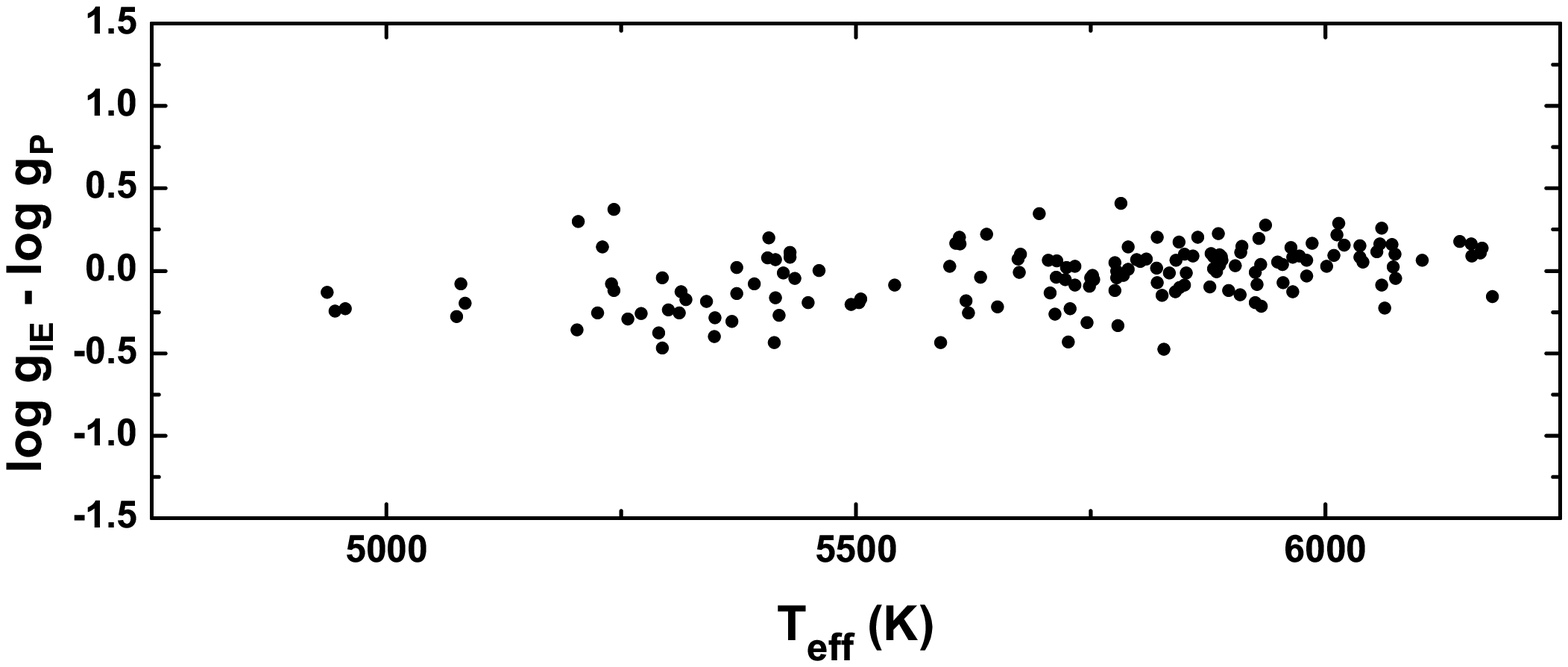}}
\caption[]{
Dependence of log $g_{IE}$ - log $g_{P}$ upon $T_{eff}$ and [Fe/H] for our sample stars.
}
\label {gi_TFe}
\end{figure}

The comparison of the determined atmospheric parameters to the results obtained
by other authors is presented in our previous studies (Mishenina et al.
\cite{mi04}, \cite{mi08}).
To additionally check our  $T_{eff}$ determinations,   we compared
the values $T_{eff}$ with  those for the recent IRFM observations  by Casagrande et al. (\cite{cas10}).
The mean difference $< \Delta (T_{eff}our - T_{eff}Cas)>$ = -6 $\pm$ 80 K.
The difference $< \Delta (T_{eff}our - T_{eff}Cas)>$
as a function of $T_{eff}$ and [Fe/H] is shown in Fig. \ref{dT_Cas}.
%The dependence of  $T_{eff}$ on [Fe/H] for our target stars is
%presented in Fig. \ref{Fe_T}.
For most stars in our study, we used the values log $g_{IE}$,  determined by
the iron ionization balance.
The dependencies  of  log $g_{IE}$ - log $g_{P}$ vs.
$T_{eff}$ and [Fe/H] are presented in Fig. \ref{gi_TFe}.
In both Figs. \ref{dT_Cas} and  \ref{gi_TFe} there are no systematic
differences. The average differences $<$(log $g_{IE}$ -
log $g_{P}$)$>$ = -0.03 $\pm$ 0.17 for our target stars with
$T_{eff} > $ 4800 K.

%\begin{figure}
%%\resizebox{\hsize}{!}
%\resizebox  {8.4cm}{3.7cm}
%{\includegraphics{Fe_T.eps}}
%\caption[]{
%Dependences  of [Fe/H] upon  $T_{eff}$ for our sample stars
%}
%\label {Fe_T}
%\end{figure}

In this paper, we also compare  our data with the results of studies
performed during recent years (Bensby et al. \cite{be05}; Reddy et al.
\cite{re06}; Mashonkina et al. \cite{ma01}; Peloso et al. \cite{pe05}),
in which  the n-capture element abundances were also determined
(see Table \ref{ncap}). As is evident from Table \ref{ncap} ,
the external accuracy of the effective temperature $\rm T_{eff} $
is within
$ \Delta \rm T_{eff} $ = $ \pm $ 100 K, the surface gravity
$\log~g$ - $ \Delta $ $ \log~g $ = $ \pm $ 0.2 dex.

\begin{table*}
\begin{center}
\caption[]{Comparison of our parameters and abundance determinations with the
results of other authors.}
\label{ncap}
\begin{tabular}{cccccccccccccccc}
\hline
($\Delta$)& T$_{\rm eff}$ & $\log~g$ & [Fe/H] & [Y/Fe] & n & [Zr/Fe] & n & [Ce/Fe] &
n & [Nd/Fe] & n & [Sm/Fe] & n & [Eu/Fe] & n \\
\hline
Bensby et al.& 19 & -0.09 & -0.03 & 0.02 & 9 & & & & & & & & & 0.01 & 7 \\
\cite{be05} &$\pm$76 & $\pm$0.19 & $\pm$0.08 & $\pm$0.10 & & & & & & & & & & $\pm$0.09 & \\
& & & & & & & & & & & & & & & \\
Reddy et al. & 92 & -0.20 & -0.01 & -0.06 & 8 & & & -0.09 & 7 & -0.11& & & &
0.04 & 7 \\
\cite{re06} &$\pm$29 & $\pm$0.24 & $\pm$0.04 & $\pm$0.12 & & & & $\pm$0.09 & & $\pm$0.12& & & & $\pm$0.11 & \\
& & & & & & & & & & & & & & & \\
Mashonkina & 19 & -0.08 & 0.02 & -0.10 & 16 & -0.02& 11 & -0.11 & 15 & -0.08&
15& & & 0.00 & 13 \\
et al. \cite{ma01} &$\pm$64 & $\pm$0.22 & $\pm$0.07 & $\pm$0.07 & & $\pm$0.10& & $\pm$0.07 & & $\pm$0.07& & &
& $\pm$0.10 & \\
& & & & & & & & & & & & & & & \\
Peloso et al.& 46 & -0.02 & -0.02 & & 5 & & & 0.01 & 5 & -0.11& 5 & 0.08& 5& &
\\
\cite{pe05} &$\pm$66 & $\pm$0.11 & $\pm$0.06 & & & & & $\pm$0.06 & & $\pm$0.13& & $\pm$0.21& & & \\
\hline
\end{tabular}
\end{center}
\end{table*}

\section{Determination of the chemical composition}

The abundances of the investigated elements Y, Zr, Ba, La, Ce, Nd, Sm, and Eu
are determined for 276 F-G-K dwarfs under LTE approximation using the atmosphere
models by Kurucz (\cite{ku93}). The choice of model for each star was made by
means of standard interpolation for $\rm T_{eff}$ and $ \log~g $.
The determination of the Y, Zr, La, Ce, Nd, and Sm abundances was carried out using
the EWs and code WIDH9 of Kurucz .

\begin{figure}
%\resizebox{\hsize}{!}
\resizebox  {8.4cm}{5.7cm}
{\includegraphics{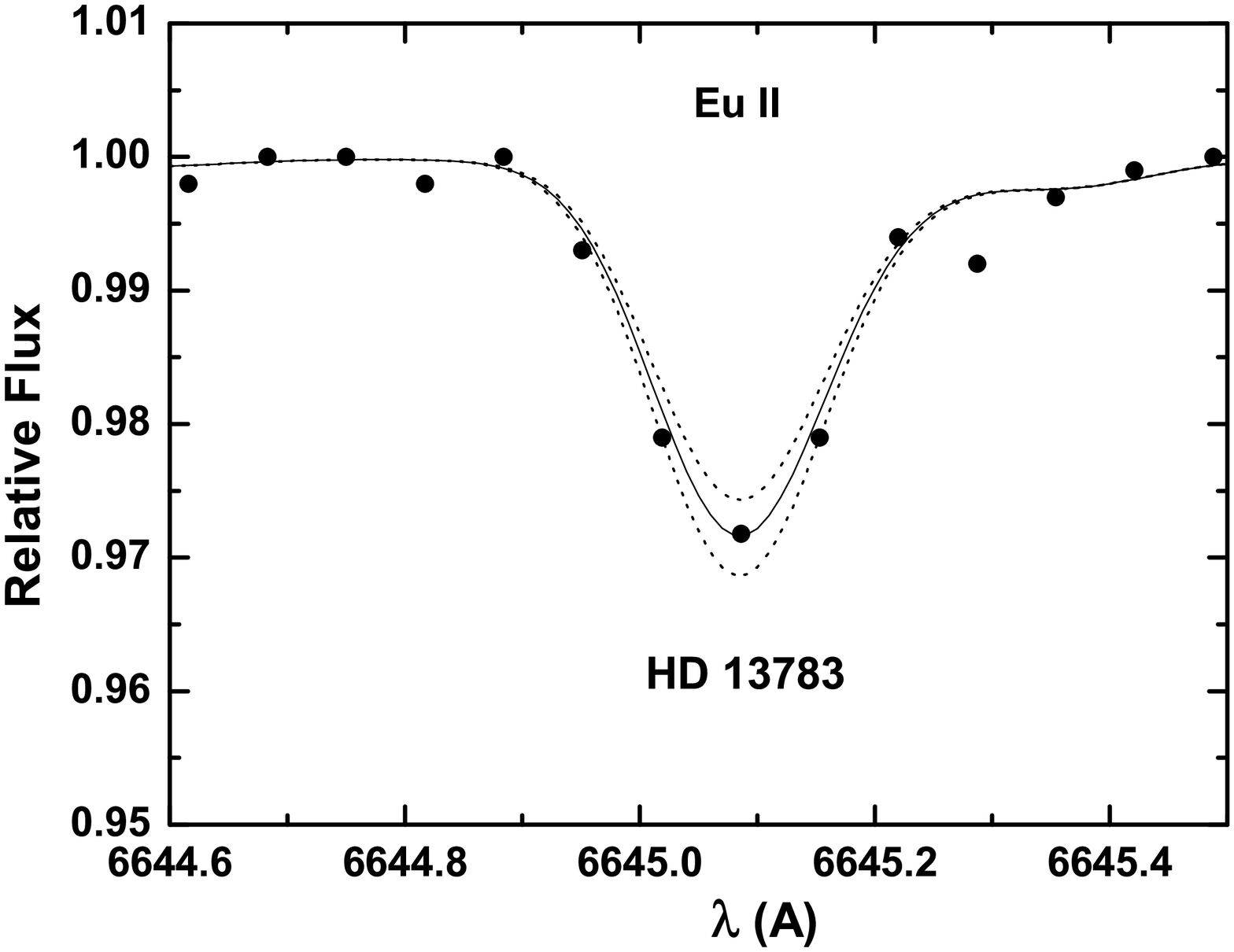}}
\caption[]{
The spectrum synthesis fitting of the Eu line to the observed profiles.
The change in Eu abundance is 0.05 dex.
}
\label {pr_Eu}
\end{figure}

\begin{figure*}
\resizebox{\hsize}{!}
{\includegraphics{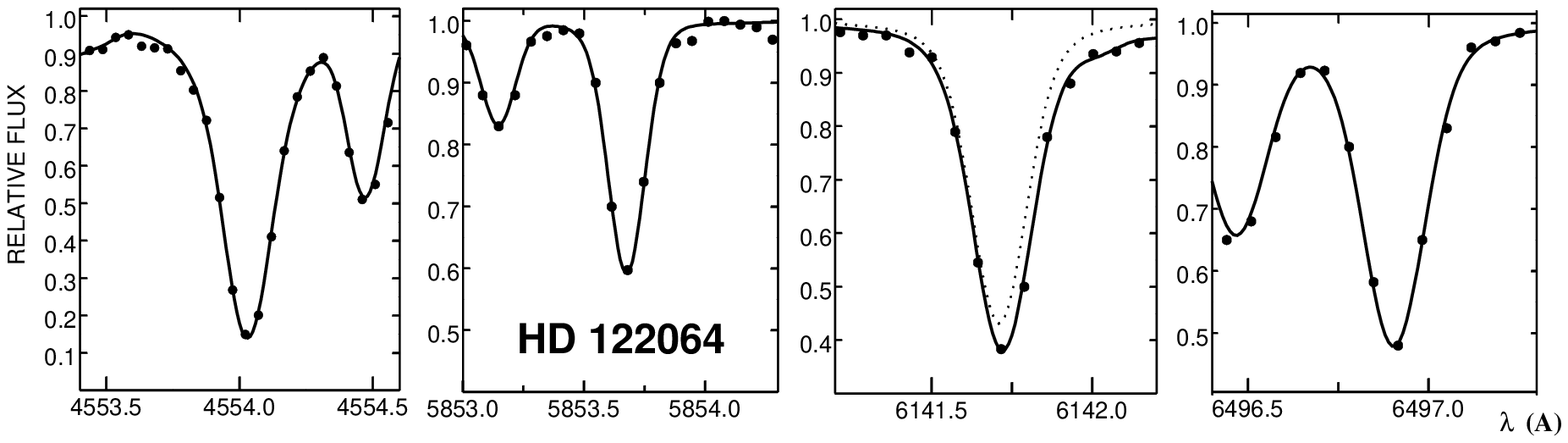}}
\resizebox{\hsize}{!}
{\includegraphics{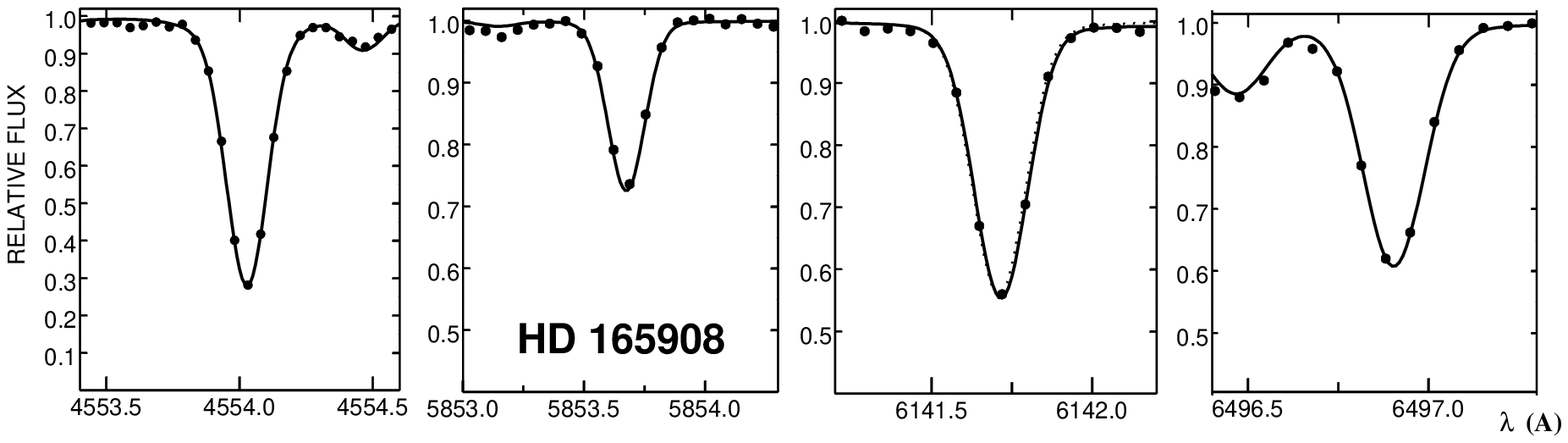}}
\caption[]{Spectrum synthesis fitting of observed profiles of Ba lines. 
Line 6141 \AA~ is blended with the iron line (the dotted line is the barium 
line profile).Computations  are presented for the same barium
abundance for  four lines in each star.} 
\label {pr_Ba}
\end{figure*}

The Eu abundance was determined using a new version of the
STARSP synthetic spectrum code (Tsymbal \cite{ts96}), with
the lines of \ion{Eu}{II} 6645 \AA~ and taking  the hyperfine
structure into account (Mashonkina  \cite{mas00}).
The spectrum synthesis fitting of the Eu and Ba lines to the observed profiles
are shown in Figs. \ref{pr_Eu},\ref{pr_Ba}.
The abundance of the investigated elements were determined  by
differential analysis relative to the Sun. Solar abundances
of Y, Zr, La, Ce, Nd, and Sm  were calculated with the solar EWs,
measured in the Moon and asteroids spectra, also obtained  with the ELODIE
spectrograph, and with the oscillator strengths log\,gf from
Kovtyukh \& Andrievsky (\cite{ko99}).
The La and Sm lines are so weak that it is possible to neglect the hyperfine splitting (HFS).
The data  for the neutron-capture element
lines (including the solar EW) are given in Table \ref{t:nlines}.

\begin {table}
\caption {Parameters of the neutron-capture elements lines and the solar equivalent widths .}
\label{t:nlines}
\begin{tabular}{lcccc}
\hline
$\rm \lambda (\AA)$ & Element& log gf & $E_{low}$&  EW (m\AA)\\
\hline
 4883.68  & \ion{Y}{ii}  &  0.02  &  1.08  &     60 \\
 4900.11  & \ion{Y}{ii}  & -0.13  &  1.03  &     55 \\
 4982.13  & \ion{Y}{ii}  & -1.26  &  1.03  &     15 \\
 5087.42  & \ion{Y}{ii}  & -0.26  &  1.08  &     50 \\
 5119.11  & \ion{Y}{ii}  & -1.29  &  0.99  &     15 \\
 5200.41  & \ion{Y}{ii}  & -0.63  &  0.99  &     39 \\
 5289.82  & \ion{Y}{ii}  & -1.83  &  1.03  &    4.7 \\
 5402.77  & \ion{Y}{ii}  & -0.55  &  1.84  &     14 \\
 5728.89  & \ion{Y}{ii}  & -1.16  &  1.84  &    4.5 \\
 5112.28  & \ion{Zr}{ii} & -0.85  &  1.66  &    8.4 \\
 5350.09  & \ion{Zr}{ii} & -0.89  &  1.76  &    6.5 \\
 5350.36  & \ion{Zr}{ii} & -0.80  &  1.81  &    6.5 \\
 4662.51  & \ion{La}{ii} & -1.24  &  0.00  &    8.4 \\
 4748.74  & \ion{La}{ii} & -0.54  &  0.92  &    5.7 \\
 5123.01  & \ion{La}{ii} & -0.85  &  0.32  &   10.5 \\
 6320.41  & \ion{La}{ii} & -1.33  &  0.17  &      6 \\
 4479.38  & \ion{Ce}{ii} &  0.42  &  0.56  &     24 \\
 4486.91  & \ion{Ce}{ii} & -0.12  &  0.29  &   16.5 \\
 4560.28  & \ion{Ce}{ii} &  0.47  &  0.90  &     16 \\
 4562.37  & \ion{Ce}{ii} &  0.28  &  0.47  &     23 \\
 4773.96  & \ion{Ce}{ii} &  0.30  &  0.92  &     11 \\
 5274.24  & \ion{Ce}{ii} &  0.31  &  1.04  &   10.5 \\
 5610.25  & \ion{Ce}{ii} &  0.12  &  1.05  &      7 \\
 4462.92  & \ion{Nd}{ii} &  0.00  &  0.55  &     19 \\
 4811.35  & \ion{Nd}{ii} & -0.89  &  0.06  &    9.6 \\
 4989.95  & \ion{Nd}{ii} & -0.36  &  0.63  &    8.5 \\
 5089.83  & \ion{Nd}{ii} & -1.09  &  0.20  &    4.8 \\
 5092.80  & \ion{Nd}{ii} & -0.66  &  0.38  &      8 \\
 5130.60  & \ion{Nd}{ii} &  0.58  &  1.30  &   15.8 \\
 5234.21  & \ion{Nd}{ii} & -0.38  &  0.55  &   10.5 \\
 5293.17  & \ion{Nd}{ii} & -0.10  &  0.82  &   10.7 \\
 5319.82  & \ion{Nd}{ii} & -0.34  &  0.55  &   11.5 \\
 4467.34  & \ion{Sm}{ii} &  0.19  &  0.65  &   13.5 \\
 4577.69  & \ion{Sm}{ii} & -0.61  &  0.25  &    5.7 \\
 4791.60  & \ion{Sm}{ii} & -0.97  &  0.10  &      4 \\
 4815.82  & \ion{Sm}{ii} & -0.89  &  0.18  &    4.8 \\
\hline
\end{tabular}
\end{table}

The solar Eu abundance was also
found by STARSP (Tsymbal \cite{ts96}) via the line of Eu II in the Moon and
asteroid spectra.
The obtained solar abundances are the following: log A(Y) = 2.24, log A( Zr) =
2.60, log A(La) = 1.22, log A(Ce) = 1.55, log A(Nd) = 1.50, log A(Sm) = 1.01,
log A(Eu) = 0.51, where log A(H) = 12.
The differential approach for determining the relative abundance  of the
element A to the abundance of iron relative to the solar ratio [A/Fe] was
applied to reduce the influence of the spectrograph characteristics, and the
errors are due to the uncertainties in the oscillator strengths and the deviations
from the LTE upon the abundance definition.

For the barium abundance determination we used four lines of \ion{Ba}{II}
(4554, 5853, 6141, and 6496 \AA) under the NLTE approximation.
The NLTE profiles of the barium lines were computed using a modified version
of the MULTI code (Carlsson \cite{ca86}). The modifications are described in
Korotin et al. (\cite{ko99}).
Our barium model contains 31 levels of \ion{Ba}{I}, 101 levels of \ion{Ba}{II}
with n $<$ 50, and the ground level of the \ion{Ba}{III} ion. In the analysis
we included 91 bound-bound transitions.

The odd barium isotopes have hyperfine splitting of their levels
and thus several HFS components for each line (Rutten \cite{rut78}).
Therefore, lines 4554 \AA~ and 6496 \AA~ were fitted by adopting the
even-to-odd abundance ratio of 82:18 (Cameron \cite{cam82}).
The HFS for lines 5853~\AA~ and 6141~\AA~ is not  significant.

Some uncertainty of the NLTE analysis of the barium spectrum is caused by the
lack of information on the photoionization cross-sections for different levels.
We used the results of the scaled Thomas-Fermi method application (Hofsaess
\cite{ho79}).

The effective collision strengths of electron excitation for the transitions
between the first levels ($\rm 6s^2S$, $\rm 5d^2D$ and $\rm 6p^2P^0$) were used
as in Schoening \& Butler (\cite{sc98}). The experimental cross-sections for
the transitions $\rm6s^2S-7s^2S$ and $\rm6s^2S-6d^2D$ were taken from
Crandall et al. (\cite{cr74}). The collisional rates for the transitions between
sublevels $\rm 5d^2D$, $\rm 6p^2P^0$ and $\rm7s^2S$, $\rm 6d^2D$, as well as
between
$\rm7s^2S$ and $\rm6d^2D$, were estimated with the help of the corresponding
formula by Sobelman et al. (\cite{so81}). For the rest of the allowed
transitions, we used the van Regemorter (\cite{re62}) formula while
the Allen (\cite{al73}) formula was used for the forbidden transitions. The
collisional ionization rate of the ground level of \ion{Ba}{II} was computed
with the appropriate formula from  Sobelman et al. (\cite{so81}).
The more detailed description of the atomic model
is given by Andrievsky et al. (\cite{an09}) and Korotin et al. (\cite{ko11}).
The adopted solar abundance of barium is equal to 2.17. The NLTE Ba abundances
for 174 stars have been determined earlier by Korotin et al. (\cite{ko11}),
for the other stars, the NLTE barium abundances are determined for the first time
in the present paper.

The values of the Mg, Si, and Ni abundance were taken from our studies
(Mishenina et al. \cite{mi04}; Mishenina et al. \cite{mi08}).
The Mg abundances were computed under the NLTE approximation. For the stars that were
investigated in Mishenina et al. (\cite{mi04}), the O and Ca abundances are
determined in the present work, and the O, Ca values for the other stars were taken
from the paper by (Mishenina et al. \cite{mi08}). The  O abundance 
was determined with a new version of the STARSP LTE spectral synthesis
code (Tsymbal \cite{ts96}). In this work we used the same line list as in
Mishenina et al. (\cite{mi08}) in the region of the [\ion{O}{I}] line
6300.3 \AA.

\section{Error analysis}

The total errors in abundances result mainly from the errors in the choice
of the parameters of the model atmospheres and in the EW
measurements (the Gaussian fitting, placement of the continuum) in the case of
Y, Zr, La, Ce, Nd, and Sm or in the fitting of the synthetic spectrum in the
case of Eu
and Ba.
Table \ref{t:mod_errors} lists the errors obtained when changing the
atmospheric parameters by $\Delta \rm T_{eff}$ = -100 K (column 1);
$\Delta \log~g $=+0.2 (column 2); $\Delta V_{t}$=+0.2 km/s (column 3);
and by assuming  uncertainty of $\pm$2 m\AA \, in the EW and 0.03 dex in the
calculated spectrum fitting. Those values were adopted taking  the
intrinsic accuracy into account for the atmospheric parameter determination, the processing
of the spectra, and the comparison of our parameter definition with those of
other authors. Those computations were performed for two stars with
different characteristics, and the total error is given in column 4.

\begin{table}
\begin{center}
\caption[]{
%For different elements, the variation in parameters by $\Delta \rm
%T_{eff}$ = -100 K , $\Delta \log~g $=+0.2, and $\Delta V_{t}$=+0.2 km/s is given
%in columns 1,2, and 3, respectively. In column 4 the total error in the same
%notation is given also assuming an uncertainty of
%$\pm$2 m\AA \, in the EW and 0.03 dex in the spectrum fitting.
Influence of stellar parameters on n-capture element abundance determination. }
\label{t:mod_errors}
\begin{tabular}{ccccc}
\hline
\multicolumn{4}{c}{HD3765 ($T_{eff}$=5079, $\log~g $=4.3, [Fe/H]=0.01)}&Total
error\\
     &   1      &  2    &  3    &   4  \\
\hline
Y    &   0	& -0.11 & 0.04  &   0.12 \\
Zr   &  -0.01	& -0.14 & 0.01  &  0.14 \\
Ba   &   0.02	& -0.04 & -0.07 &  0.10 \\
La   &  -0.01	& -0.15 & -0.01 &  0.15 \\
Ce   &   0.01	& -0.10 & 0.02  &  0.10 \\
Nd   &   0.03	& -0.13 & 0.02  &  0.14 \\
Sm   &   0.03	& -0.13 & -0.01 &  0.14 \\
Eu   &   0.01	& -0.09 & -0.01 &  0.10 \\
\hline
\hline
\\
\multicolumn{5}{c}{HD165401 ($T_{eff}$=5877, $\log~g $=4.3,  [Fe/H]=-0.36)}\\
	 &1	 &2	 &3	 &   4    \\
\hline
Y	 &0.02	 &-0.07	 &0.02  & 0.08 \\
Zr	 &0.02	 &-0.08	 &0.01  & 0.08 \\
Ba	 &0.06	 &-0.03	 &-0.06 & 0.09  \\
La	 &0.03	 &-0.12	 &0.00  & 0.13 \\
Ce	 &0.03	 &-0.07	 &0.01  & 0.08 \\
Nd	 &0.04	 &-0.08	 &0.01  & 0.09 \\
Sm	 &0.04	 &-0.09	 &0.00  & 0.10 \\
Eu	 &-0.01	 &-0.08	 &0.00  & 0.08 \\
\hline
\end{tabular}
\end{center}
\end{table}

As seen in Table \ref{t:mod_errors}, the total uncertainty
reaches 0.14 - 0.15 dex in the abundance determination  for the stars with low
temperatures, and its values are 0.08 - 0.13 dex for the hotter stars.
The standard deviation, obtained by comparing our [Fe/H] determinations to
those from other authors (Table \ref{ncap}), shows that
we are consistent with them at the level lower than 0.11 dex.

\section{Results and discussion}

\begin{figure}
%\resizebox  {8.4cm}{6.cm}{\includegraphics{O_Fe.eps}}
\resizebox  {8.4cm}{6.cm}
{\includegraphics{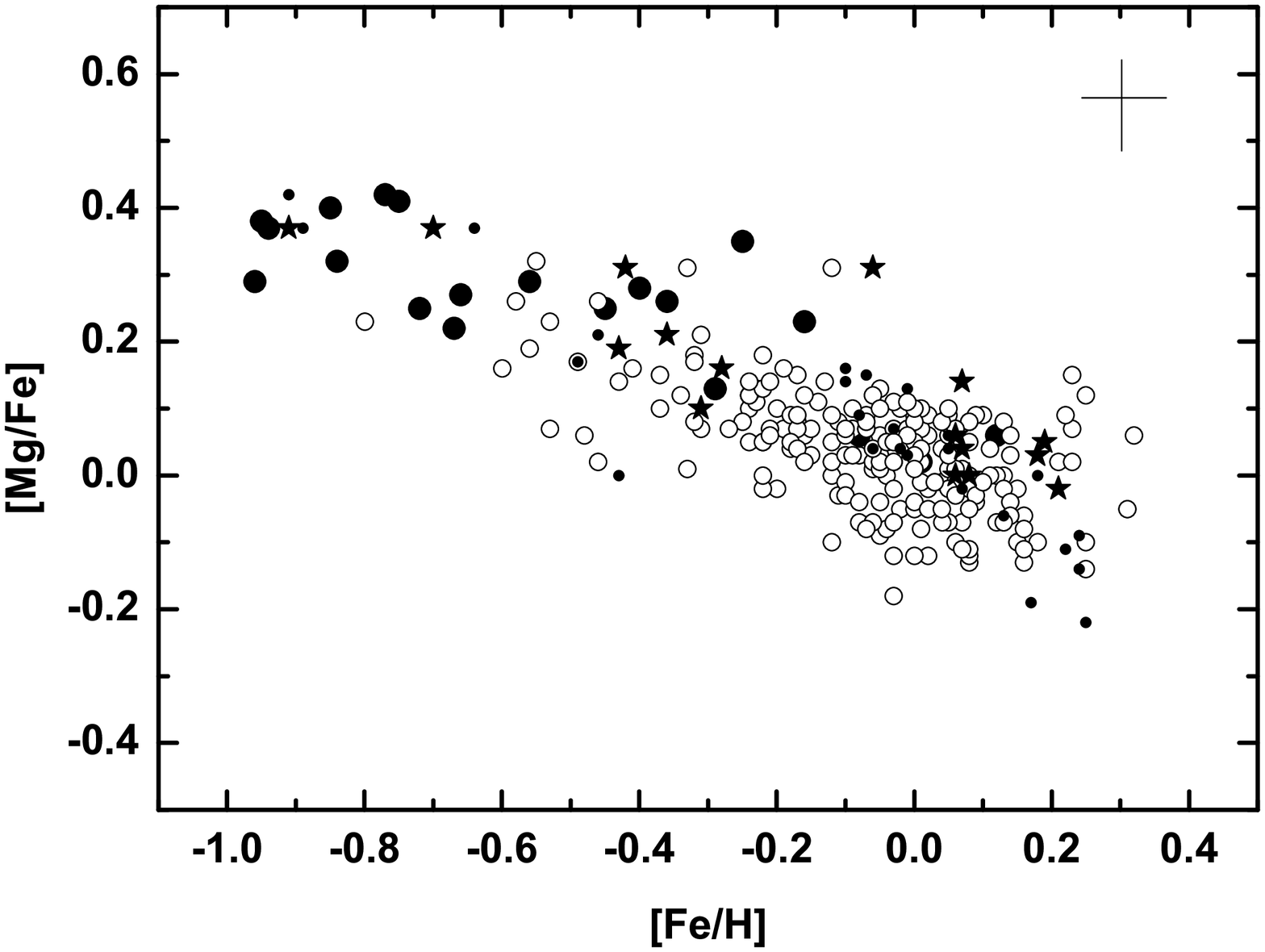}}
\resizebox  {8.4cm}{6.cm}
{\includegraphics{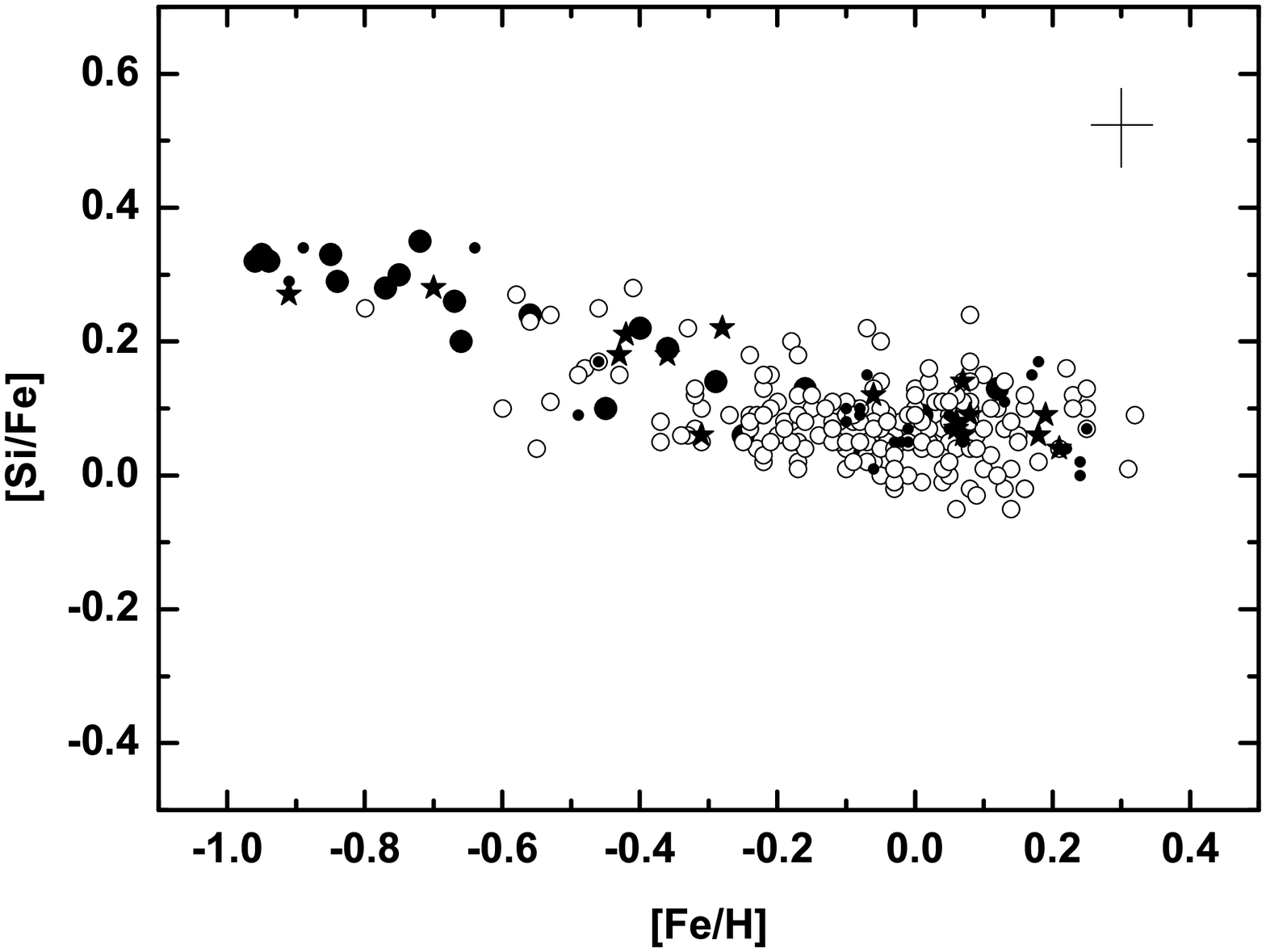}}
%\resizebox  {8.4cm}{6.cm}{\includegraphics{CaFe.eps}}
\caption[]{
Dependences of [Mg/Fe] and [Si/Fe] on [Fe/H] for the stars of the
thick disk (filled symbol), of the thin disk (open circle), the Hercules stream
(asterisks), and unclassified stars (small circle).
}
\label {alpha_fe}
\end{figure}

\begin{figure}
%\resizebox{\hsize}{!}
\resizebox  {8.4cm}{6.cm}
{\includegraphics{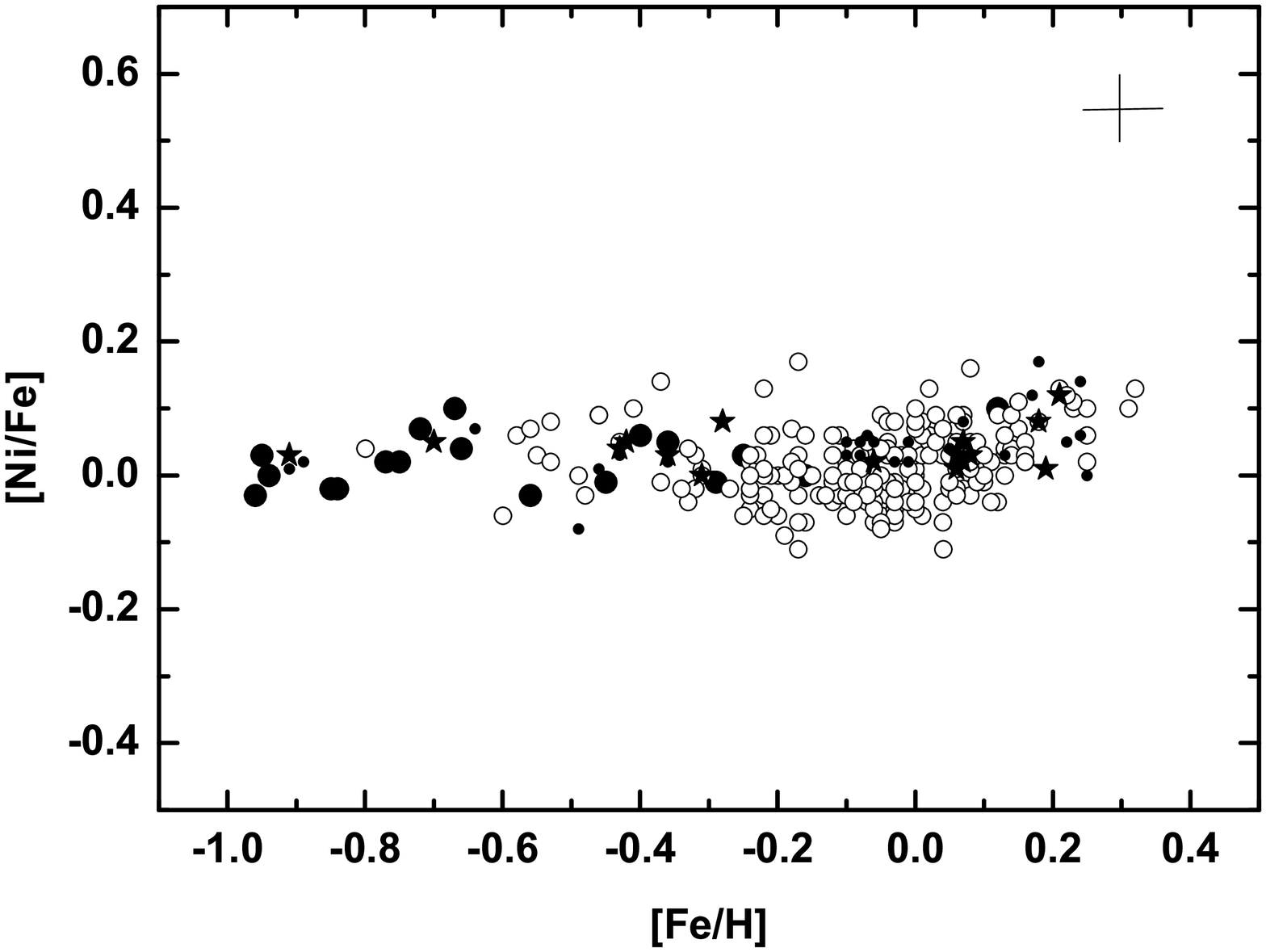}}
\caption[]{
Dependences of [Ni/Fe] on [Fe/H], the notation is the same as in
Fig.\ref{alpha_fe}
}
\label {ni_fe}
\end{figure}

\begin{figure}
%\resizebox{\hsize}{!}
\resizebox  {8.4cm}{6.cm}
{\includegraphics{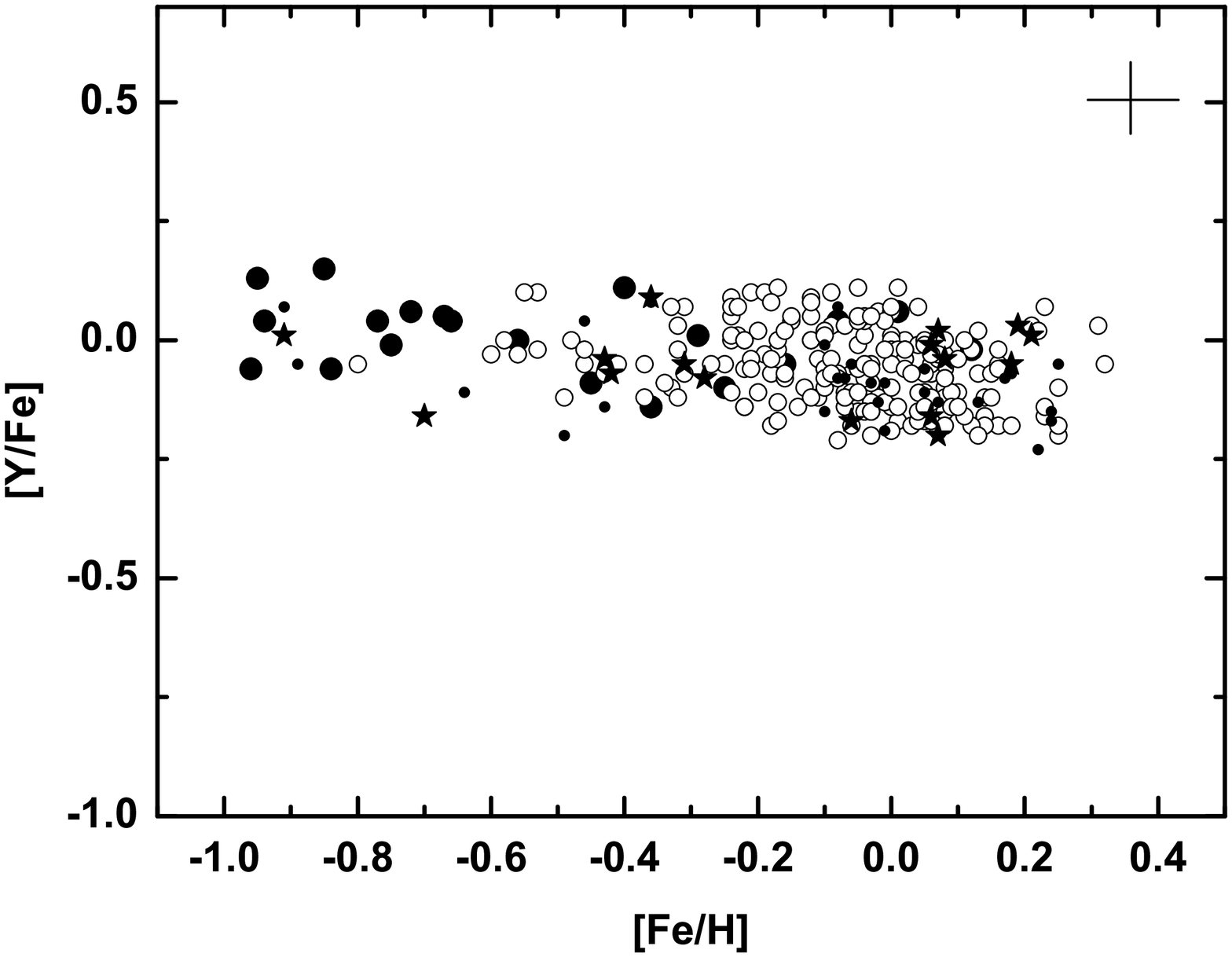}}
\resizebox  {8.4cm}{6.cm}
{\includegraphics{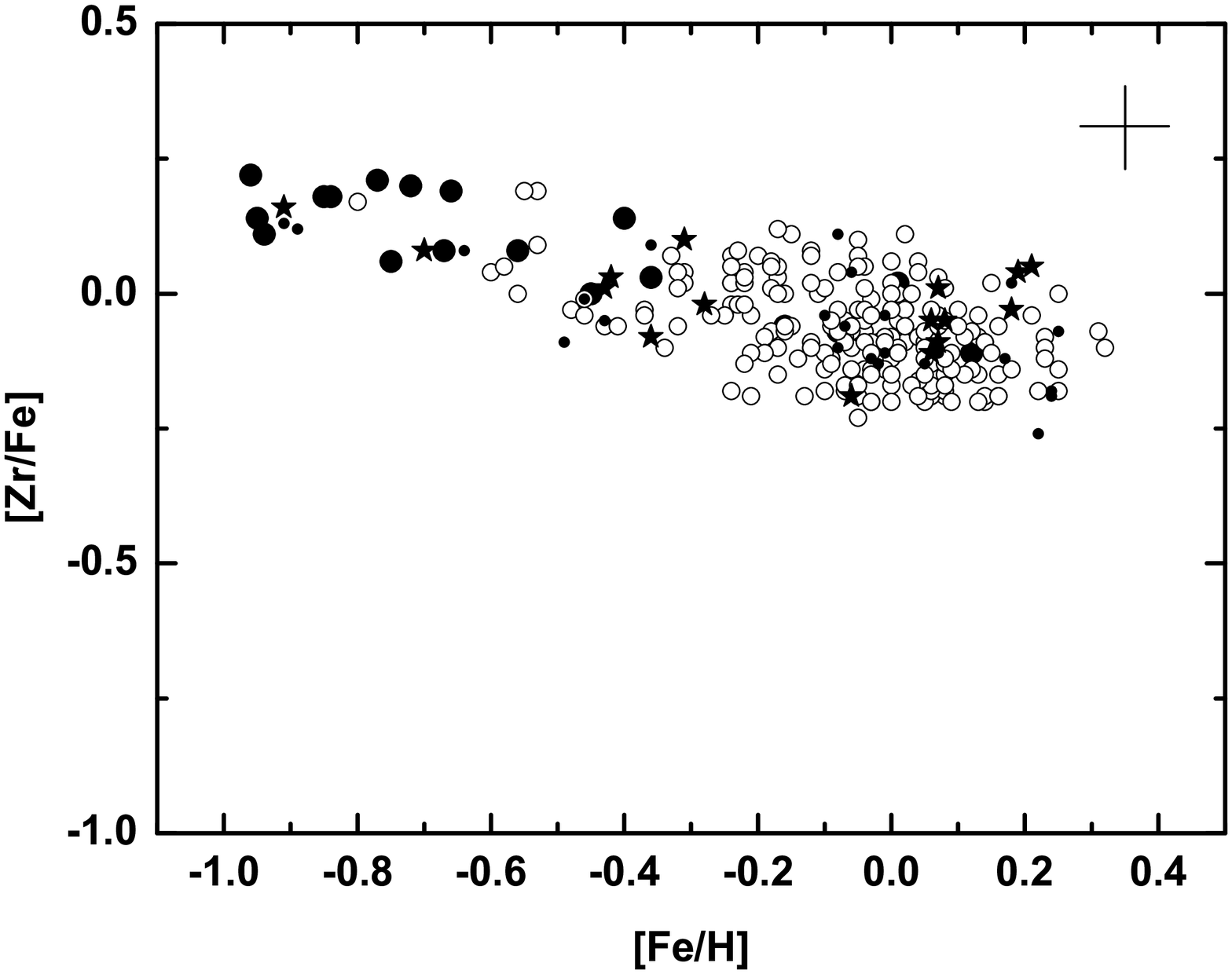}}
\caption[]{
Dependences of [Y/Fe] and [Zr/Fe] on [Fe/H], the notation is the same as in
Fig.\ref{alpha_fe}.
}
\label {N50_fe}
\end{figure}

\begin{figure}
%\resizebox{\hsize}{!}
\resizebox  {8.4cm}{6.cm}
{\includegraphics{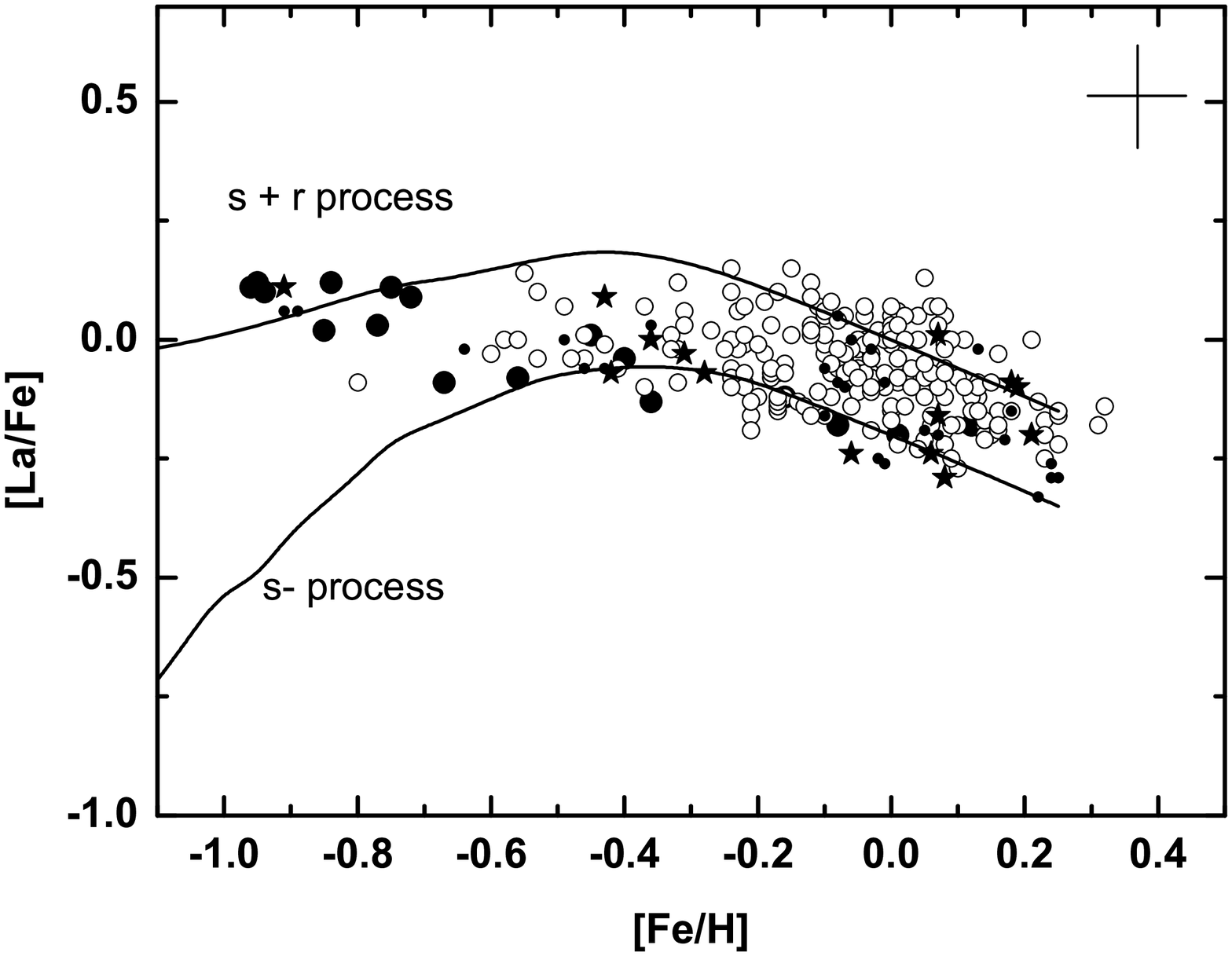}}
\resizebox  {8.4cm}{6.cm}
{\includegraphics{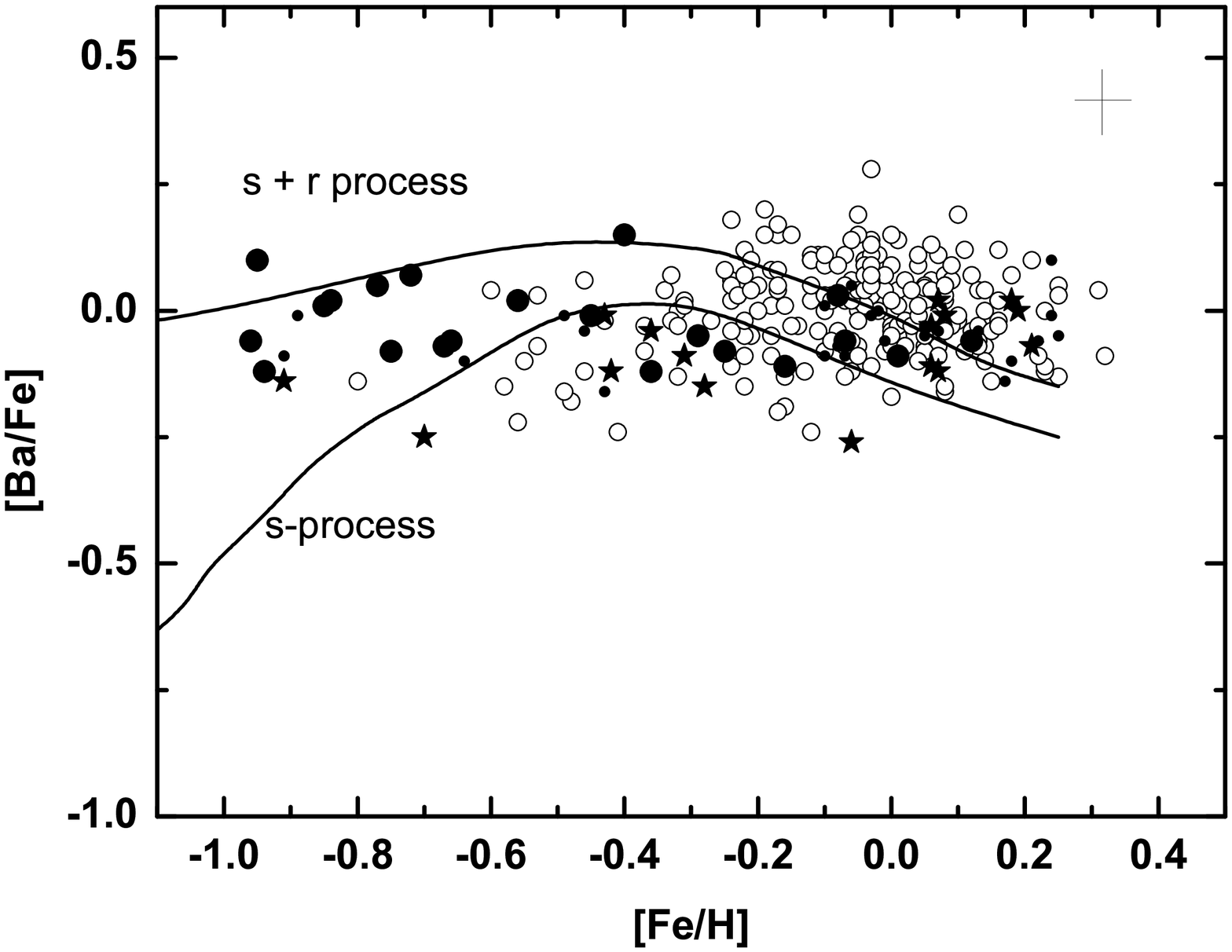}}
\resizebox  {8.4cm}{6.cm}
{\includegraphics{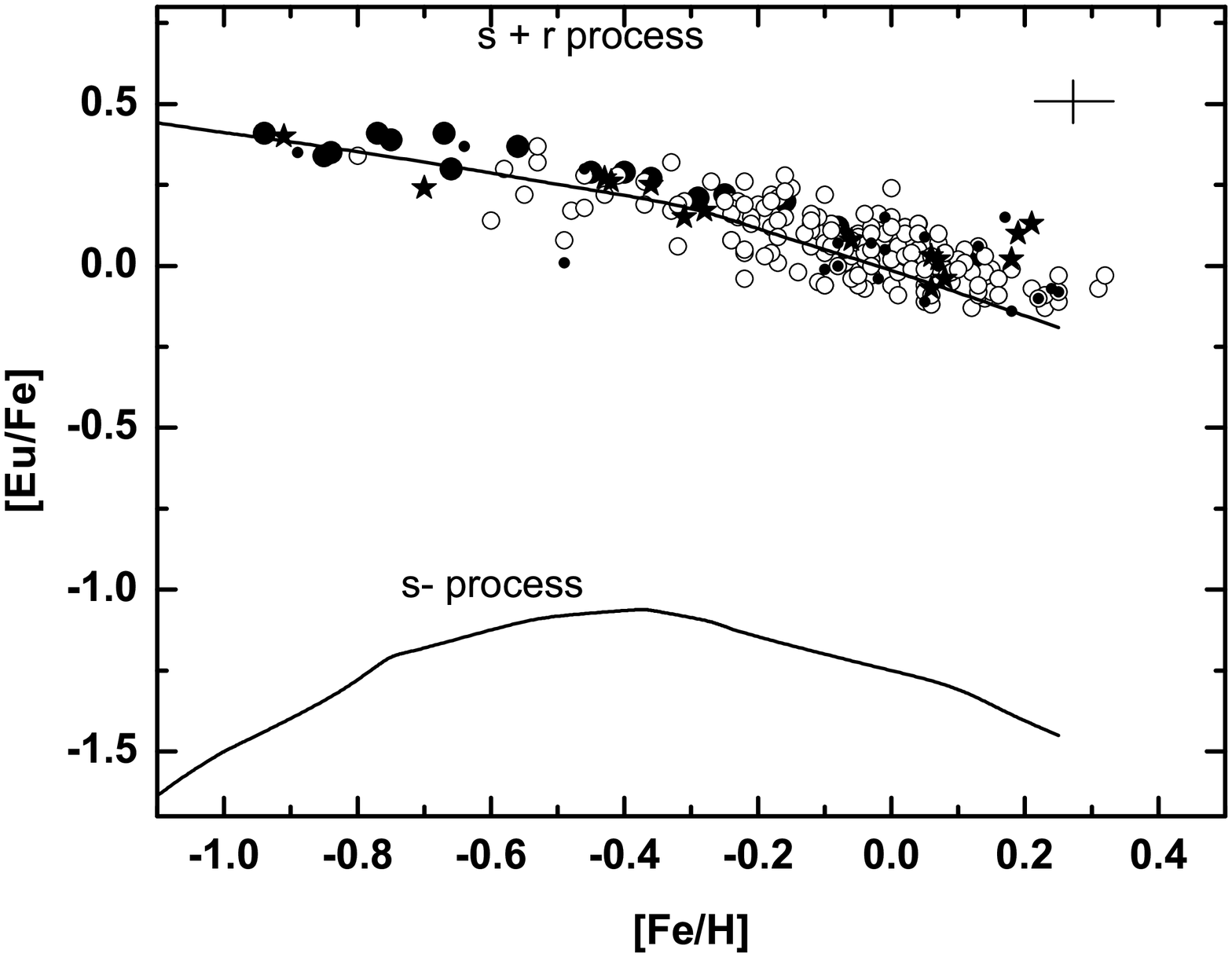}}
\caption[]{
Dependences of [La/Fe], [Ba/Fe], [Eu/Fe] on [Fe/H], the notation is the same as
in Fig.\ref{alpha_fe}
The model calculations by Serminato et al. (\cite{se09})  for the thin disk are marked with a solid line.
}
\label {N82_and_eu_fe}
\end{figure}

\begin{figure}
%\resizebox{\hsize}{!}
\resizebox  {8.4cm}{6.cm}
{\includegraphics{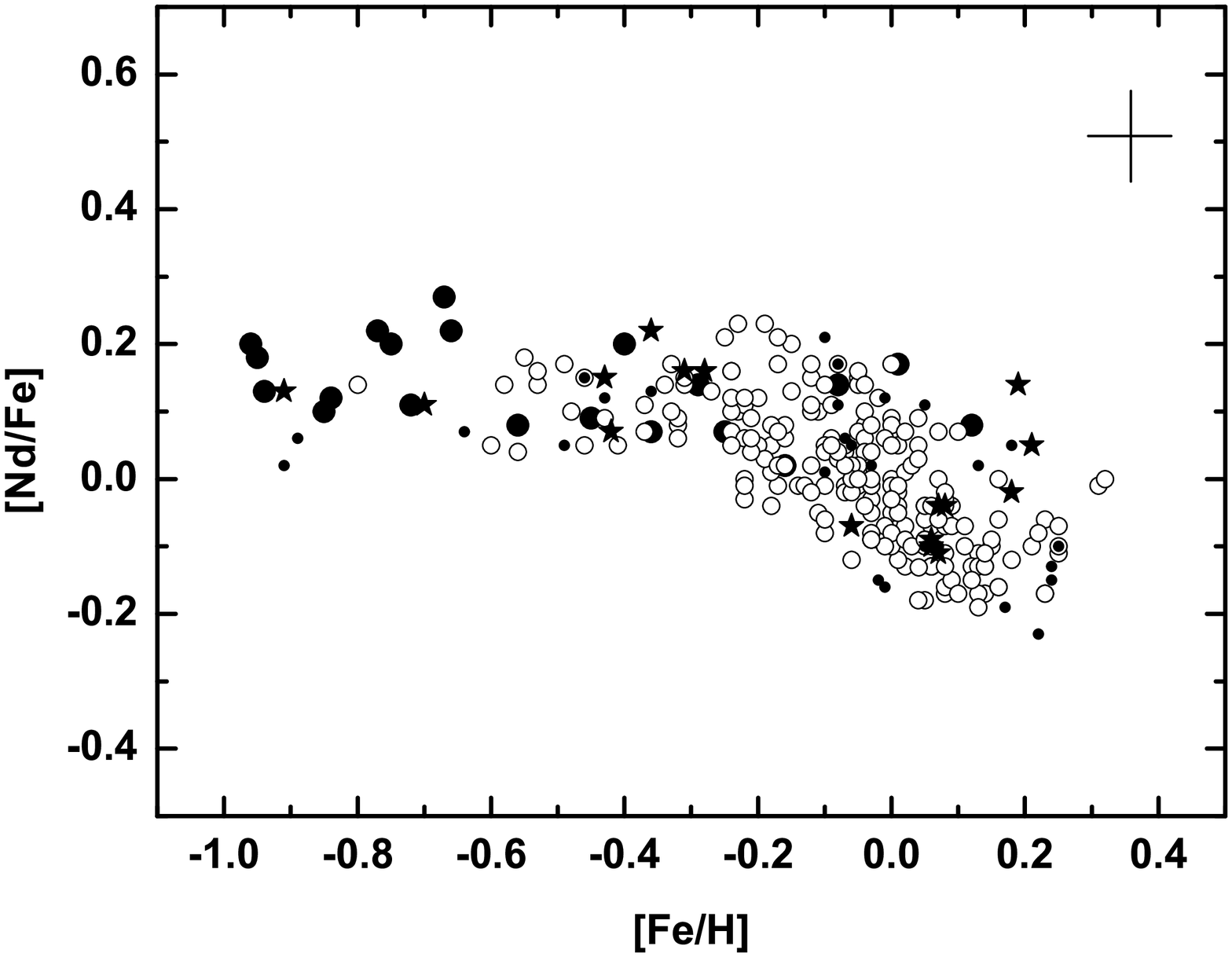}}
\resizebox  {8.4cm}{6.cm}
{\includegraphics{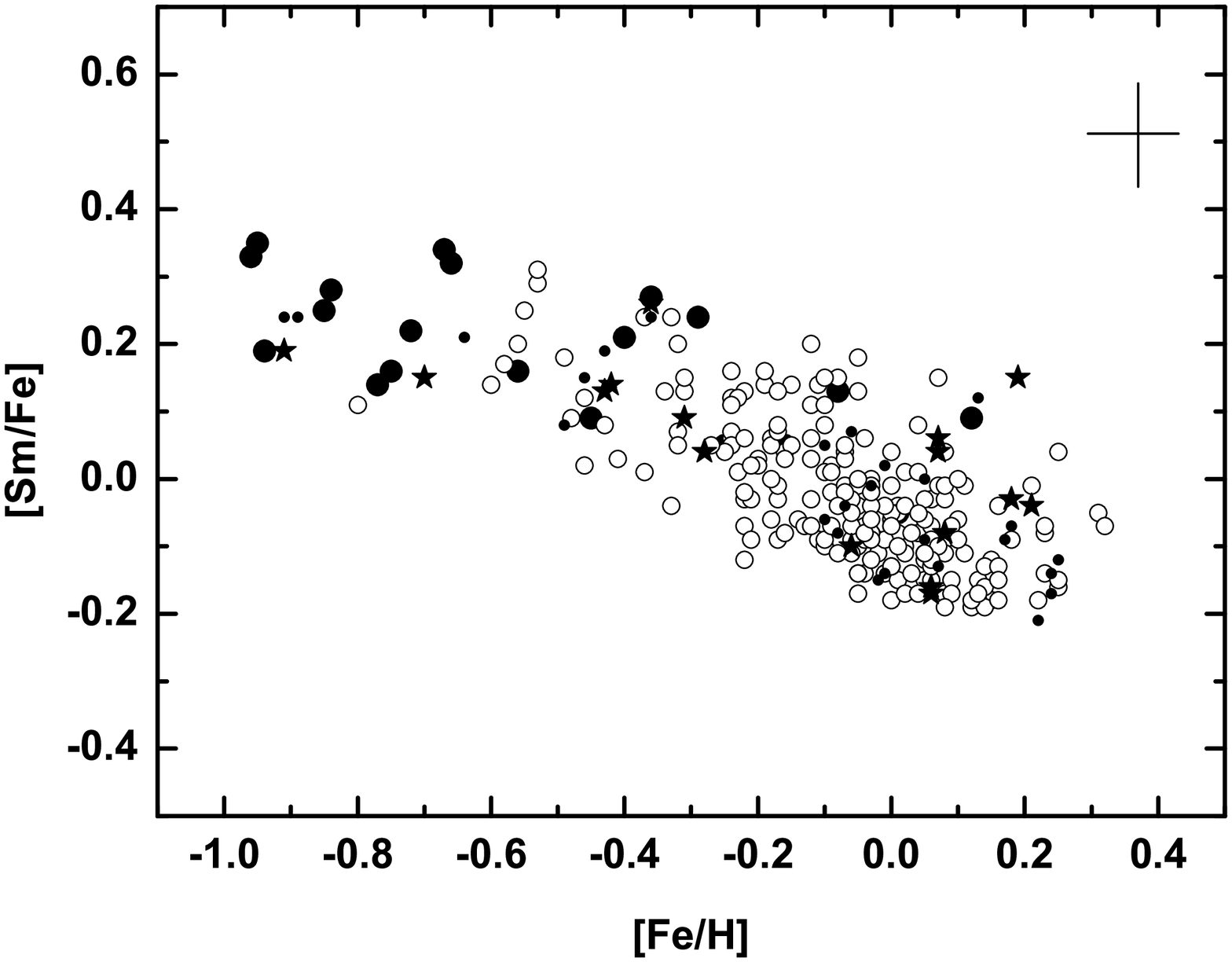}}
\resizebox  {8.4cm}{6.cm}
{\includegraphics{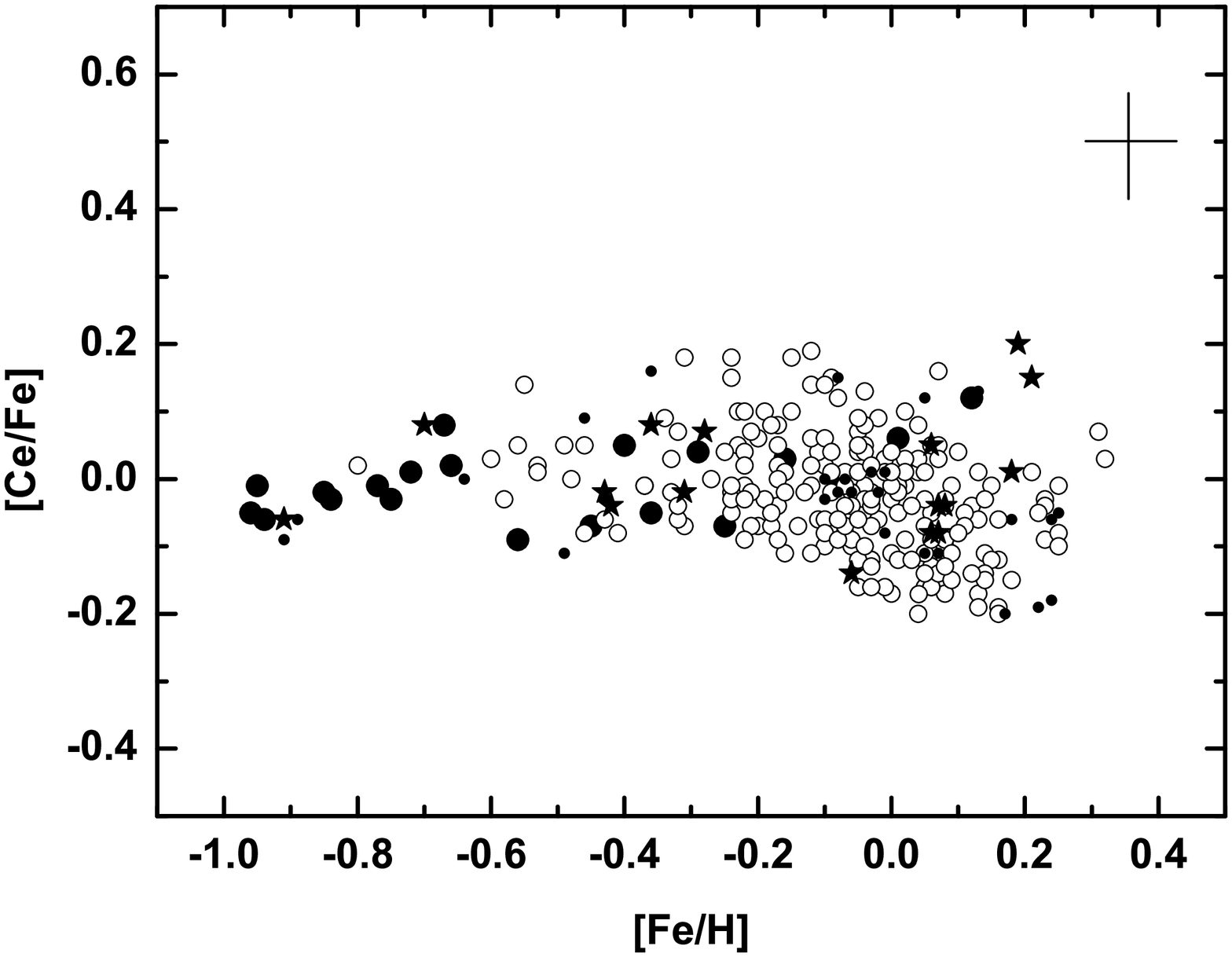}}
\caption[]{
Dependences of [Nd/Fe], [Sm/Fe], and [Ce/Fe] on [Fe/H], where the notation is the same as
in Fig.\ref{alpha_fe}
}
\label {N82_fe1}
\end{figure}

To  discuss our results better, we report in
Fig. \ref{alpha_fe} available measurements for the $\alpha$-elements Mg and Si
of the stars in the present stellar sample (Mishenina et al. \cite{mi04};
Mishenina et al. \cite{mi08}).
The  iron-group element Ni is shown
in Fig. \ref{ni_fe}; heavy elements Y
and Zr (elements representative of the neutron magic peak N = 50) in Fig.
\ref{N50_fe}; Ba, La, Eu ($s$-process elements representative of the neutron
magic peak N = 82, and Eu indicative of the $r$-process contribution) and Ce,
Nd, Sm (also representative of the neutron magic peak N = 82) in Figs.
\ref{N82_and_eu_fe} and \ref{N82_fe1}, respectively.
The complete elemental abundance data are given in Tables \ref{t_abund} and 
\ref{t_abund2}.
The stars are marked according to their classification
(see \S \ref{sec: obs_proc_sel}): full circles indicate the thick disk stars,
open circles the thin disk stars, asterisks the Hercules stream
stars and small circles are unclassified stars.

As for instance in Bensby et al. (\cite{be05}),
Reddy et al. (\cite{re06}), Nissen \& Shuster (\cite{ni08}), and 
Feltzing et al. (\cite{fe09}), our stellar sample also includes  the thick disk stars
at the solar metallicity and a metal-poor tail of the thin disk stars down to
[Fe/H] $\sim$ -0.80, allowing the analysis
of different stellar populations across a wide range of metallicity.

\subsection{The $\alpha$-elements Mg and Si}

In general, for Mg and Si in Fig. \ref{alpha_fe} (but see also
heavier elements in Figs. \ref{ni_fe}, \ref{N50_fe}, and \ref{N82_and_eu_fe},
and other $\alpha$-elements), the stars of the thick disk show narrower dispersion
than those of the thin disk. That
could suggest that thick disk stars are formed from a more homogeneous material.
The small number of the thick disk stars in our sample does not allow us to shed
more light on that point of view. We refer to \S \ref{subsec: age-met} for a detailed
discussion about the nature of the thick and thin disk.

From Fig. \ref{alpha_fe}, at the solar metallicity, the stars
of the thin and thick disks tend to have similar chemical signatures.
Then, with decreasing metallicity, all $\alpha$-element signatures
increase, confirming previous results by e.g., Bensby et al. (\cite{be03}),
Soubiran \& Girard (\cite{so05}), Mishenina et al. (\cite{mi04}),
Reddy et al. (\cite{re06}).

That trend of [$\alpha$/Fe] versus metallicity can be understood well once the
stellar nucleosynthesis feed-back to the galactic chemical evolution is
considered. Indeed, below metallicity [Fe/H] $\lesssim$ $-$1 the only relevant
astrophysical source contributing to the chemical evolution of the $\alpha$-elements
and iron are core collapse Supernovae (CC-SN, see for instance Timmes et al.
(\cite{timmes95}) for the GCE calculations). Therefore, the roughly constant ratio
O-Mg-Si-Ca/Fe observed for instance, in the halo stars (however, see the possible
increase in [O/Fe] at low metallicity, e.g., Mishenina et al.
\cite{mishenina:00}) reflects a mixed contribution from CC-SN of different
masses and (low) metallicities.
On the other hand, for [Fe/H] $\gtrsim$ $-$1, thermonuclear SN (SNIa,
Hillebrandt et al. \cite{hillebrandt:00}) from remnants of low- and
intermediate- mass stars has time to contribute to the chemical enrichment
of the disk (see e.g., Matteucci et al. \cite{matteucci:06}), mostly feeding
the iron-group elements and the $\alpha$-elements, except for oxygen and magnesium
(Thielemann et al. \cite{thielemann:86}; Travaglio et al. \cite{travaglio:11};
Kusakabe et al. \cite{kusakabe:11}).
Therefore, the slope of the [$\alpha$/Fe] ratio and the amount of departure from
the solar ratio toward lower metallicities reflect the differential contribution
from the CC-SN and SNIa to Fe and to the $\alpha$-elements.

The [Mg/Fe] observations in the thick disk stars show higher values than in the thin disk stars,
as well as  in the metallicity range where the two disks overlap
(-0.50 $<$ Fe/H] $<$ 0).
The small number of the Mg measurements for the thick disk stars in our sample is not
statistically significant, but does agree with previous, more extended
studies. The ratio [Mg/Fe] for the stars of the Hercules stream spans all
values of both disks.

Several studies (Bensby et al. \cite{be03}; Fuhrmann \cite{fu04}; Mishenina et
al. \cite{mi04}; Soubiran \& Girard \cite{so05}; Reddy et al. \cite{re06}; and
Bensby et al. \cite{be07}) have shown a magnesium
abundance behavior with a ``break'' of the correlation between [Mg/Fe] and
[Fe/H] at [Fe/H] $\sim$ -0.3. Indeed, above [Fe/H] $\sim$ -0.3, all stars with
the thick disk kinematics show the Mg chemical signature typical of the thin disk.
Reddy et al. (\cite{re06}) also identify a small sample of thick disk stars with
the thin disk abundance signature and [Fe/H] $\lesssim$ -0.3, and defined all stars
with thin disk [Mg/Fe] and thick disk kinematics as the TKTA stars, belonging to an
independent subgroup. Therefore, in this scenario, the metallicity of the thick
disk would not exceed [Fe/H] $\sim$ -0.3.
On the other hand, other works (e.g., Mishenina et al. \cite{mi04}) did not use
such a distinction, simply assuming the existence of a ``knee'' in the [Mg/Fe] trend
toward [Fe/H] $\sim$ -0.2 in the thick disk, making   the thick overlap and thin
disks abundance signature.
Owing to the small number of the thick disk stars in the present sample, we cannot
shed more light on this matter, even if we could define  the TKTA stars or definitively
defining them as the thick disk objects.
Therefore, in this work, we consider them as the thick disk members, according
to their kinematics alone.
Standard thick disk stars show a dominant CCSN signature, whereas the TKTA-like
stars are affected by a larger contribution from the SNIa, feeding Fe efficiently 
but not Mg. Therefore, they could have an abundance signature similar to the
thin disk, not because they share some peculiar history compared to the rest of
the thick disk objects, but simply because their pristine signature is more
affected by SNIa.
According to this picture, the TKTA stars should also have [O/Fe] typical of the
thin disk, since as  mentioned   oxygen is also not made in large amounts
in SNIa. Furthermore, most of the TKTA stars in the Reddy et al. (\cite{re06}) sample
should be younger, than standard thick disk stars with the same
metallicity more likely carrying a stronger SNIa signature than do older objects. More
TKTA-like stars need to be identified to draw a definitive picture.

[Si/Fe] versus [Fe/H] shows  behavior similar to that of [Mg/Fe],
but with  a smaller slope and with narrower dispersion.
As we also have mentioned before, this is because  the SNIa contribute
more efficiently to the chemical evolution of the $\alpha$-elements heavier
than O and Mg, smoothing the effect of the strong iron production from those
objects.

\subsection{The  iron-group element Ni}

In Fig. \ref{ni_fe}, we show [Ni/Fe] versus [Fe/H], which is flat and roughly
solar for the whole metallicity range, for the thin disk, the thick disk, and the Hercules
stream stars. A slightly increasing trend may be
possible for [Ni/Fe], for metallicity [Fe/H] $\gtrsim$ 0.1.
The increasing trend of [Ni/Fe]  toward higher  metallicities than the solar ones
is confirmed by other works, e.g. Neves et al. \cite{neves:09}.
This is because  Ni/Fe in the SNIa ejecta depends on the metallicity
of the progenitor (e.g., Timmes et al. \cite{timmes:03},
Bravo et al. \cite{bravo:10}, Travaglio et al. \cite{travaglio:05}).
In particular, the ejecta of unstable $^{56}$Ni form the bulk of
the produced iron. Its production tends to decrease with the increasing 
metallicity, whereas most of Ni is produced in the NSE conditions, and its production
is quite constant with metallicity. Therefore, the consequent [Ni/Fe] is expected
to increase in the disk with [Fe/H].
A flat trend for [Ni/Fe] at solar metallicity may be explained
if the average Ni/Fe ratio in the CC-SN ejecta is similar to the SNIa signature.
No significant slope or dispersion is observed in that case.
The Ni/Fe ratio in SNIa ejecta may change quite significantly from different
theoretical predictions, from a ratio that is two to three times higher  than the solar one (e.g., Thielemann et
al. \cite{thielemann:86}; Travaglio et al. \cite{travaglio:11})
to a ratio close to the solar one (e.g. Thielemann et al. \cite{thielemann:04}).
Present observations seem to support  those  predictions more.

\subsection{AMR for the thick and thin disks}
\label{subsec: age-met}

In the previous sections, we  discussed the abundance signature of
the $\alpha$-elements and Ni. All those elements that include Fe are primary. Their
yields from the CCSN or SNIa do not depend significantly on the initial metallicity
of the parent star. As  mentioned above, the abundance dispersion (besides
observational errors) and the [El/Fe] slope
are given by the differential contribution (i.e., by the different elemental
ratio in the ejecta) between the CCSN and SNIa.

This is not the case for the $s$-process.
Therefore, before discussing observations for heavy elements, we revise 
 the age-metallicity relation in the thick and thin disks in this
section.
As  is well known, the age-metallicity relation for the thick disk stars show a
signature of decreasing age with increasing metallicity, with some dispersion (e.g., Bensby
et al. \cite{be07}).

On the other hand, the stars belonging to the thin disk tend to show wide
metallicity dispersion, in particular around the time of formation of the Sun,
and there is no clear trend in age versus metallicity. Indeed, as written in
Bensby et al. (\cite{be07}), ``the most metal-rich thin-disk stars evidently are
not the youngest ones''.
A possible proposed scenario to explain that missing trend is an infall of fresh
material in the thin disk around the time of  the Sun's formation, causing a
spread in metallicity for the stars of the same age and,  more in general, a dilution
of metals available in the interstellar medium at that time produced by previous
stellar generations (Edvardsson et al. \cite{ed93}; Feltzing et al.
\cite{feltzing01}; Haywood \cite{haywood06}).

Such a contribution could have had a small impact in the [El/Fe]
ratio in the disk at that stage, but the average [El/H] abundance
was probably modified (and possibly reduced) by the dilution
with fresh material, including the [Fe/H].
Furthermore, the yields of nucleosynthesis processes that are affected by the
initial metallicity of the star (secondary)
will be affected by such a dispersion.
For instance, the $s$-process yields from the AGB stars or massive stars,
born in the thin disk about 5 Gyr ago, will carry the signature of  nonuniform
pristine metal content, affecting the abundance signature in the youngest
generations of evolving and unevolved stars.

\begin{figure}
%\resizebox{\hsize}{!}
\resizebox  {8.4cm}{6.cm}
{\includegraphics{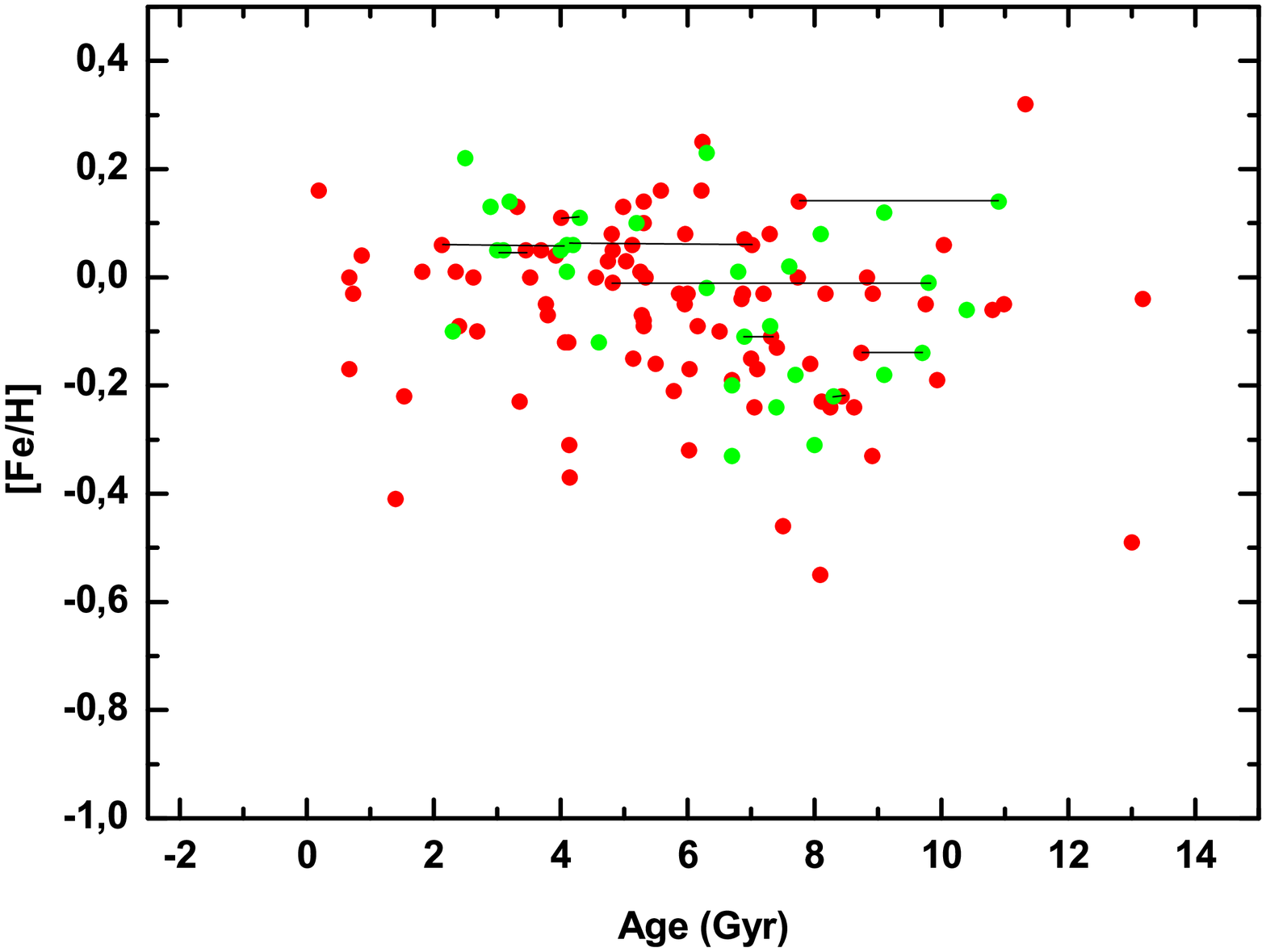}}
%\resizebox  {8.4cm}{6.cm}
%{\includegraphics{Age_Fe.eps}}
\caption[]{
Dependence of [Fe/H] on age
%and that of the Age vs. [Fe/H] (lower panel)
for the thin disk stars in our sample according
%to Girardi et al. (\cite{girardi12}) (black points),
to Holmberg et al. (\cite{holm09}) (green points) and
Mowlavi et al. (\cite{mowlavi12}) (red points).
Common stars in  in the two samples are connected
by a continous line.
}
\label {Age_Fe}
\end{figure}

In Fig. \ref{Age_Fe} we show the age versus metallicity
relation for the thin disk stars in our sample as derived from different
analyses based on two sets of stellar tracks,
by Mowlavi et al. (\cite{mowlavi12}) and Holmberg et al. (\cite{holm09}).
%, top panel.
%Girardi et al. (\cite{girardi12}),
In the first case, to estimate the age we use
%the free-access code PARAM
%\footnote{http://stev.oapd.inaf.it/cgi-bin/param}
%developed by Girardi, based on the
%Bayesian estimation method by Jorgenson \& Lindegren (\cite{jor05}) and
%slightly modified by da Silva et al. (\cite{dasilva:06}), and 
%In the second case, we used
the python k-d tree based
interpolation technique.
% for the stellar tracks
%by  Mowlavi et al. \cite{mowlavi12}.
Keeping into account the observed metallicity for each star,
in the figure we include only the objects fitted by the correct
isochrones set with pairwise euclidean distances
in two-dimensional space (given by T$_{\rm eff}$/T$_{\odot}$ and L/L$_{\odot}$)
smaller than 0.02.

In the second case 
(Holmberg et al. \cite{holm09}
\footnote{http://cdsarc.u-strasbg.fr/viz-bin/Cat?V/130}), ages are 
based on the
theoretical isochrones from Padova (Girardi et al. \cite{girardi00})
and the photometric metallicities used . We include only stars with reported
errors in the age estimation smaller than 25\%.
%In Fig. \ref{Age_Fe}, in the lower panel, we show the three sources again, but
%in this case we report also the errors for the age estimation by
%Girardi et al. (\cite{girardi12}).
The spread of predictions using different set of isochrones
is mainly due to the stellar uncertainties.
In the figure ten stars are fitted by both
Holmberg et al. (\cite{holm09}) and Mowlavi et al. (\cite{mowlavi12})
within the mentioned criteria. For six of them the age estimation
is consistent within 2 Gyr, whereas for the remaining stars there is
 larger discrepancy (namely,
HD 28447, HD 70923, HD 178428, and HD 75767).

\begin{figure}
%\resizebox{\hsize}{!}
\resizebox  {8.4cm}{6.cm}
{\includegraphics{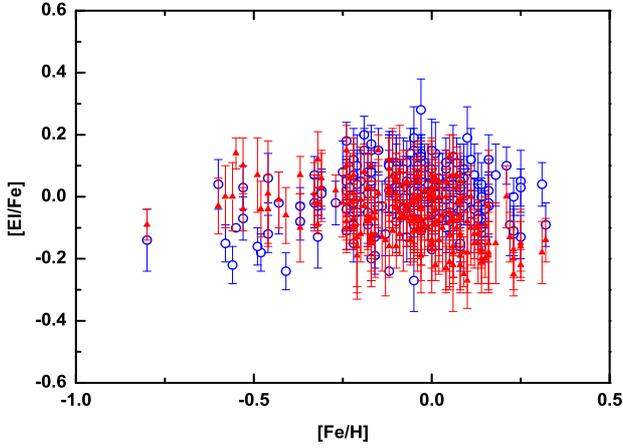}}
\caption[]{
The [Ba/Fe] and [La/Fe] trends with [Fe/H] and error determination bars  for
each star (open blue circles and red triangles, respectively).
}
\label {fig:bala_fe}
\end{figure}

\begin{figure}
%\resizebox{\hsize}{!}
\resizebox  {8.4cm}{6.cm}
{\includegraphics{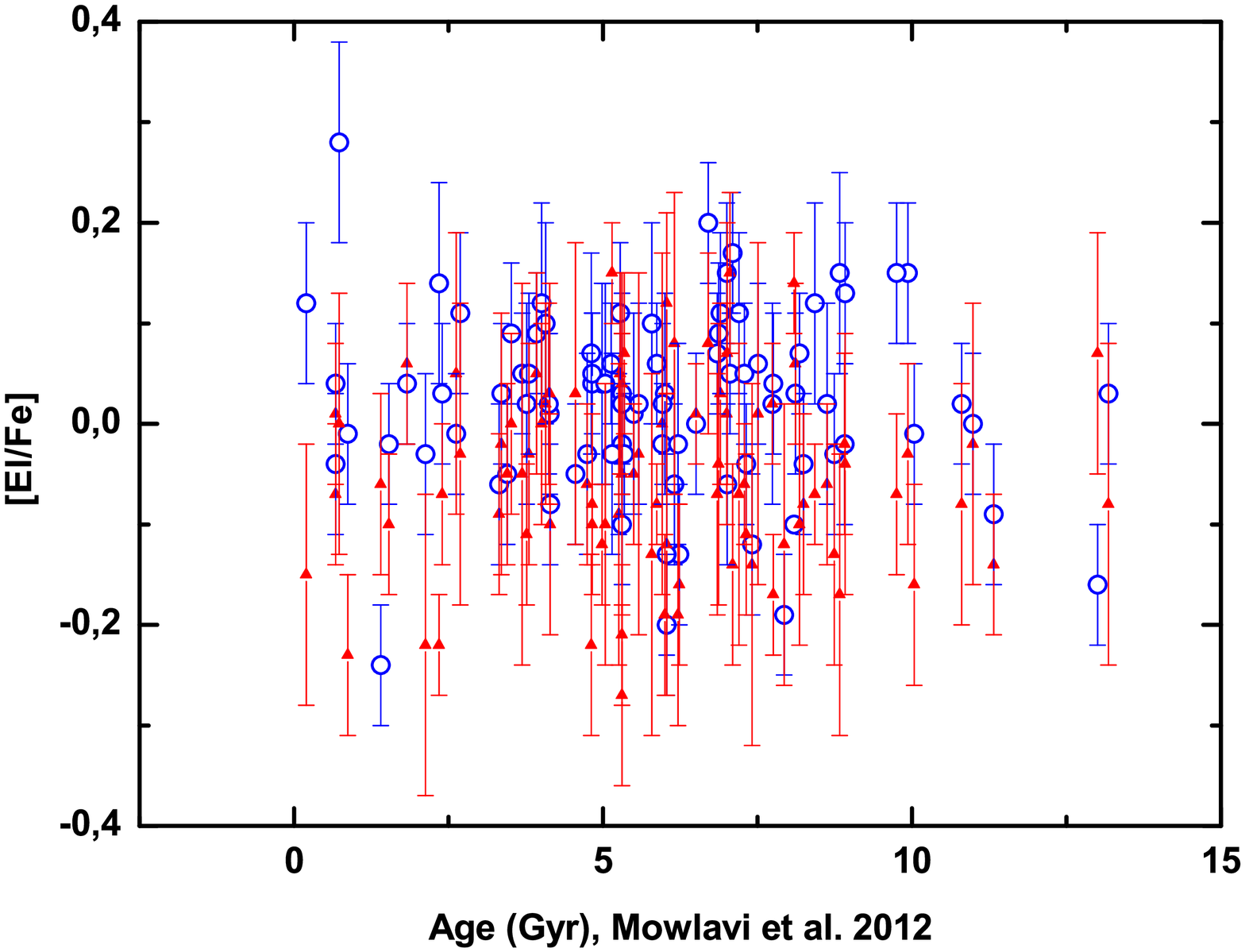}}
\resizebox  {8.4cm}{6.cm}
{\includegraphics{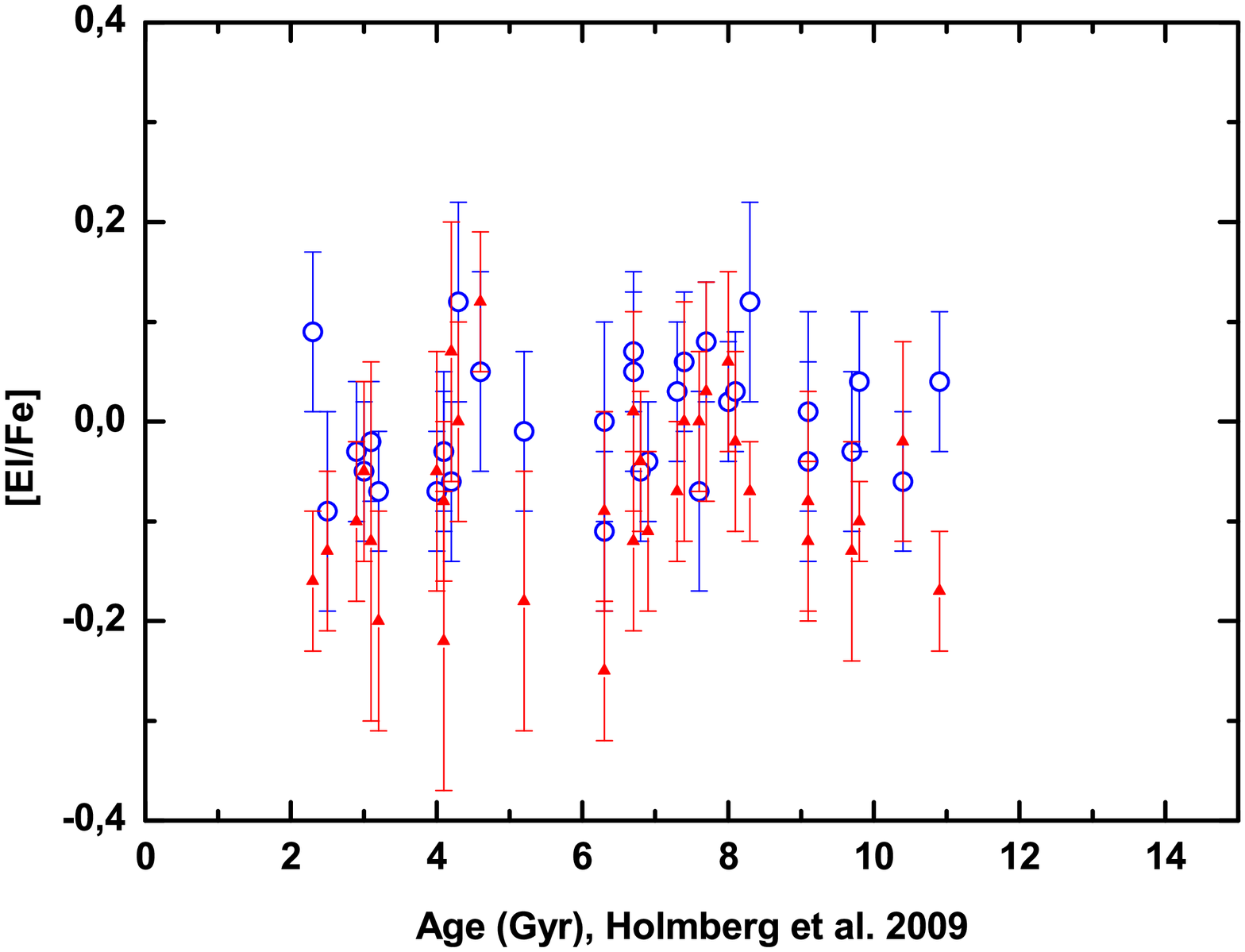}}
\caption[]{
[Ba/Fe] and [La/Fe] trends versus age determined by us
according to Mowlavi et al. (\cite{mowlavi12}) (top panel) and according to
%Girardi et al. (\cite{girardi12})
Holmberg et al. (\cite{holm09}) (bottom panel). The notation is the same as in  Fig. \ref{fig:bala_fe}.
%For the errors, please see Fig. \ref{fig:bala_fe}.
}
\label {fig:bala_age}
\end{figure}

According to the results obtained for our sample thin disk stars with different stellar track compilations,
we confirm wide dispersion of the age-metallicity relation, in
particular at the time of the Sun's formation  
(e.g., Bensby et al. \cite{be07}).
%For Girardi et al. (\cite{girardi12}) a slight trend of the Age versus
%metallicity could be seen for the stars younger than 3-4 Gyr, with a saturation
%toward the solar metallicity. However, the high concentration of objects around 4
%Gyr compared to the Age spread makes difficult to confirm it. Furthermore,
Using the tracks from Holmberg et al. (\cite{holm09})
and Mowlavi et al. (\cite{mowlavi12}), the metallicity dispersion does not decrease for younger stars,
and such a trend is not observed.
Therefore, within the present uncertainties, a specific trend for the age
versus metallicity cannot be observed (in agreement with Bensby et al.
\cite{be07}), as  is instead possible for the thick disk stars. We do
not consider here the thick disk stars since our sample is too small to derive any
specific conclusion, although we may confirm the age-metallicity trend for the
thick disk (Soubiran \& Girard  \cite{so05}).
Such a result for local metallicity dispersion and AMR is also
confirmed by recent reanalysis of the metallicity distribution function of
the solar neighborhood over the Geneva-Copenhagen survey
(Casagrande et al. \cite{cas11}). Forthcoming results from the RAVE survey (Steinmetz et al. \cite{stein06})
will probably improve present scenario and shed more light on the
age-metallicity dispersion and its trend for the thin disk stars.

In several previous works (Haywood \cite{haywood06}, etc), the uncertainty
related to the age definition by fitting stellar luminosity and surface
temperature has been largely discussed.
Stellar model uncertainties are affecting theoretical
calculations within isochrones sets. Different choices for macrophysics
(e.g., convection criteria and mass loss) and microphysics
(e.g., opacities, equation of state, nuclear physics reaction rates)
may introduce significant discrepancies between different stellar
theoretical predictions.
However, despite offsets from one isochrone set to the other
and large uncertainties of 2 Gyr or more for several stars,
the different AMR behavior for the thin and thick disks is robust,
compared to different sets of the stellar models.

\subsection{Heavy neutron-capture elements}

The Y and Zr elements belong to the neutron magic peak N = 50.
Several processes are likely to be responsible for their nucleosynthesis in stars.
During early stages of the chemical evolution of the Galaxy they can be made by the
$r$-process (reproducing 8 and 15 per cent of their solar abundance,
respectively, Travaglio et al. \cite{tr04}).
Another component active in the early Universe has been identified in
several stars (e.g., Truran et al. \cite{truran02}; Honda et al. \cite{ho06};
Chiappini et al. \cite{ch11}), unrelated to the main $r$-process.
Several scenarios have been proposed, such as  charged particle reactions
in the SN explosive nucleosynthesis (e.g., Hoffman et al. \cite{hoffman96};
Qian \& Wasserburg \cite{qi08}) and in neutrino winds
(Frohlich et al. \cite{froelich06}; Farouqi et al. \cite{farouqi09};
Arcones et al. \cite{arcones11}) or the $s$-process in fast rotating massive
stars (Pignatari et al. \cite{pignatari08}; Frischknecht et al.
\cite{fricknekt12}). It is also a matter of debate whether the process(es)
possibly responsible for the Sr-Y-Zr enrichment of those old stars is(are)
active until the solar metallicities as a primary process. Indeed, Travaglio et al.
(\cite{tr04}) identified a similar missing component in the solar system
$s$-process distribution (lighter element primary process, or LEPP). In the
latter case, the LEPP would be responsible for about 20\% of solar Y and Zr.
Finally, the rest of Y and Zr is made by the $s$-process mostly in the AGB stars
with some minor contribution from massive stars.
In both cases, their $s$-process production from the AGB stars becomes relevant
quite late for the chemical evolution, reaching a contribution peak at about
[Fe/H] $\sim$ $-$0.4 with an approximately constant (for Y) or slight decrease
(for Zr) for higher metallicities ([Fe/H] $\gtrsim$ $-$0.4,
Travaglio et al. \cite{tr04}).

The light $s$-process elements Y and Zr show different trends in 
[Fe/H]. The [Y/Fe] ratio versus [Fe/H] shows a more or less flat trend in
our stellar sample.
Similar results are obtained by Reddy et al. (\cite{re06}).
Bensby et al. (\cite{be05}) show a [Y/Fe] that is lower by 0.1$-$0.15 dex
for [Fe/H] $\lesssim$ -0.3.
The [Zr/Fe] ratio versus [Fe/H] increases  by about
0.2 dex with decreasing metallicity. Indeed, within some dispersion of abundances in the thin disk stars,
the average abundances of the thick disk stars [Fe/H] $\lesssim$ -0.3 are on average
larger than for stars at the solar metallicity.
Such differences between Y and Zr could be because Y
receives a larger $s$-process contribution than Zr in particular in
the thin disk, compensating more efficiently for the iron made by SNIa.
Indeed, the GCE calculations by Travaglio et al. (\cite{tr04})
can account at least qualitatively for such a variation,
because of the higher $s$-process contribution to Y than to Zr
(according to Travaglio et al. \cite{tr04}, 74\% and 67\%, respectively), and
because as we mentioned, the Zr $s$-process yields from the AGB stars start decreasing
earlier
than Sr and Y with increasing  metallicity, as above [Fe/H] $\gtrsim$ -0.3.
(see e.g., Travaglio et al. \cite{tr04}).

Barium and La are the $s$-process elements at the neutron magic peak N= 82, with a
smaller contribution from the $r$-process.
The trend [Ba/Fe] versus [Fe/H] shows a significant dispersion and, on average, an
underabundance of $\lesssim$ 0.2 dex
in the thick disk compared to the thin disk.
The [La/Fe] ratio  tends to show a slightly decreasing trend
to [Fe/H] $>$ 0 for the thin disk and the Hercules stream stars,
but with a large dispersion that is  the same as for Ba.
To compare  the behavior of two $s$-process elements more easily, we show in
Fig. \ref{fig:bala_fe} together
the [Ba/Fe] and [La/Fe] versus [Fe/H], including observational uncertainties,
only for the thin disk stars.
From the figure, both elements show a dispersion of about 0.4 dex,
ranging between $-$0.2 $\lesssim$ [El/Fe] $\lesssim$ +0.2, as well as  similar
trends that are difficult to disentangle.

Europium is mainly formed by the $r$-process, showing a marked
trend with [Fe/H] and a slight overabundance for the thick disk stars.
Since the bulk of Eu is created in massive stars, the [Eu/Fe] ratio
is expected to decrease once Fe from the SNIa starts to play a role in
the chemical evolution of the disk. Therefore, the thick disk 
shows a higher [Eu/Fe] on average, than does the thin disk.
A similar general trend is observed for Nd and Sm, since both of them,
like Eu, receive a major contribution from the $r$-process.

Finally, the Ce abundance seems to behave similarly in all substructures, and it
shows a relevant dispersion that  agrees with Reddy et al. (\cite{re06}). A
slightly decreasing trend could be seen for [Fe/H] $>$ $-$0.2 in the thin disk,
but  it is not followed by all the stars. Therefore, the similar consideration as made for Ba
and La still holds for Ce.
In summary, compared to the thick disk, the thin disk stars are created by
a more complex combination of
contributions from massive stars, $r$-process, SNIa, and AGB stars.

\begin{figure}
\resizebox{\hsize}{!}
{\includegraphics{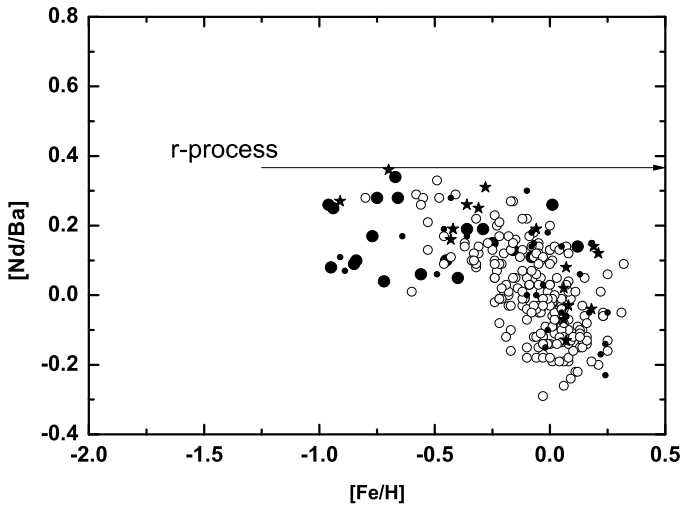}}
\caption[]{
Dependence of [Nd/Ba] on [Fe/H], the notation is the same as in
Fig.\ref{alpha_fe}
}
\label {fig51}
\end{figure}

On the other hand, typical
disk stars with metallicity [Fe/H] $\gtrsim$ -1
(i.e., once the iron contribution from the SNIa starts to be observed),
may start carrying an evident signature of the $s$-process yields from the AGB stars,
which are the parents of thermonuclear supernovae. That is clearly shown in Fig.
\ref{fig51}, where [Nd/Ba] and [Nd/Eu]  versus [Fe/H] for stars in our sample are given
 in comparison with the [Nd/Ba]$_{\rm r}$ and [Nd/Eu]$_{\rm r}$ observed
by Mashonkina et al. (\cite{ma04}). On average, the stars that are already in the thick disk
show relatively low [Nd/Ba] and [Nd/Eu], which is higher than those  observed for the
pure $r$-process signature, as expected from the $s$-process contribution by the
AGB stars with the thick disk metallicities.

In Fig. \ref{N82_and_eu_fe}, we compare the observations for La, Ba and Eu with GCE calculations
by Serminato et al. (\cite{se09}). In particular, in the Serminato et al.
(\cite{se09}) simulations, the contribution from the $s$-process in low-mass AGB
stars and from the $r$-process are included.
For  those elements, both neutron capture processes need to be considered
in order to obtain the solar abundance.
A small scatter is observed for Eu, consistent with the
GCE predictions once both $r$-process and small $s$-process contributions are
included.
On the other hand, both [La/Fe] and [Ba/Fe] show  large dispersion for the thick
and thin disk stars, possibly increasing towards the solar and super-solar
metallicities.
No clear trend can be identified with such a dispersion, which by definition
cannot be reproduced by the single-zone GCE calculations of Serminato et al.
(\cite{se09}). Also the GCE decreasing trend in [Ba/Fe] and [La/Fe] for [Fe/H]
$>$ $-$0.3 is not clearly identifiable, even if several stars fall along the Ba
and La theoretical curves.

Compared to the $\alpha$-elements, the $s$-process elements La and Ba show
similar dispersion at high metallicities, but the GCE evolutionary trends are
not clearly reproduced. The reason is that the $\alpha$-elements (as well as Fe) are
primary, and their chemical evolution is affected by the age and, only marginally, by
the metallicity. Therefore, the use of decent stellar yields from core-collapse
SNe and SNIa allows  the chemical evolution
trends of Fe and $\alpha$-elements to be  predicted with reasonable accuracy, 
once the weight of two main yield
donors is known for the Sun. In our case, Fe provided a phenomenological
indicator of such a weight, and it is not surprising that one-dimensional GCE
models may reproduce
 the [$\alpha$/Fe] trend of the thick and thin disks well.

On the other hand, Ba and La are mostly created by the $s$-process in the AGB stars,
whose contribution to the interstellar medium depends on metallicity,
as clearly shown by, e.g., Travaglio et al. (\cite{tr04}), and on the age:
i.e., on the different life timescale of the stars with different initial mass.
If the age versus metallicity relation of a stellar system is linear (e.g., the
thick disk in its lowest metallicity population), then the $s$-process
abundances are expected to show a similar dependence on the age and metallicity, and
therefore simple theoretical GCE calculations may provide a reasonable fitting
of observations. As we discussed in the previous section, this is not the case
for the thin disk, where a wide spread of metallicity is observed for the stars of
the same age, and there is no clear age versus metallicity evolution trend.
A  wide abundance dispersion therefore observed , independently for the event that causes such a spread, 
 and simple GCE models may lose their
predictive power for the $s$-process elements. To consistently compare
the stellar abundances with GCE predictions, a preliminary selection of stars with
the same location in the thin disk and that fall on the same age - metallicity
slope should be performed. Such a sample of stars would be representative of a
specific subgroup. A comprehensive GCE study of the thin disk would be given by
taking  those different populations into account, and therefore,
would require multidimensional GCE simulations (e.g., Minchev et al. 
\cite{Minchev:12}).

Similar conclusions may be obtained for Ce in Fig. \ref{N82_fe1},
which is an $s$-process element as Ba and La. Sm receives a comparable contribution
from the $s$- and $r$- processes, and as expected, the [Sm/Fe] ratio shows a
clearer decreasing trend with increasing metallicity.

Summing up, the trend of [Eu/Fe] is reproduced  well by the GCE simulations, since Eu
is made mostly by the $r$-process (that is primary), and the chemical evolution
models are well constrained in the Fe evolution.
The [$\alpha$/Fe] show some dispersion in the disk stars, due to the differential
contribution from the core-collapse SNe and SNIa to the initial stellar abundances.
Since their production in the primary, the GCE calculations can
reproduce a general trend quite well,  once the age of the stellar system together
with basic evolution properties (e.g., IMF) are given,
Finally, the $s$-process elements Ba, La, and Ce are not primary, and due to the
lack of an age versus metallicity relation in the late disk their dispersion and
evolutionary trends become more difficult to predict.

In Fig. \ref{fig:bala_age} we show [Ba/Fe] and [La/Fe] for the stars in our
sample with respect to the age estimated from stellar tracks by Mowlavi et al.
(\cite{mowlavi12}) and Girardi et al. (\cite{girardi12}).
With the abundance dispersion and uncertainties,
 we find it difficult for both references to clearly identify in our sample of stars
any increase in [Ba/Fe] for youngest stars, as suggested by
Bensby et al. (\cite{be07}) for the thin disk, or by D'Orazi et al.
(\cite{dorazi:09}) and Maiorca et al. (\cite{maiorca:12}) for open clusters
(however, see D'Orazi et al. \cite{dorazi:12} where possible observational
issues are discussed for metal-rich open clusters).
The same conclusion is obtained for [La/Fe].
The dispersion of [Ba/Fe] and [La/Fe] for the stars with the same age may be due
to different chemical enrichment histories inside the thin disk.

\section{Conclusions}

We present and examine the abundance of
the iron peak element Ni, and of the neutron-capture elements Y, Zr, Ba, La,
Eu, Nd, Sm and Ce for 276 stars belonging to different substructures
of the Galaxy, separated according to kinematic criteria.
For most of the stars in this sample, the abundances of neutron-capture elements
have not been measured before.

Concerning Ni, all stellar structures show a flat trend up to [Fe/H] $\sim$ 0.1
with an [Ni/Fe] close to the solar one, with a slight increase for the super-solar
metallicities.
That implies that both CCSN and SNIa ejecta should have an Ni/Fe yield ratio close
to the solar one  and that the relative contribution to the Ni and Fe inventory in the
solar system from these two different astrophysical sites should be similar,
with the SNIa producing about two-thirds of the solar Fe and Ni.
For the stars with [Fe/H] over $\sim$ 0.1, the observed
increasing trend of [Ni/Fe] can be explained by the decrease in the Fe yields
from the SNIa with the increasing  metallicity.

Considering four different sets of theoretical stellar tracks, we showed that
under large uncertainties it is not possible to define a clear age -
metallicity trend in the thin disk stars from our sample, as  already pointed out
 for the thin disk. That will not affect  the
chemical evolution of the primary elements such as the $\alpha$ elements and Ni  too much,
instead of that, we  expect it to cause a noticeable dispersion for the elements
 whose production can  change significantly with metallicity.

We discussed the differences between the Y and Zr trends. In particular, the [Zr/Fe]
ratio is slightly higher in the thick disk compared to the thin disk ($\sim$ 0.2
dex). On the other hand, the [Y/Fe] shows  a flat trend in first approximation for
the observed metallicity range. Such a difference may be due to a larger primary
contribution to Zr compared to Y, by the $r$-process and  the LEPP component,
as predicted by the theoretical calculations.

Ba, La, and Ce are mostly produced by the $s$-process in the AGB stars, with their yields
significantly affected by the initial metal content in the range of the
metallicity considered. In our stellar sample, the thin disk stars show a dispersion
of about 0.4 dex at the solar-like metallicity. For [Fe/H] $\gtrsim$ 0.1, they may
start showing a decreasing trend, at least for the bulk of the stars, noticeable
in particular for La.
We cannot confirm any particular trend by  [Ba/Fe] and [La/Fe] versus the age, also
due to the large uncertainties in age determination.

Eu is mainly made by the primary $r$-process. We confirmed its decreasing trend
with metallicity, which was also observed in previous works. In particular, in our stellar
sample we found a really small dispersion, and it is well reproduced by the GCE
calculations.

Finally, for the metallicities typical of  the thick disk, [Sm/Fe] and [Nd/Fe] show a
higher ratio  than the solar one, due to the larger $r$-process contribution compared
to Ba, La, and Ce (comparable to the $s$-process contribution for Nd, and around
70 \% for Sm).
Within uncertainties and intrinsic dispersion, most of the stars show  a
decreasing trend for Sm moving from the typical thick disk metallicities to the thin disk ones,
such as  for Eu. For Nd, we found a  more similar trend to the $s$-process elements
discussed before, with [Nd/Fe] decreasing with metallicity for [Fe/H] $\gtrsim$
0, and at the same time an increase in the abundance dispersion.

\begin{figure}
\resizebox{\hsize}{!}
{\includegraphics{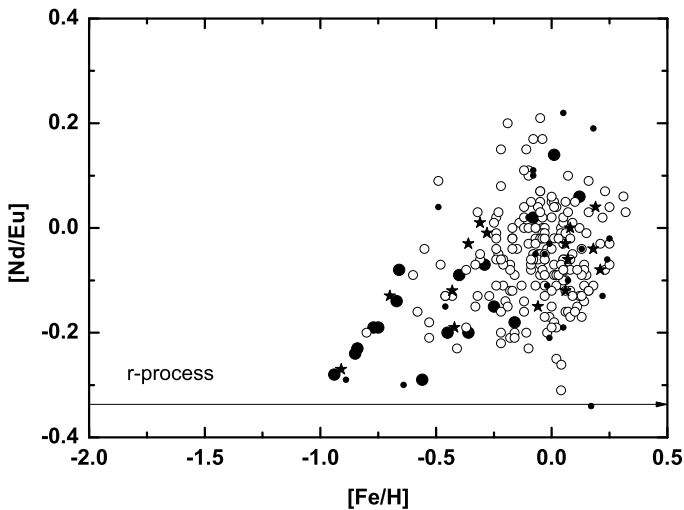}}
\caption[]{
Dependence of [Nd/Eu] on [Fe/H], the notation is the same as in
Fig.\ref{alpha_fe}
}
\label {fig53}
\end{figure}

\begin{acknowledgements}

The authors thank the anonymous referee for a careful reading of the manuscript.
TM thanks Prof. I.L. Andronov for some suggestions.  
This work was supported by the Swiss National Science Foundation
(SCOPES project No.~IZ73Z0-128180). MP thanks  the
Ambizione grant of the SNSF (Switzerland)  for  support. FKT, MP, and CC also thank to
Eurogenesis. TM is grateful  the Laboratoire d'Astrophysique de Bordeaux
for their kind hospitality.

\end{acknowledgements}

\begin {longtable}{rccccccccc}
\label{t_abund}\\
\caption{The stellar parameters and $\alpha$-element and Ni abundances in
studied stars }\\
\hline\hline
HD    & Teff &  log g & [Fe/H] & Probab   &[O/Fe] &[Mg/Fe] &[Si/Fe] &[Ca/Fe] &[Ni/Fe]\\
\hline
\endfirsthead
\caption{continued.}\\
\hline
HD    & Teff &  log g & [Fe/H]  & Probab  &[O/Fe] &[Mg/Fe] &[Si/Fe] &[Ca/Fe] &[Ni/Fe]\\
\hline
\endhead
\hline
\endfoot
\hline
Thick    &      &        &       &  &       &        &        &        &       \\
\hline
HD    & Teff &  log g & [Fe/H] & Probab   &[O/Fe] &[Mg/Fe] &[Si/Fe] &[Ca/Fe] &[Ni/Fe]\\
\hline
\object{HD000245} & 5400 & 3.40 & -0.84 &0.95 &0.33 & 0.32 & 0.29 & 0.38 & -0.02\\
\object{HD003765} & 5079 & 4.30 & 0.01  &0.9  &   & 0.02 & 0.09 & 0.06 & 0.04\\
\object{HD005351} & 4378 & 4.60 & -0.21 &0.95 &0.14 & 0.09 & 0.22 &    & 0.07\\
\object{HD006582} & 5240 & 4.30 & -0.94 &0.69 &0.51 & 0.37 & 0.32 & 0.32 & 0.00\\
\object{HD013783} & 5350 & 4.10 & -0.75 &0.72 &   & 0.41 & 0.30 & 0.25 & 0.02\\
\object{HD018757} & 5741 & 4.30 & -0.25 &0.92 &0.08 & 0.35 & 0.06 & 0.19 & 0.03\\
\object{HD022879} & 5972 & 4.50 & -0.77 &0.93 &0.48 & 0.42 & 0.28 & 0.25 & 0.02\\
\object{HD065583} & 5373 & 4.60 & -0.67 &0.96 &0.48 & 0.22 & 0.26 & 0.20 & 0.10\\
\object{HD076932} & 5840 & 4.00 & -0.95 &0.89 &0.51 & 0.38 & 0.33 & 0.38 & 0.03\\
\object{HD106516} & 6165 & 4.40 & -0.72 &0.96 &0.50 & 0.25 & 0.35 & 0.36 & 0.07\\
\object{HD110897} & 5925 & 4.20 & -0.45 &0.62 &   & 0.25 & 0.10 & 0.06 & -0.01\\
\object{HD135204} & 5413 & 4.00 & -0.16 &0.97 &   & 0.23 & 0.13 & 0.18 & 0.00\\
\object{HD152391} & 5495 & 4.30 & -0.08 &0.95 &0.07 & 0.06 & 0.04 & 0.04 & 0.01\\
\object{HD157089} & 5785 & 4.00 & -0.56 &0.79 &0.41 & 0.29 & 0.24 & 0.23 & -0.03\\
\object{HD157214} & 5820 & 4.50 & -0.29 &0.97 &   & 0.13 & 0.14 & 0.21 & -0.01\\
\object{HD159062} & 5414 & 4.30 & -0.40 &0.57 &   & 0.28 & 0.22 & 0.17 & 0.06\\
\object{HD165401} & 5877 & 4.30 & -0.36 &0.97 &   & 0.26 & 0.19 & 0.15 & 0.05\\
\object{HD190360} & 5606 & 4.40 & 0.12  &0.63 &   & 0.06 & 0.13 & -0.02 & 0.10\\
\object{HD201889} & 5600 & 4.10 & -0.85 &0.93 &   & 0.40 & 0.33 & 0.40 & -0.02\\
\object{HD201891} & 5850 & 4.40 & -0.96 &0.89 &0.40 & 0.29 & 0.32 & 0.18 & -0.03\\
\object{HD204521} & 5809 & 4.60 & -0.66 &0.6  &   & 0.27 & 0.20 & 0.15 & 0.04\\
\hline
Thin     &      &        &   &      &       &        &        &        &       \\
\hline
\object{HD000166} & 5514 & 4.60 & 0.16 &  0.99 & -0.23 & -0.13 & -0.02 & 0.03 & 0.02\\
\object{HD001562} & 5828 & 4.00 & -0.32&  0.98 & 0.16 & 0.18 & 0.07 & 0.23 & -0.02\\
\object{HD001835} & 5790 & 4.50 & 0.13 &  0.99 & -0.07 & 0.00 & 0.07 & 0.02 & 0.03\\
\object{HD003651} & 5277 & 4.50 & 0.15 &  0.98 & -0.32 & -0.02 & 0.05 & 0.01 & 0.11\\
\object{HD004256} & 5020 & 4.30 & 0.08 &  0.9  & -0.25 & 0.08 & 0.17 & 0.12 & 0.16\\
\object{HD004307} & 5889 & 4.00 & -0.18&  0.99 & 0.17 & 0.06 & 0.05 & 0.06 & 0.00\\
\object{HD004614} & 5965 & 4.40 & -0.24&  0.99 & 0.33 & 0.10 & 0.09 & 0.02 & 0.02\\
\object{HD005294} & 5779 & 4.10 & -0.17&  0.99 & -0.02 & 0.08 & 0.02 & 0.11 & -0.03\\
\object{HD006660} & 4759 & 4.60 & 0.08 &  0.92 & -0.11 & -0.02 & 0.14 & 0.06 & 0.05\\
\object{HD007590} & 5962 & 4.40 & -0.10&  0.99 & -0.07 & 0.06 & 0.05 & 0.22 & -0.06\\
\object{HD007924} & 5165 & 4.40 & -0.22&  0.99 & 0.10 & 0.18 & 0.09 & 0.07 & 0.01\\
\object{HD008648} & 5790 & 4.20 & 0.12 &  0.99 & -0.07 & 0.00 & 0.10 & 0.02 & 0.09\\
\object{HD009407} & 5666 & 4.45 & 0.05 &  0.99 &   & 0.01 & 0.05 & 0.03 & 0.05\\
\object{HD009826} & 6074 & 4.00 & 0.10 &  0.98 & -0.15 & 0.09 & 0.15 & 0.04 & -0.01\\
\object{HD010086} & 5696 & 4.30 & 0.13 &  0.98 & -0.13 & -0.02 & -0.02 & 0.05 & 0.00\\
\object{HD010307} & 5881 & 4.30 & 0.02 &  0.92 & -0.03 & 0.06 & 0.05 & -0.02 & 0.06\\
\object{HD010476} & 5242 & 4.30 & -0.05&  0.98 & 0.00 & 0.03 & 0.08 & -0.02 & -0.01\\
\object{HD010780} & 5407 & 4.30 & 0.04 &  0.99 & -0.13 & 0.04 & -0.01 & -0.02 & 0.03\\
\object{HD011007} & 5980 & 4.00 & -0.20&  0.96 & 0.16 & 0.10 & 0.06 & 0.07 & 0.00\\
\object{HD011373} & 4783 & 4.65 & 0.08 &  0.99 &   & -0.13 & 0.11 & -0.04 & 0.03\\
\object{HD012846} & 5766 & 4.50 & -0.24&  0.99 & 0.07 & 0.14 & 0.08 & 0.08 & 0.00\\
\object{HD013507} & 5714 & 4.50 & -0.02&  0.99 & 0.06 & -0.05 & 0.05 & 0.01 & -0.02\\
\object{HD014374} & 5449 & 4.30 & -0.09&  0.99 & 0.05 & 0.07 & 0.08 & 0.08 & 0.01\\
\object{HD016160} & 4829 & 4.60 & -0.16&  0.94 & -0.01 & 0.12 &    & 0.11 &    \\
\object{HD017674} & 5909 & 4.00 & -0.14&  0.99 & 0.15 & 0.11 & 0.06 & 0.08 & -0.03\\
\object{HD017925} & 5225 & 4.30 & -0.04&  0.99 & -0.06 & 0.04 & 0.05 & 0.10 & 0.03\\
\object{HD018632} & 5104 & 4.40 & 0.06 &  0.98 & -0.18 & 0.01 & 0.09 & 0.12 & 0.05\\
\object{HD018803} & 5665 & 4.55 & 0.14 &  0.99 & -0.26 & -0.04 & 0.01 & -0.02 & 0.04\\
\object{HD019019} & 6063 & 4.00 & -0.17&  0.99 & 0.05 & 0.15 & 0.05 & 0.17 & -0.11\\
\object{HD019373} & 5963 & 4.20 & 0.06 &  0.96 & -0.06 & -0.03 & 0.08 & -0.01 & 0.06\\
\object{HD020630} & 5709 & 4.50 & 0.08 &  0.99 & -0.08 & -0.12 & -0.02 & 0.04 & -0.03\\
\object{HD022049} & 5084 & 4.40 & -0.15&  0.99 & 0.04 & 0.03 & 0.10 & 0.07 & 0.00\\
\object{HD022484} & 6037 & 4.10 & -0.03&  0.95 & 0.14 & 0.02 & 0.05 &   & -0.01\\
\object{HD022556} & 6155 & 4.20 & -0.17&  0.99 & 0.14 & 0.07 & 0.12 & 0.03 & 0.03\\
\object{HD024053} & 5723 & 4.40 & 0.04 &  0.99 & -0.03 & -0.01 & 0.07 & -0.01 & -0.04\\
\object{HD024238} & 4996 & 4.30 & -0.46&  0.98 & 0.42 & 0.26 & 0.25 & 0.19 & 0.09\\
\object{HD024496} & 5536 & 4.30 & -0.13&  0.98 & -0.14 & 0.14 & 0.10 & 0.12 & -0.03\\
\object{HD025665} & 4967 & 4.70 & 0.01 &  0.98 & -0.01 & -0.08 & 0.04 & -0.02 & 0.08\\
\object{HD025680} & 5843 & 4.50 & 0.05 &  0.99 & 0.01 & -0.07 & 0.00 & 0.08 & -0.02\\
\object{HD026923} & 5920 & 4.40 & -0.03&  0.99 & 0.11 & 0.11 & -0.02 & 0.10 & -0.05\\
\object{HD028005} & 5980 & 4.20 & 0.23 &  0.97 & -0.27 & 0.15 & 0.12 & -0.05 & 0.10\\
\object{HD028447} & 5639 & 4.00 & -0.09&  0.99 & 0.07 & 0.07 & 0.10 & 0.01 & 0.04\\
\object{HD029150} & 5733 & 4.30 & 0.00 &  0.99 &   & 0.09 & 0.05 & 0.00 & 0.00\\
\object{HD029310} & 5852 & 4.20 & 0.08 &  0.98 & -0.03 & 0.00 & 0.09 & 0.05 & -0.02\\
\object{HD029645} & 6009 & 4.00 & 0.14 &  0.97 & -0.09 & 0.03 & 0.08 & -0.03 & 0.06\\
\object{HD030495} & 5820 & 4.40 & -0.05&  0.99 & 0.09 & 0.06 & 0.07 & 0.04 & -0.01\\
\object{HD033632} & 6072 & 4.30 & -0.24&  0.99 & 0.24 & 0.12 & 0.07 & 0.06 & -0.05\\
\object{HD034411} & 5890 & 4.20 & 0.10 &  0.7  & 0.07 & -0.01 & 0.07 & -0.03 & 0.03\\
\object{HD037008} & 5016 & 4.40 & -0.41&  0.89 & 0.24 & 0.16 & 0.28 & 0.18 & 0.10\\
\object{HD037394} & 5296 & 4.50 & 0.09 &  0.98 & -0.21 & -0.04 & -0.03 & 0.02 & -0.01\\
\object{HD038858} & 5776 & 4.30 & -0.23&  0.97 & 0.16 & 0.11 & 0.04 & 0.10 & 0.01\\
\object{HD039587} & 5955 & 4.30 & -0.03&  0.99 &   & 0.07 & 0.06 & 0.02 & -0.07\\
\object{HD040616} & 5881 & 4.00 & -0.22&  0.97 &   & 0.05 & 0.02 & 0.08 & -0.03\\
\object{HD041330} & 5904 & 4.10 & -0.18&  0.96 & 0.22 & 0.04 & 0.08 & 0.06 & -0.01\\
\object{HD041593} & 5312 & 4.30 & -0.04&  0.99 & 0.03 & -0.08 & 0.05 & 0.05 & -0.03\\
\object{HD042618} & 5787 & 4.50 & -0.07&  0.98 & 0.07 & 0.07 & 0.03 & 0.04 & -0.01\\
\object{HD042807} & 5719 & 4.40 & -0.03&  0.98 & 0.03 & 0.03 & -0.01 & 0.07 & -0.05\\
\object{HD043587} & 5927 & 4.10 & -0.11&  0.99 & 0.09 & 0.08 & 0.08 & 0.07 & 0.06\\
\object{HD043856} & 6143 & 4.10 & -0.19&  0.99 &   & 0.07 & 0.08 & 0.08 & 0.00\\
\object{HD043947} & 6001 & 4.30 & -0.24&  0.99 & 0.23 & 0.12 & 0.08 & 0.06 & -0.03\\
\object{HD045088} & 4959 & 4.30 & -0.21&  0.99 &   & 0.14 & 0.15 & 0.18 & 0.06\\
\object{HD047752} & 4613 & 4.60 & -0.05&  0.98 & 0.05 & -0.09 & 0.14 & -0.10 & 0.09\\
\object{HD048682} & 5989 & 4.10 & 0.05 &  0.99 &   & 0.03 & 0.08 & 0.10 & -0.02\\
\object{HD050281} & 4712 & 3.90 & -0.20&  0.99 & 0.15 & -0.02 & 0.11 & 0.14 & -0.06\\
\object{HD050692} & 5911 & 4.50 & -0.10&  0.99 & 0.10 & 0.07 & 0.01 & 0.06 & -0.01\\
\object{HD051419} & 5746 & 4.10 & -0.37&  0.99 & 0.30 & 0.15 & 0.08 & 0.12 & -0.01\\
\object{HD051866} & 4934 & 4.40 & 0.00 &  0.63 &   & 0.08 & 0.09 & 0.02 & 0.07\\
\object{HD053927} & 4860 & 4.64 & -0.22&  0.97 &   & -0.02 & 0.09 &   & 0.06\\
\object{HD054371} & 5670 & 4.20 & 0.06 &  0.99 & -0.06 & 0.07 & -0.05 & 0.09 & -0.02\\
\object{HD055575} & 5949 & 4.30 & -0.31&  0.94 &   & 0.21 & 0.10 & 0.05 & 0.01\\
\object{HD058595} & 5707 & 4.30 & -0.31&  0.99 & 0.16 & 0.07 & 0.05 & 0.12 & 0.00\\
\object{HD059747} & 5126 & 4.40 & -0.04&  0.99 &   & 0.05& 0.05& 0.06& -0.0\\
\object{HD061606} & 4956 & 4.40 & -0.12&  0.99 & 0.04 & -0.10 & 0.08 & 0.01 & -0.04\\
\object{HD062613} & 5541 & 4.40 & -0.10&  0.97 & 0.08 & 0.07 & 0.04 & -0.01 & 0.02\\
\object{HD063433} & 5693 & 4.35 & -0.06&  0.99 & 0.10 & 0.02 & 0.02 & 0.07 & -0.05\\
\object{HD064468} & 5014 & 4.20 & 0.00 &  0.94 &   & 0.10 & 0.13 & 0.17 & 0.08\\
\object{HD064815} & 5864 & 4.00 & -0.33&  0.78 & 0.01 & 0.31 & 0.22 & 0.25 & 0.04\\
\object{HD065874} & 5936 & 4.00 & 0.05 &  0.99 & 0.01 & 0.09 & 0.08 & 0.05 & 0.06\\
\object{HD066573} & 5821 & 4.60 & -0.53&  0.97 &   & 0.23 & 0.24 & 0.19 & 0.08\\
\object{HD068638} & 5430 & 4.40 & -0.24&  0.96 & 0.18 & 0.05 & 0.07 & 0.10 & -0.02\\
\object{HD070923} & 5986 & 4.20 & 0.06 &  0.96 & 0.03 & 0.07 & 0.09 & -0.01 & 0.07\\
\object{HD071148} & 5850 & 4.20 & 0.00 &  0.92 &   & 0.03 & 0.05 & 0.00 & 0.01\\
\object{HD072760} & 5349 & 4.10 & 0.01 &  0.99 & 0.05 & 0.02 & -0.01 & 0.07 & -0.06\\
\object{HD072905} & 5884 & 4.40 & -0.07&  0.99 & 0.01 & 0.09 & 0.04 & 0.07 & -0.04\\
\object{HD073344} & 6060 & 4.10 & 0.08 &  0.98 &   & 0.08 & 0.10 & 0.08 & 0.03\\
\object{HD073667} & 4884 & 4.40 & -0.58&  0.96 & 0.46 & 0.26 & 0.27 & 0.18 & 0.06\\
\object{HD075732} & 5373 & 4.30 & 0.25 &  0.99 & -0.12 & 0.12 & 0.10 & 0.04 & 0.10\\
\object{HD075767} & 5823 & 4.20 & -0.01&  0.99 & -0.11 & 0.11 & 0.00 & 0.08 & -0.04\\
\object{HD076151} & 5776 & 4.40 & 0.05 &  0.99 &   & 0.06 & 0.09 & -0.01 & 0.05\\
\object{HD079969} & 4825 & 4.40 & -0.05&  0.95 &   & 0.01 & 0.07 & 0.00 & 0.00\\
\object{HD082106} & 4827 & 4.10 & -0.11&  0.99 & 0.02 & -0.03 & 0.05 & 0.07 & -0.03\\
\object{HD082443} & 5334 & 4.40 & -0.03&  0.99 & 0.09 & -0.02 & 0.01 & 0.11 & -0.06\\
\object{HD087883} & 5015 & 4.40 & 0.00 &  0.99 & 0.08 & 0.03 & 0.12 & 0.07 & 0.10\\
\object{HD088072} & 5778 & 4.30 & 0.00 &  0.98 & -0.21 & -0.05 & 0.09 & -0.01 & 0.03\\
\object{HD089251} & 5886 & 4.00 & -0.12&  0.99 &   & 0.05 & 0.07 & 0.06 & 0.03\\
\object{HD089269} & 5674 & 4.40 & -0.23&  0.98 & 0.21 & 0.13 & 0.09 & 0.05 & 0.03\\
\object{HD091347} & 5931 & 4.40 & -0.43&  0.98 &   & 0.14 & 0.15 & 0.06 & 0.05\\
\object{HD094765} & 5077 & 4.40 & -0.01&  0.99 & 0.04 & 0.06 & 0.05 & 0.05 & 0.03\\
\object{HD095128} & 5887 & 4.30 & 0.01 &  0.99 & -0.07 & 0.07 & 0.05 & -0.01 & 0.03\\
\object{HD097334} & 5869 & 4.40 & 0.06 &  0.98 & -0.23 & -0.01 & 0.04 & 0.10 & -0.03\\
\object{HD097658} & 5136 & 4.50 & -0.32&  0.99 & 0.25 & 0.08 & 0.13 & 0.01 & 0.03\\
\object{HD098630} & 6060 & 4.00 & 0.22 &  0.98 & -0.14 & 0.09 & 0.16 & 0.01 & 0.12\\
\object{HD101177} & 5932 & 4.10 & -0.16&  0.9  & 0.21 & 0.06 & 0.04 & 0.09 & -0.07\\
\object{HD102870} & 6055 & 4.00 & 0.13 &  0.99 & -0.03 & 0.08 & 0.07 & -0.02 & 0.04\\
\object{HD105631} & 5416 & 4.40 & 0.16 &  0.98 & -0.28 & -0.06 & -0.02 & 0.02 & 0.05\\
\object{HD107705} & 6040 & 4.20 & 0.06 &  0.99 & -0.04 & -0.10 & 0.09 & 0.00 & 0.01\\
\object{HD108954} & 6037 & 4.40 & -0.12&  0.99 & 0.10 & 0.09 & 0.07 & 0.06 & -0.01\\
\object{HD109358} & 5897 & 4.20 & -0.18&  0.99 & 0.11 & 0.09 & 0.05 & 0.06 & 0.02\\
\object{HD110463} & 4950 & 4.50 & -0.05&  0.99 & -0.02 & 0.03 & 0.00 & 0.13 & -0.07\\
\object{HD110833} & 5075 & 4.30 & 0.00 &  0.99 & -0.01 & -0.04 & 0.10 & 0.02 & 0.05\\
\object{HD111395} & 5648 & 4.60 & 0.10 &  0.99 &   & -0.01 & 0.01 & 0.03 & 0.00\\
\object{HD112758} & 5203 & 4.20 & -0.56&  0.85 & 0.43 & 0.19 & 0.23 & 0.22 & 0.07\\
\object{HD114710} & 5954 & 4.30 & 0.07 &  0.99 & -0.05 & -0.07 & 0.05 & -0.05 & -0.02\\
\object{HD115383} & 6012 & 4.30 & 0.11 &  0.99 & -0.03 & 0.04 & 0.10 & 0.06 & 0.02\\
\object{HD115675} & 4745 & 4.45 & 0.02 &  0.9  &   & -0.05 & 0.14 & 0.12 & 0.13\\
\object{HD116443} & 4976 & 3.90 & -0.48&  0.98 & 0.32 & 0.06 & 0.16 & 0.14 & -0.03\\
\object{HD116956} & 5386 & 4.55 & 0.08 &  0.99 & -0.31 & -0.11 & 0.04 & 0.05 & 0.01\\
\object{HD117043} & 5610 & 4.50 & 0.21 &  0.93 & -0.15 & 0.02 & 0.04 & -0.08 & 0.13\\
\object{HD119802} & 4763 & 4.00 & -0.05&  0.99 &   & -0.04 & 0.07 & 0.09 & -0.06\\
\object{HD122064} & 4937 & 4.50 & 0.07 &  0.99 & 0.06 & 0.01 & 0.14 & -0.06 & 0.09\\
\object{HD124642} & 4722 & 4.65 & 0.02 &  0.99 & 0.05 & -0.12 & 0.16 & -0.01 & 0.03\\
\object{HD125184} & 5695 & 4.30 & 0.31 &  0.96 & -0.30 & -0.05 & 0.01 & -0.11 & 0.10\\
\object{HD126053} & 5728 & 4.20 & -0.32&  0.96 & 0.19 & 0.17 & 0.12 & 0.08 & -0.02\\
\object{HD127506} & 4542 & 4.60 & -0.08&  0.96 &   & -0.07 & 0.08 &   & 0.03\\
\object{HD128311} & 4960 & 4.40 & 0.03 &  0.99 &   & -0.01 & 0.11 & 0.01 & 0.05\\
\object{HD130307} & 4990 & 4.30 & -0.25&  0.98 & 0.02 & 0.08 & 0.05 & -0.02 & -0.06\\
\object{HD130948} & 5943 & 4.40 & -0.05&  0.99 & 0.14 & 0.02 & 0.04 & 0.09 & -0.08\\
\object{HD131977} & 4683 & 3.70 & -0.24&  0.98 & 0.15 & 0.12 & 0.18 & 0.19 & 0.03\\
\object{HD135599} & 5257 & 4.30 & -0.12&  0.99 & 0.08 & 0.00 & 0.05 & 0.04 & -0.01\\
\object{HD137107} & 6037 & 4.30 & 0.00 &  0.98 & 0.00 & 0.05 & 0.06 & 0.06 & -0.05\\
\object{HD139777} & 5771 & 4.40 & 0.01 &  0.98 &   & -0.01 & 0.05 & 0.11 & -0.02\\
\object{HD139813} & 5408 & 4.50 & 0.00 &  0.98 &   & 0.01 & 0.09 & 0.09 & -0.04\\
\object{HD140538} & 5675 & 4.50 & 0.02 &  0.99 & 0.06 & -0.02 & 0.05 & 0.09 & 0.05\\
\object{HD141004} & 5884 & 4.10 & -0.02&  0.91 &   & 0.10 & 0.08 & 0.02 & 0.03\\
\object{HD141272} & 5311 & 4.40 & -0.06&  0.97 &   & 0.04 & 0.07 & 0.05 & -0.02\\
\object{HD142267} & 5856 & 4.50 & -0.37&  0.96 & 0.25 & 0.10 & 0.05 & 0.07 & 0.14\\
\object{HD144287} & 5414 & 4.50 & -0.15&  0.94 &   & 0.07 & 0.12 & 0.10 & 0.00\\
\object{HD145675} & 5406 & 4.50 & 0.32 &  0.99 & -0.14 & 0.06 & 0.09 & -0.04 & 0.13\\
\object{HD146233} & 5799 & 4.40 & 0.01 &  0.99 & -0.12 & 0.04 & 0.05 & -0.08 & 0.03\\
\object{HD149661} & 5294 & 4.50 & -0.04&  0.99 & 0.06 & 0.02 & 0.09 & 0.05 & 0.05\\
\object{HD149806} & 5352 & 4.55 & 0.25 &  0.99 & -0.25 & -0.14 & 0.07 & -0.02 & 0.02\\
\object{HD151541} & 5368 & 4.20 & -0.22&  0.98 & 0.23 & 0.05 & 0.03 & 0.07 & -0.06\\
\object{HD153525} & 4810 & 4.70 & -0.04&  0.99 & 0.11 & -0.08 & 0.08 &   & 0.08\\
\object{HD154345} & 5503 & 4.30 & -0.21&  0.91 &   & 0.07 & 0.10 & 0.12 & 0.00\\
\object{HD156668} & 4850 & 4.20 & -0.07&  0.98 & -0.05 & 0.08 & 0.22 & 0.06 & 0.05\\
\object{HD156985} & 4790 & 4.60 & -0.18&  0.99 &   & 0.05 & 0.20 &   & 0.07\\
\object{HD158633} & 5290 & 4.20 & -0.49&  0.66 &   & 0.17 & 0.15 & 0.11 & 0.00\\
\object{HD160346} & 4983 & 4.30 & -0.10&  0.99 & 0.10 & -0.01 & 0.09 & 0.11 & 0.00\\
\object{HD161098} & 5617 & 4.30 & -0.27&  0.86 & 0.24 & 0.07 & 0.09 & 0.12 & -0.02\\
\object{HD164922} & 5392 & 4.30 & 0.04 &  0.88 & -0.01 & 0.08 & 0.11 & 0.12 & 0.07\\
\object{HD165173} & 5505 & 4.30 & -0.05&  0.99 & 0.06 & 0.13 & 0.10 & 0.10 & 0.04\\
\object{HD165341} & 5314 & 4.30 & -0.08&  0.99 & 0.02 & 0.03 & 0.10 & 0.15 & 0.03\\
\object{HD165476} & 5845 & 4.10 & -0.06&  0.94 & 0.05 & 0.01 & 0.04 & 0.00 & -0.04\\
\object{HD165670} & 6178 & 4.00 & -0.10&  0.99 & 0.07 & -0.03 & 0.11 & 0.07 & 0.01\\
\object{HD165908} & 5925 & 4.10 & -0.60&  0.99 & 0.29 & 0.16 & 0.10 & 0.20 & -0.06\\
\object{HD166620} & 5035 & 4.00 & -0.22&  0.98 & 0.13 & 0.13 & 0.13 & 0.24 & 0.00\\
\object{HD171314} & 4608 & 4.65 & 0.07 &  0.98 & -0.16 & -0.11 & 0.11 & 0.00 & 0.08\\
\object{HD174080} & 4764 & 4.55 & 0.04 &  0.98 & -0.04 & -0.07 & 0.01 & 0.18 & -0.11\\
\object{HD175742} & 5030 & 4.50 & -0.03&  0.98 &   & -0.12 & 0.01 & 0.06 & -0.04\\
\object{HD176377} & 5901 & 4.40 & -0.17&  0.97 &   & 0.04 & 0.18 & 0.04 & 0.17\\
\object{HD176841} & 5841 & 4.30 & 0.23 &  0.98 & -0.10 & 0.07 &   & -0.02 &   \\
\object{HD178428} & 5695 & 4.40 & 0.14 &  0.99 & -0.26 & -0.06 & -0.05 & -0.01 & 0.03\\
\object{HD180161} & 5473 & 4.50 & 0.18 &  0.97 & -0.35 & -0.10 & 0.02 & 0.02 & 0.08\\
\object{HD182488} & 5435 & 4.40 & 0.07 &  0.99 & 0.01 & 0.05 & 0.10 & 0.03 & 0.09\\
\object{HD183341} & 5911 & 4.30 & -0.01&  0.88 & -0.05 & 0.07 & 0.09 & 0.03 & 0.01\\
\object{HD184385} & 5536 & 4.45 & 0.12 &  0.99 & -0.12 & -0.07 & 0.00 & 0.02 & -0.04\\
\object{HD185144} & 5271 & 4.20 & -0.33&  0.94 & 0.30 & 0.01 & 0.06 & 0.09 & -0.04\\
\object{HD185414} & 5818 & 4.30 & -0.04&  0.99 & -0.05 & 0.05 & 0.08 & 0.02 & 0.04\\
\object{HD186408} & 5803 & 4.20 & 0.09 &  0.98 &   & 0.09 & 0.06 & 0.02 & 0.03\\
\object{HD186427} & 5752 & 4.20 & 0.02 &  0.98 &   & 0.09 & 0.07 & -0.06 & 0.05\\
\object{HD187897} & 5887 & 4.30 & 0.08 &  0.99 &   & 0.05 & 0.08 & -0.05 & 0.02\\
\object{HD189087} & 5341 & 4.40 & -0.12&  0.99 & 0.14 & 0.02 & 0.11 & 0.01 & 0.00\\
\object{HD189733} & 5076 & 4.40 & -0.03&  0.99 & -0.19 & 0.02 & -0.01 & 0.06 & 0.03\\
\object{HD190007} & 4724 & 4.50 & 0.16 &  0.99 & -0.16 & -0.11 & 0.10 & 0.13 & 0.03\\
\object{HD190406} & 5905 & 4.30 & 0.05 &  0.98 & -0.12 & -0.02 & 0.02 & 0.09 & 0.03\\
\object{HD190470} & 5130 & 4.30 & 0.11 &  0.99 &   & 0.00 & 0.03 & -0.04 & -0.04\\
\object{HD190771} & 5766 & 4.30 & 0.13 &  0.99 &   & -0.07 & 0.14 & 0.04 & 0.06\\
\object{HD191533} & 6167 & 3.80 & -0.10&  0.97 & 0.08 & 0.03 & 0.09 & 0.12 & -0.03\\
\object{HD191785} & 5205 & 4.20 & -0.12&  0.62 & 0.15 & 0.31 & 0.07 & 0.15 & 0.06\\
\object{HD195005} & 6075 & 4.20 & -0.06&  0.99 & 0.12 & 0.12 & 0.08 & 0.11 & -0.07\\
\object{HD195104} & 6103 & 4.30 & -0.19&  0.99 & 0.21 & 0.16 & 0.07 & 0.10 & -0.09\\
\object{HD197076} & 5821 & 4.30 & -0.17&  0.98 & 0.16 & 0.07 & 0.09 & 0.11 & 0.01\\
\object{HD199960} & 5878 & 4.20 & 0.23 &  0.98 & -0.44 & 0.02 & 0.10 & 0.02 & 0.11\\
\object{HD200560} & 5039 & 4.40 & 0.06 &  0.98 &   & -0.03 & 0.12 & 0.07 & 0.09\\
\object{HD202108} & 5712 & 4.20 & -0.21&  0.99 & 0.15 & 0.06 & 0.05 & 0.02 & -0.05\\
\object{HD202575} & 4667 & 4.60 & -0.03&  0.99 & 0.06 & -0.07 & 0.04 &   & -0.02\\
\object{HD203235} & 6071 & 4.10 & 0.05 &  0.98 &   & 0.10 & 0.11 & 0.06 & -0.01\\
\object{HD205702} & 6020 & 4.20 & 0.01 &  0.98 & 0.03 & 0.10 & 0.08 & 0.11 & 0.06\\
\object{HD206860} & 5927 & 4.60 & -0.07&  0.99 &   & -0.08 & 0.02 & 0.06 & -0.03\\
\object{HD208038} & 4982 & 4.40 & -0.08&  0.99 & -0.19 & -0.04 & 0.05 & 0.01 & 0.01\\
\object{HD208313} & 5055 & 4.30 & -0.05&  0.98 & -0.02 & 0.10 & 0.20 & 0.04 & 0.04\\
\object{HD208906} & 5965 & 4.20 & -0.80&  0.98 &   & 0.23 & 0.25 & 0.15 & 0.04\\
\object{HD210667} & 5461 & 4.50 & 0.15 &  0.98 & -0.31 & -0.10 & 0.06 & 0.01 & 0.07\\
\object{HD210752} & 6014 & 4.60 & -0.53&  0.68 &   & 0.07 & 0.11 & 0.07 & 0.02\\
\object{HD211472} & 5319 & 4.40 & -0.04&  0.99 & -0.07 & 0.00 & 0.05 & 0.00 & -0.01\\
\object{HD214683} & 4747 & 4.60 & -0.46&  0.99 &   & 0.02 & 0.17 &   & 0.09\\
\object{HD216259} & 4833 & 4.60 & -0.55&  0.99 &   & 0.32 & 0.04 &   & 0.03\\
\object{HD216520} & 5119 & 4.40 & -0.17&  0.99 & 0.20 & 0.09 & 0.01 & 0.02 & -0.07\\
\object{HD217014} & 5778 & 4.20 & 0.14 &  0.97 &   & 0.06 & 0.08 & -0.06 & 0.09\\
\object{HD217813} & 5845 & 4.30 & 0.03 &  0.99 & -0.30 & -0.01 & 0.04 & -0.01 & 0.09\\
\object{HD218868} & 5534 & 4.60 & 0.25 &  0.97 & -0.37 & -0.10 & 0.13 & -0.04 & 0.06\\
\object{HD219538} & 5114 & 4.40 & -0.09&  0.99 & 0.09 & 0.03 & 0.09 & 0.06 & -0.01\\
\object{HD219623} & 5949 & 4.20 & 0.04 &  0.97 &   & -0.05 & 0.11 & 0.07 & -0.07\\
\object{HD220140} & 5144 & 4.60 & -0.03&  0.99 &   & -0.18 & 0.03 & 0.16 & -0.02\\
\object{HD220182} & 5364 & 4.50 & -0.03&  0.98 & 0.03 & 0.05 & 0.01 & 0.05 & 0.08\\
\object{HD220221} & 4868 & 4.50 & 0.16 &  0.99 & -0.22 & -0.08 & 0.12 & 0.05 & 0.02\\
\object{HD221851} & 5184 & 4.40 & -0.09&  0.98 & 0.07 & 0.10 & 0.02 & 0.07 & -0.04\\
\object{HD222143} & 5781 & 4.30 & 0.09 &  0.99 & -0.09 & -0.03 & 0.04 & 0.09 & 0.05\\
\object{HD224465} & 5745 & 4.50 & 0.08 &  0.98 & -0.05 & -0.05 & 0.24 & 0.04 & 0.02\\
\object{HD263175} & 4734 & 4.50 & -0.16&  0.78 & 0.04 & 0.02 & 0.08 &   & 0.06\\
\object{BD+01 2063} & 4859 & 4.40 & -0.22 & 0.99  &  0.10 & 0.00 & 0.15 & -0.02 & 0.13\\
\object{BD+12 4499} & 4678 & 4.70 & 0.00 &  0.97  &  0.03 & -0.12 &    & 0.04 & -0.01\\
\hline
Hercules &      &        &         &       &        &        &        &       \\
\hline
\object{HD013403} & 5724 & 4.00 & -0.31 & 0.65& 0.28 & 0.10 & 0.06 & -0.11 & 0.00\\
\object{HD019308} & 5844 & 4.30 & 0.08  & 0.67& 0.04 & 0.00 & 0.09 & 0.09 & 0.03\\
\object{HD023050} & 5929 & 4.40 & -0.36 & 0.76& 0.33 & 0.21 & 0.18 & -0.23 & 0.03\\
\object{HD030562} & 5859 & 4.00 & 0.18  & 0.85& -0.11 & 0.03 & 0.06 & 0.24 & 0.08\\
\object{HD064606} & 5250 & 4.20 & -0.91 & 0.85& & 0.37 & 0.27 & -0.57 & 0.03\\
\object{HD068017} & 5651 & 4.20 & -0.42 & 0.77& 0.35 & 0.31 & 0.21 & -0.23 & 0.05\\
\object{HD081809} & 5782 & 4.00 & -0.28 & 0.58& 0.28 & 0.16 & 0.22 & -0.18 & 0.08\\
\object{HD107213} & 6156 & 4.10 & 0.07  & 0.78& & 0.14 & 0.14 & 0.19 & 0.02\\
\object{HD139323} & 5204 & 4.60 & 0.19  & 0.81& & 0.05 & 0.09 & 0.25 & 0.01\\
\object{HD139341} & 5242 & 4.60 & 0.21  & 0.75& & -0.02 & 0.04 & 0.19 & 0.12\\
\object{HD144579} & 5294 & 4.10 & -0.70 & 0.84& & 0.37 & 0.28 & -0.39 & 0.05\\
\object{HD159222} & 5834 & 4.30 & 0.06  & 0.71& & 0.00 & 0.08 & 0.13 & 0.03\\
\object{HD159909} & 5749 & 4.10 & 0.06  & 0.88& & 0.06 & 0.07 & 0.07 & 0.01\\
\object{HD215704} & 5418 & 4.20 & 0.07  & 0.71& & 0.04 & 0.06 & 0.06 & 0.05\\
\object{HD218209} & 5705 & 4.50 & -0.43 & 0.81& 0.22 & 0.19 & 0.18 & -0.35 & 0.04\\
\object{HD221354} & 5242 & 4.10 & -0.06 & 0.89& 0.09 & 0.31 & 0.12 &    & 0.02\\
\hline  
Nonclas & & & & & & & & \\
\hline  
\object{HD004628} & 4905 & 4.60 & -0.36 & & 0.36 & 0.21 & 0.18 & 0.07 & 0.02\\
\object{HD004635} & 5103 & 4.40 & 0.07  & & -0.34 & -0.02 & 0.05 & 0.01 & 0.08\\
\object{HD010145} & 5673 & 4.40 & -0.01 & & 0.03 & 0.13 & 0.07 & -0.04 & 0.05\\
\object{HD012051} & 5458 & 4.55 & 0.24  & & -0.41 & -0.09 & 0.00 & -0.04 & 0.06\\
\object{HD013974} & 5590 & 3.80 & -0.49 & & 0.21 & 0.17 & 0.09 & 0.19 & -0.08\\
\object{HD017660} & 4713 & 4.75 & 0.17  & & -0.13 & -0.19 & 0.15 & 0.11 & 0.12\\
\object{HD020165} & 5145 & 4.40 & -0.08 & & 0.11 & 0.09 & 0.10 & 0.04 & 0.05\\
\object{HD024206} & 5633 & 4.50 & -0.08 & & 0.05 & 0.05 & 0.09 & -0.05 & 0.03\\
\object{HD032147} & 4945 & 4.40 & 0.13  & & -0.08 & -0.06 & 0.11 & 0.05 & 0.03\\
\object{HD045067} & 6058 & 4.00 & -0.02 & & 0.01 & 0.04 & 0.05 & 0.03 &   \\
\object{HD084035} & 4808 & 4.80 & 0.25  & & -0.25 & -0.22 & 0.07 & 0.16 & 0.00\\
\object{HD086728} & 5725 & 4.30 & 0.22  & &   & -0.11 & 0.04 & 0.06 & 0.05\\
\object{HD090875} & 4788 & 4.50 & 0.24  & &   & -0.14 & 0.02 & 0.22 & 0.14\\
\object{HD117176} & 5611 & 4.00 & -0.03 & & 0.08 & 0.07 & 0.05 & 0.02 & 0.02\\
\object{HD117635} & 5230 & 4.30 & -0.46 & & & 0.21 & 0.17 & 0.19 & 0.01\\
\object{HD154931} & 5910 & 4.00 & -0.10 & & 0.04 & 0.14 & 0.08 & 0.12 & 0.05\\
\object{HD159482} & 5620 & 4.10 & -0.89 & & 0.50 & 0.37 & 0.34 & 0.36 & 0.02\\
\object{HD168009} & 5826 & 4.10 & -0.01 & & & 0.03 & 0.05 & 0.04 & 0.02\\
\object{HD173701} & 5423 & 4.40 & 0.18  & & -0.09 & 0.00 & 0.17 & -0.08 & 0.17\\
\object{HD182736} & 5430 & 3.70 & -0.06 & & & 0.04 & 0.01 & 0.00 & 0.05\\
\object{HD184499} & 5750 & 4.00 & -0.64 & & 0.40 & 0.37 & 0.34 & 0.26 & 0.07\\
\object{HD184768} & 5713 & 4.20 & -0.07 & & 0.11 & 0.15 & 0.15 & 0.03 & 0.06\\
\object{HD186104} & 5753 & 4.20 & 0.05  & & & 0.06 & 0.07 & 0.03 & 0.04\\
\object{HD215065} & 5726 & 4.00 & -0.43 & & & 0.00 & 0.18 & 0.23 & 0.03\\
\object{HD219134} & 4900 & 4.20 & 0.05  & & -0.26 & 0.04 & 0.09 & 0.01 & 0.04\\
\object{HD219396} & 5733 & 4.00 & -0.10 & & & 0.16 & 0.10 & 0.10 & 0.03\\
\object{HD224930} & 5300 & 4.10 & -0.91 & & 0.44 & 0.42 & 0.29 & 0.34 & 0.01\\
\hline  
\end {longtable}

\begin {longtable}{rcccccccc}
\label{t_abund2}\\
\caption{ n-capture element abundances in studied stars }\\
\hline\hline
HD & [Y/Fe]& [Zr/Fe]& [Ba/Fe]& [La/Fe]& [Ce/Fe] & [Nd/Fe]& [Sm/Fe]& [Eu/Fe]\\
\hline
\endfirsthead
\caption{continued.}\\
\hline
HD & [Y/Fe]& [Zr/Fe]& [Ba/Fe]& [La/Fe]& [Ce/Fe] & [Nd/Fe]& [Sm/Fe]& [Eu/Fe]\\
\hline
\endhead
\hline
\endfoot
\hline
Thick & & & & & & & & \\
\hline
HD & [Y/Fe]& [Zr/Fe]& [Ba/Fe]& [La/Fe]& [Ce/Fe] & [Nd/Fe]& [Sm/Fe]& [Eu/Fe]\\
\hline
\object{HD000245} & -0.06 & 0.18 & 0.02 & 0.17 & -0.03 & 0.12 & 0.28 & 0.35 \\
\object{HD003765} & 0.06 & 0.02 & -0.09 & -0.20 & 0.06 & 0.17 & -0.05 & 0.03 \\
\object{HD005351} & 0.13 & 0.13 & -0.33 & 0.00 & -0.04 & 0.09 & 0.05 & 0.08 \\
\object{HD006582} & 0.04 & 0.11 & -0.12 & 0.10 & -0.06 & 0.13 & 0.19 & 0.41 \\
\object{HD013783} & -0.01 & 0.06 & -0.08 & 0.11 & -0.03 & 0.20 & 0.16 & 0.39 \\
\object{HD018757} & -0.14 & -0.08 & -0.08 & -0.11 & -0.07 & 0.07 & 0.05 & 0.22 \\
\object{HD022879} & 0.04 & 0.25 & 0.05 & 0.03 & -0.01 & 0.22 & 0.14 & 0.41 \\
\object{HD065583} & 0.05 & 0.08 & -0.07 & -0.09 & 0.08 & 0.27 & 0.34 & 0.41 \\
\object{HD076932} & 0.13 & 0.14 & 0.10 & 0.15 & -0.01 & 0.18 & 0.35 &    \\
\object{HD106516} & 0.06 & 0.24 & 0.07 & 0.09 & 0.01 & 0.11 & 0.22 &    \\
\object{HD110897} & -0.09 & 0.00 & -0.01 & 0.01 & -0.07 & 0.09 & 0.09 & 0.29 \\
\object{HD135204} & -0.05 & -0.06 & -0.11 & -0.12 & 0.03 & 0.02 & 0.05 & 0.20 \\
\object{HD152391} & 0.04 & -0.07 & 0.03 & -0.18 & 0.00 & 0.14 & 0.13 & 0.12 \\
\object{HD157089} & 0.00 & 0.08 & 0.02 & -0.08 & -0.09 & 0.08 & 0.16 & 0.37 \\
\object{HD157214} & 0.01 & 0.09 & -0.05 & 0.00 & 0.04 & 0.14 & 0.24 & 0.21 \\
\object{HD159062} & 0.11 & 0.14 & 0.15 & -0.04 & 0.05 & 0.20 & 0.21 & 0.29 \\
\object{HD165401} & -0.14 & 0.03 & -0.12 & -0.13 & -0.05 & 0.07 & 0.27 & 0.27 \\
\object{HD190360} & -0.02 & -0.11 & -0.06 & -0.18 & 0.12 & 0.08 & 0.09 & 0.02 \\
\object{HD201889} & 0.15 & 0.18 & 0.01 & 0.02 & -0.02 & 0.10 & 0.25 & 0.34 \\
\object{HD201891} & -0.06 & 0.27 & -0.06 & 0.15 & -0.05 & 0.20 & 0.33 &    \\
\object{HD204521} & 0.04 & 0.19 & -0.06 &    & 0.02 & 0.22 & 0.32 & 0.30 \\
\hline
Thin & & & & & & & & \\
\hline
\object{HD000166} & -0.05 & -0.15 & 0.12 &  -0.15 & -0.12 & 0.00 & -0.13 & -0.09 \\
\object{HD001562} & -0.02 & -0.06 & 0.00 &  -0.02 & 0.07 & 0.08 & 0.07 &    \\
\object{HD001835} & 0.02 & -0.08 & 0.04  &  -0.12 & 0.01 & -0.11 &    & 0.06 \\
\object{HD003651} & -0.12 & -0.11 & -0.14&  -0.2 & -0.12 & -0.09 & -0.18 & -0.08 \\
\object{HD004256} & -0.15 & -0.19 & -0.16&  -0.19 & -0.04 & -0.04 & -0.03 & \\
\object{HD004307} & -0.07 & -0.07 & 0.08 &  0.03 & -0.04 & 0.05 & 0.06 & 0.12 \\
\object{HD004614} & 0.05 & -0.02 & 0.02  &  -0.06 & -0.05 & 0.07 & 0.06 & 0.08 \\
\object{HD005294} & -0.04 & -0.10 & 0.15 &  -0.15 & 0.02 & -0.01 & -0.09 & 0.01 \\
\object{HD006660} & -0.10 & -0.13 & -0.15&  -0.12 & -0.17 & -0.11 & -0.09 & -0.03 \\
\object{HD007590} & -0.10 & -0.14 & 0.11 &  -0.03 & -0.10 & 0.04 & 0.01 & 0.07 \\
\object{HD007924} & 0.00 & 0.03 & -0.05  & -0.01 & 0.10 & 0.12 & 0.06 & 0.04 \\
\object{HD008648} & -0.02 & -0.07 & -0.04&  -0.12 & -0.04 & -0.13 & -0.19 & -0.13 \\
\object{HD009407} & -0.11 & -0.10 & -0.02&  -0.12 & -0.12 & -0.06 & -0.06 & -0.03 \\
\object{HD009826} & -0.11 &    & -0.02   &  -0.27 & -0.08 &    & -0.06 &    \\
\object{HD010086} & -0.18 & -0.15 & -0.06&  -0.09 & -0.17 & -0.17 & -0.18 & -0.08 \\
\object{HD010307} & 0.00 & -0.09 & -0.02 &  -0.08 & -0.01 & -0.13 & -0.10 & 0.12 \\
\object{HD010476} & -0.01 & -0.07 & 0.00 &  -0.02 & 0.05 & 0.15 & -0.10 & -0.06 \\
\object{HD010780} & -0.01 & 0.06 & 0.09  & 0.05 & 0.01 & 0.05 & -0.08 & 0.05 \\
\object{HD011007} & -0.11 & 0.07 & 0.05  &  -0.12 & -0.07 & 0.05 & 0.03 & 0.19 \\
\object{HD011373} & 0.00 & 0.01 & -0.04  &  -0.01 & -0.12 & -0.03 & -0.01 & -0.01 \\
\object{HD012846} & -0.11 & -0.18 & -0.04&  -0.08 & -0.01 & 0.05 & 0.11 & 0.16 \\
\object{HD013507} & 0.06 & -0.09 & 0.11  &  0.02 & 0.09 & 0.12 & -0.09 & 0.16 \\
\object{HD014374} & 0.10 & -0.06 & 0.02  &  -0.12 & 0.15 & 0.05 & -0.02 & 0.13 \\
\object{HD016160} & -0.08 & -0.07 & -0.19&  -0.12 & -0.11 & 0.08 & -0.08 & 0.28 \\
\object{HD017674} & -0.14 & -0.12 & -0.03&  -0.13 & -0.07 & -0.01 & -0.06 & -0.02 \\
\object{HD017925} & 0.01 & -0.07 & 0.03  &  -0.08 & 0.13 & 0.08 & -0.03 & 0.08 \\
\object{HD018632} & -0.13 & -0.19 & -0.04&  -0.18 & -0.13 & -0.10 & -0.09 & -0.04 \\
\object{HD018803} & -0.16 & -0.20 & 0.00 &  -0.18 & -0.14 & -0.13 & -0.19 & -0.02 \\
\object{HD019019} & -0.02 & 0.03 & 0.17  &  -0.14 & -0.04 & 0.07 & -0.03 &    \\
\object{HD019373} & -0.07 & -0.08 & -0.03&  -0.22 & -0.05 & -0.07 & -0.10 & 0.03 \\
\object{HD020630} & -0.12 & -0.18 & 0.07 &  -0.22 & -0.09 & -0.06 & -0.17 & \\
\object{HD022049} & 0.04 & 0.11 & 0.15   &  0.01 & 0.18 & 0.20 & 0.14 & 0.24 \\
\object{HD022484} & -0.13 & -0.10 & 0.03 &  -0.19 & -0.02 & -0.09 & -0.09 & 0.02 \\
\object{HD022556} & 0.00 & 0.05 & 0.04   &  -0.07 & 0.05 & 0.17 & -0.01 & 0.21 \\
\object{HD024053} & 0.07 & -0.08 & 0.11  &  0.00 & 0.03 & 0.09 & 0.08 & 0.10 \\
\object{HD024238} & -0.05 & -0.04 & -0.12&  -0.04 & -0.08 & 0.05 & 0.02 & 0.18 \\
\object{HD024496} & -0.10 & -0.19 & -0.12&  -0.14 & -0.02 & -0.01 & -0.07 & 0.10 \\
\object{HD025665} & -0.17 & -0.11 & -0.03&  -0.09 & -0.05 & -0.12 & -0.15 & 0.06 \\
\object{HD025680} & -0.07 & -0.20 & 0.05 &  -0.05 & -0.11 & -0.04 & -0.03 & 0.02 \\
\object{HD026923} & -0.05 & -0.04 & 0.28 &  0.00 & -0.12 & -0.01 & -0.01 & 0.00 \\
\object{HD028005} & 0.07 & -0.08 & 0.00  &  -0.17 & -0.03 & -0.06 & -0.14 & -0.13 \\
\object{HD028447} & 0.03 & -0.12 & 0.03  &  -0.07 & -0.02 & 0.06 & 0.02 & 0.13 \\
\object{HD029150} & 0.06 & -0.10 & -0.03 &  -0.01 & -0.03 & 0.00 & -0.01 & 0.04 \\
\object{HD029310} & -0.06 & -0.13 & 0.02 &     & -0.09 & -0.17 &    &    \\
\object{HD029645} & -0.12 & -0.09 & -0.07&  -0.20 & -0.11 & -0.17 & -0.18 & -0.10 \\
\object{HD030495} & 0.11 & 0.10 & 0.19   &  -0.06 & 0.07 & 0.14 & 0.18 & 0.07 \\
\object{HD033632} & 0.09 & 0.07 & 0.18   &  0.10 & 0.15 & 0.16 & 0.16 & 0.18 \\
\object{HD034411} & -0.14 & -0.06 & -0.01&  -0.18 & -0.08 & -0.17 & -0.09 & -0.01 \\
\object{HD037008} & -0.05 & -0.06 & -0.24&  -0.06 & -0.08 & 0.05 & 0.03 & 0.28 \\
\object{HD037394} & -0.11 & -0.20 & 0.06 &  -0.25 & -0.11 & -0.07 & -0.15 & -0.02 \\
\object{HD038858} & 0.01 & -0.02 & 0.03  &  -0.02 & 0.05 & 0.10 & 0.01 & 0.15 \\
\object{HD039587} & -0.10 & -0.01 & 0.14 &  -0.03 & -0.04 & -0.03 & 0.00 & -0.03 \\
\object{HD040616} & -0.06 & 0.04 & 0.12  &  -0.07 & -0.01 & 0.11 & 0.13 & -0.04 \\
\object{HD041330} & -0.18 & 0.01 & 0.01  &  -0.08 & -0.07 & 0.01 & -0.06 & 0.22 \\
\object{HD041593} & 0.04 & 0.01 & 0.10   &  0.00 & -0.01 & 0.10 & -0.14 & -0.07 \\
\object{HD042618} & -0.11 & -0.18 & 0.02 &  -0.04 & -0.08 & -0.02 & -0.01 & 0.09 \\
\object{HD042807} & -0.09 & -0.09 & 0.11 &  -0.07 & -0.07 & -0.01 & -0.08 & 0.05 \\
\object{HD043587} & -0.12 & 0.00 & -0.04 &  -0.11 & -0.06 & -0.05 & -0.09 & 0.15 \\
\object{HD043856} & -0.03 & -0.08 & 0.15 &  -0.03 & -0.03 & 0.03 & 0.14 & 0.18 \\
\object{HD043947} & 0.00 & 0.02 & 0.06   &  0.00 & -0.02 & 0.16 & 0.07 & 0.20 \\
\object{HD045088} & -0.06 & -0.19 & 0.04 &  -0.16 & -0.07 & 0.06 & -0.03 & 0.13 \\
\object{HD047752} & 0.04 & 0.05 & -0.02  &  0.00 & 0.01 & 0.01 & -0.01 & 0.10 \\
\object{HD048682} & -0.17 & -0.17 & -0.08&  -0.21 & -0.16 & -0.18 & -0.15 & -0.08 \\
\object{HD050281} & 0.02 &    & 0.00     &  -0.01 & 0.06 & 0.12 & 0.02 &    \\
\object{HD050692} & -0.08 & -0.18 & 0.03 &  -0.01 & -0.08 & -0.01 & 0.11 & 0.22 \\
\object{HD051419} & -0.05 & -0.03 & -0.08&  -0.10 & -0.01 & 0.07 & 0.01 & 0.26 \\
\object{HD051866} & -0.14 & -0.17 & -0.07&  -0.15 & -0.11 & -0.10 & -0.13 & 0.02 \\
\object{HD053927} & -0.14 & -0.13 & -0.02&  -0.10 & -0.09 & -0.03 & -0.12 & 0.19 \\
\object{HD054371} & -0.17 & -0.18 & -0.01&  -0.16 & -0.16 & -0.13 & -0.15 & 0.03 \\
\object{HD055575} & -0.07 & 0.04 & 0.02  &  0.06 & -0.07 & 0.14 & 0.13 & 0.20 \\
\object{HD058595} & 0.07 & 0.02 & 0.01   &  0.03 & 0.18 & 0.15 & 0.15 & 0.20 \\
\object{HD059747} & -0.08 & -0.14 & 0.09 &  -0.10 & 0.04 & 0.04 & -0.08 & 0.02 \\
\object{HD061606} & 0.00 &    & 0.02     &  0.03 & 0.19 & 0.15 & 0.20 & 0.13 \\
\object{HD062613} & 0.02 & 0.01 & 0.00   &  0.01 & 0.02 & 0.05 & 0.08 & -0.06 \\
\object{HD063433} & -0.16 & -0.18 & 0.02 &  -0.08 & -0.09 & -0.02 & -0.07 & 0.03 \\
\object{HD064468} & -0.19 & -0.12 & -0.17&  -0.14 & -0.17 & -0.08 & -0.13 & \\
\object{HD064815} & 0.07 & 0.07 & 0.07   &  0.01 & 0.03 & 0.17 & 0.24 & 0.32 \\
\object{HD065874} & 0.00 & -0.09 & -0.07 &  -0.05 & 0.01 & -0.10 & -0.11 & -0.11 \\
\object{HD066573} & 0.10 & 0.19 & -0.07  &  0.10 & 0.02 & 0.14 & 0.29 & 0.32 \\
\object{HD068638} & 0.07 & 0.02 & 0.05   &  0.15 & 0.18 & 0.10 & 0.12 & 0.08 \\
\object{HD070923} & -0.07 & -0.09 & -0.06&  0.07 & -0.08 & -0.13 & -0.07 & -0.12 \\
\object{HD071148} & -0.02 & -0.08 & -0.01&  0.05 & 0.03 & -0.01 & -0.06 & -0.06 \\
\object{HD072760} & -0.04 & -0.01 & 0.04 &  0.06 & 0.00 & -0.02 & -0.04 & 0.05 \\
\object{HD072905} & 0.03 & -0.05 & 0.11  &  0.05 & 0.01 & -0.01 & 0.04 & 0.01 \\
\object{HD073344} & -0.08 & -0.15 & -0.02&  0.05 & -0.15 & -0.16 & -0.09 & -0.04 \\
\object{HD073667} & 0.00 & 0.05 & -0.15  &  0.00 & -0.03 & 0.14 & 0.17 & 0.30 \\
\object{HD075732} & -0.20 & 0.00 & -0.13 &  -0.16 & -0.01 & -0.07 & 0.04 & -0.11 \\
\object{HD075767} & -0.13 & -0.14 & 0.04 &  -0.10 & -0.16 & -0.10 & -0.09 & \\
\object{HD076151} & -0.01 & -0.07 & -0.03&  0.13 & -0.03 & -0.04 & -0.12 & -0.06 \\
\object{HD079969} & -0.15 & -0.23 &      &  -0.08 & -0.16 & 0.02 & -0.14 & 0.07 \\
\object{HD082106} & -0.04 &    & 0.11    &  0.07 & 0.04 & 0.10 & 0.14 & -0.05 \\
\object{HD082443} & -0.13 & -0.20 & 0.13 &  -0.04 & -0.12 & -0.05 & -0.08 & 0.12 \\
\object{HD087883} & -0.05 & -0.15 & -0.05&  0.03 & -0.01 & 0.08 & -0.05 & 0.02 \\
\object{HD088072} & 0.01 & 0.06 & -0.03  &  0.07 & -0.01 & 0.17 & -0.06 & 0.15 \\
\object{HD089251} & 0.09 & -0.09 & 0.05  &  0.12 & -0.01 & 0.02 & 0.11 & 0.16 \\
\object{HD089269} & 0.07 & 0.08 & 0.03   &  0.06 & 0.10 & 0.23 & 0.12 & 0.20 \\
\object{HD091347} & -0.07 & -0.06 & -0.02&  -0.01 & -0.06 & 0.09 & 0.08 & 0.22 \\
\object{HD094765} & -0.02 & -0.09 & 0.07 &  0.00 & 0.01 & 0.06 & -0.04 & \\
\object{HD095128} & -0.04 & 0.00 & -0.05 &  -0.04 & -0.01 & 0.05 & -0.05 & 0.00 \\
\object{HD097334} & -0.12 & -0.17 & 0.13 &  -0.08 & -0.16 & -0.13 & -0.13 & -0.01 \\
\object{HD097658} & -0.12 & 0.01 & -0.03 &  -0.09 & -0.04 & 0.06 & 0.05 & 0.19 \\
\object{HD098630} & 0.02 & -0.18 & -0.09 &  -0.13 & -0.05 & -0.08 & -0.18 & -0.10 \\
\object{HD101177} & 0.02 & 0.00 & 0.01   &  -0.05 & 0.01 & 0.06 & 0.04 & 0.15 \\
\object{HD102870} & -0.07 & -0.04 & -0.03&  -0.10 & -0.06 & -0.13 & -0.15 & -0.09 \\
\object{HD105631} & -0.18 & -0.19 & -0.02&  -0.19 & -0.19 & -0.16 & -0.18 & -0.04 \\
\object{HD107705} & -0.05 & -0.13 & 0.06 &     & 0.05 & -0.13 & -0.03 & -0.05 \\
\object{HD108954} & 0.05 & 0.02 & 0.11   &  0.01 & 0.02 & 0.10 & -0.03 & 0.06 \\
\object{HD109358} & -0.04 & 0.06 & -0.05 &  -0.07 & -0.05 & -0.04 & 0.00 & 0.04 \\
\object{HD110463} & 0.05 & -0.03 & 0.04  &  0.00 & 0.05 & 0.16 & 0.00 & 0.09 \\
\object{HD110833} & 0.00 & 0.02 & -0.04  &  0.01 & 0.02 & 0.09 & -0.07 &    \\
\object{HD111395} & -0.04 & -0.03 & 0.19 &  -0.08 & 0.04 & 0.07 & 0.00 & 0.02 \\
\object{HD112758} & -0.03 & 0.00 & -0.22 &  0.00 & 0.05 & 0.04 & 0.20 &    \\
\object{HD114710} & -0.01 & -0.06 & 0.11 &  0.03 & 0.05 & 0.00 & -0.01 & -0.03 \\
\object{HD115383} & 0.00 & -0.08 & 0.12  &  0.00 & -0.05 & -0.10 & -0.01 & 0.05 \\
\object{HD115675} & -0.02 & -0.03 & -0.07&  -0.14 & 0.01 & 0.07 & -0.04 & 0.03 \\
\object{HD116443} & 0.00 & -0.03 & -0.18 &  -0.04 & 0.00 & 0.10 & 0.09 & 0.17 \\
\object{HD116956} & -0.08 & -0.17 & 0.05 &  -0.06 & -0.03 & -0.02 & -0.11 & 0.04 \\
\object{HD117043} & 0.03 & -0.04 & 0.10  &  0.00 & 0.01 & -0.10 & -0.01 & -0.07 \\
\object{HD119802} & -0.06 & 0.07 & 0.02  &  -0.11 & 0.04 & -0.01 & -0.04 & -0.06 \\
\object{HD122064} & -0.19 &    & -0.07   &  0.07 & 0.16 &    & 0.15 & 0.07 \\
\object{HD124642} & -0.06 & -0.06 & -0.02&  -0.08 & -0.09 & -0.09 & -0.17 & 0.10 \\
\object{HD125184} & 0.03 & -0.07 & 0.04  &  -0.18 & 0.07 & -0.01 & -0.05 & -0.07 \\
\object{HD126053} & 0.03 & 0.04 & -0.13  &  0.12 & -0.06 & 0.09 & 0.20 & 0.06 \\
\object{HD127506} & -0.07 & -0.03 & -0.04&  -0.06 & -0.09 & 0.03 & -0.11 & 0.08 \\
\object{HD128311} & -0.07 & 0.00 & -0.03 &  -0.06 & -0.04 & 0.02 & -0.08 & 0.04 \\
\object{HD130307} & -0.05 & -0.04 & 0.08 &  -0.02 & 0.04 & 0.21 & 0.04 & 0.20 \\
\object{HD130948} & -0.06 & -0.19 & 0.15 &  -0.07 & -0.12 & 0.00 & -0.14 & 0.07 \\
\object{HD131977} & 0.01 & 0.05 & -0.11  &  -0.1 & -0.03 & 0.12 & 0.05 & 0.18 \\
\object{HD135599} & 0.00 & 0.08 & 0.10   &  0.02 & 0.06 & 0.11 & 0.03 & 0.11 \\
\object{HD137107} & -0.02 & 0.00 & 0.09  &  0.00 & 0.04 & -0.05 & -0.18 &    \\
\object{HD139777} & -0.14 & -0.11 & 0.14 &  -0.22 & -0.12 & -0.05 & -0.10 & -0.09 \\
\object{HD139813} & -0.10 & -0.20 & 0.15 &  -0.17 & 0.04 & -0.03 & 0.04 & 0.12 \\
\object{HD140538} & -0.02 & 0.11 & 0.06  &  0.05 & 0.10 & 0.01 & 0.01 & 0.12 \\
\object{HD141004} & -0.12 & -0.11 & 0.00 &  -0.09 & -0.06 & -0.09 & -0.11 & 0.11 \\
\object{HD141272} & -0.18 & -0.06 & 0.14 &  -0.14 & -0.04 & 0.02 & -0.03 & 0.08 \\
\object{HD142267} & -0.12 & -0.04 & -0.03&  0.07 & -0.01 & 0.11 & 0.24 & 0.19 \\
\object{HD144287} & 0.05 & -0.06 & -0.03 &  0.15 & 0.10 & 0.13 & 0.05 &    \\
\object{HD145675} & -0.05 & -0.10 & -0.09&  -0.14 & 0.03 & 0.00 & -0.07 & -0.03 \\
\object{HD146233} & -0.02 & -0.05 & 0.01 &  0.03 & -0.02 & -0.01 & -0.08 & 0.08 \\
\object{HD149661} & -0.07 & 0.05 & 0.01  &  0.05 & 0.08 & 0.02 & -0.09 & 0.03 \\
\object{HD149806} & -0.18 & -0.18 & 0.05 &  -0.22 & -0.08 & -0.11 & -0.16 & -0.08 \\
\object{HD151541} & -0.14 & 0.02 & -0.15 &  0.01 & 0.04 & 0.06 & -0.07 & 0.26 \\
\object{HD153525} & -0.05 & -0.03 & 0.04 &  -0.01 & -0.03 & 0.14 & 0.06 & 0.16 \\
\object{HD154345} & 0.10 & -0.04 & -0.05 &  -0.19 & 0.07 & 0.04 & 0.02 & 0.15 \\
\object{HD156668} & -0.12 & -0.09 & -0.13&  -0.07 & -0.07 & 0.05 & -0.02 & 0.05 \\
\object{HD156985} & 0.08 & 0.05 & -0.09  &  -0.06 & 0.08 & 0.08 & 0.05 & 0.20 \\
\object{HD158633} & -0.12 &    & -0.16   &  0.07 & 0.05 & 0.17 & 0.18 & 0.08 \\
\object{HD160346} & -0.06 &    & -0.08   &  0.05 & 0.06 & 0.14 & -0.10 & 0.04 \\
\object{HD161098} & -0.05 & -0.04 & -0.02&  0.02 & 0.00 & 0.13 & 0.05 & 0.26 \\
\object{HD164922} & -0.15 & 0.04 & -0.10 &     & 0.08 & 0.03 & 0.01 & 0.10 \\
\object{HD165173} & -0.13 &    & -0.07   &     & 0.09 & 0.07 & 0.13 & 0.09 \\
\object{HD165341} & -0.21 & 0.04 & 0.03  &  0.04 & 0.12 & 0.17 & 0.15 & 0.00 \\
\object{HD165476} & -0.07 & -0.07 & -0.06&  -0.02 & -0.10 & 0.00 & -0.09 &    \\
\object{HD165670} & -0.06 &    & 0.10    &  0.06 & -0.06 & -0.08 & 0.15 &    \\
\object{HD165908} & -0.03 & 0.04 & 0.04  &  -0.03 & 0.03 & 0.05 & 0.14 & 0.14 \\
\object{HD166620} & -0.14 &    & -0.09   &  0.07 & 0.02 & 0.00 & -0.03 & 0.16 \\
\object{HD171314} & -0.03 & -0.04 & -0.09&  -0.1 & -0.14 & -0.06 & -0.10 & 0.10 \\
\object{HD174080} & -0.11 & -0.16 & -0.01&  -0.23 & -0.17 & -0.13 & -0.05 & 0.13 \\
\object{HD175742} & -0.20 & -0.14 & 0.06 &  -0.08 & -0.13 & -0.08 & -0.09 & 0.05 \\
\object{HD176377} & -0.17 & -0.15 & 0.05 &  -0.13 & -0.09 & 0.02 & 0.07 & 0.14 \\
\object{HD176841} & -0.16 & -0.10 & -0.12&  -0.2 & -0.09 & -0.17 & -0.08 & -0.09 \\
\object{HD178428} & -0.18 & -0.19 & 0.04 &  -0.17 & -0.15 & -0.11 & -0.16 & 0.03 \\
\object{HD180161} & -0.18 & -0.14 & 0.07 &  -0.15 & -0.15 & -0.12 & -0.09 & -0.01 \\
\object{HD182488} & -0.04 & 0.03 & -0.07 &     & 0.03 & 0.07 &    & -0.03 \\
\object{HD183341} & 0.04 & -0.08 & -0.08 &  -0.07 & 0.03 & -0.07 & -0.14 & 0.10 \\
\object{HD184385} & -0.18 & -0.14 & 0.07 &  -0.15 & -0.14 & -0.15 & -0.18 & -0.02 \\
\object{HD185144} & -0.10 &    & -0.02   &  -0.02 & -0.02 & 0.10 & -0.04 & 0.17 \\
\object{HD185414} & -0.15 & -0.09 & 0.07 &  -0.07 & -0.10 & -0.04 & -0.08 & 0.04 \\
\object{HD186408} & -0.11 & -0.11 & -0.03&  0.00 & -0.01 & -0.04 & -0.07 & -0.05 \\
\object{HD186427} & -0.04 & -0.03 & -0.07&  0.00 & 0.03 & -0.07 & -0.11 & 0.02 \\
\object{HD187897} & -0.03 & -0.05 & 0.03 &  -0.02 & -0.10 & -0.07 & 0.04 &    \\
\object{HD189087} & 0.08 & 0.07 & 0.10   &  0.09 & 0.14 & 0.17 & 0.06 & 0.06 \\
\object{HD189733} & -0.15 & -0.15 & -0.11&  -0.11 & -0.04 & 0.00 & -0.04 & 0.05 \\
\object{HD190007} & -0.06 & -0.19 & -0.03&  -0.18 & -0.20 & -0.16 & -0.15 & -0.04 \\
\object{HD190406} & -0.16 & -0.15 & 0.05 &  -0.08 & -0.14 & -0.09 & -0.11 & -0.03 \\
\object{HD190470} & -0.16 & -0.15 & -0.08&  -0.1 & -0.07 & -0.07 & -0.11 & 0.01 \\
\object{HD190771} & -0.20 & -0.20 & -0.07&  -0.15 & -0.19 & -0.19 & -0.17 & -0.06 \\
\object{HD191533} & 0.01 & -0.11 & 0.09  &  -0.16 & 0.14 & -0.06 & -0.09 & -0.06 \\
\object{HD191785} & -0.12 & -0.10 & -0.24&  -0.16 & -0.11 & -0.02 & -0.07 & 0.14 \\
\object{HD195005} & -0.09 & -0.10 & 0.06 &  -0.06 & -0.02 & -0.12 & -0.11 &    \\
\object{HD195104} & 0.10 & -0.11 & 0.20  &  0.08 & 0.10 & 0.23 & 0.16 & 0.03 \\
\object{HD197076} & 0.11 & 0.12 & 0.08   &  0.10 & 0.08 & 0.21 & 0.13 & 0.21 \\
\object{HD199960} & -0.14 & -0.12 & -0.11&  -0.25 & -0.04 & -0.17 & -0.07 &    \\
\object{HD200560} & -0.03 & -0.03 & 0.04 &  -0.07 & -0.09 & -0.04 & -0.12 & -0.09 \\
\object{HD202108} & -0.04 & -0.11 & 0.10 &  -0.13 & -0.02 & 0.09 & -0.09 & 0.15 \\
\object{HD202575} & 0.05 & -0.04 & 0.09  &  -0.04 & 0.02 & 0.08 & -0.02 & 0.10 \\
\object{HD203235} & -0.14 & -0.12 & -0.05&  -0.05 & -0.07 & -0.09 & -0.11 & -0.01 \\
\object{HD205702} & 0.11 & -0.10 & -0.03 &  -0.08 & -0.02 & -0.04 &    & -0.02 \\
\object{HD206860} & -0.12 & -0.17 & 0.05 &  -0.03 & -0.04 & 0.02 & 0.05 & \\
\object{HD208038} & -0.08 & -0.07 & 0.09 &  -0.02 & -0.06 & 0.04 & -0.04 & 0.10 \\
\object{HD208313} & -0.11 & -0.17 & -0.09&  -0.08 & -0.06 & 0.00 & -0.17 & -0.03 \\
\object{HD208906} & -0.05 & 0.17 & -0.14 &  -0.09 & 0.02 & 0.14 & 0.11 & 0.34 \\
\object{HD210667} & -0.07 & 0.02 & -0.04 &  -0.09 & -0.01 & -0.10 & -0.12 & -0.01 \\
\object{HD210752} & -0.02 & 0.09 & 0.03  &  -0.04 & 0.01 & 0.16 & 0.31 & 0.37 \\
\object{HD211472} & 0.05 & -0.05 & 0.10  &  0.07 & 0.05 & 0.06 & -0.07 &    \\
\object{HD214683} & -0.02 & -0.01 & 0.06 &  0.01 & 0.05 & 0.15 & 0.12 & 0.28 \\
\object{HD216259} & 0.10 & 0.19 & -0.10  &  0.14 & 0.14 & 0.18 & 0.25 & 0.22 \\
\object{HD216520} & -0.13 & 0.00 & -0.20 &  -0.12 & 0.00 & 0.07 & 0.08 & 0.09 \\
\object{HD217014} & -0.16 & -0.09 & -0.10&  -0.21 & -0.03 & -0.13 & -0.13 & -0.05 \\
\object{HD217813} & -0.18 & -0.17 & 0.04 &  -0.10 & -0.12 & -0.10 & -0.14 & -0.01 \\
\object{HD218868} & -0.10 & -0.14 & 0.03 &  -0.15 & -0.10 & -0.10 & -0.15 & -0.03 \\
\object{HD219538} & -0.04 & -0.05 & -0.06&  0.08 & 0.04 & 0.11 & 0.01 & 0.06 \\
\object{HD219623} & -0.17 & -0.19 & 0.01 &  -0.23 & -0.20 & -0.18 & -0.17 & 0.13 \\
\object{HD220140} & -0.15 & -0.11 & 0.05 &  0.00 & -0.16 & -0.09 & -0.12 & 0.00 \\
\object{HD220182} & -0.06 & -0.20 & 0.07 &  -0.10 & -0.03 & 0.04 & -0.06 & 0.10 \\
\object{HD220221} & -0.02 & -0.06 & 0.02 &  -0.03 & -0.06 & -0.06 & -0.04 & -0.09 \\
\object{HD221851} & -0.07 & -0.13 & 0.02 &  -0.05 & 0.01 & 0.05 & -0.07 & 0.11 \\
\object{HD222143} & -0.14 & -0.14 & 0.09 &  -0.18 & -0.15 & -0.15 & -0.17 & -0.02 \\
\object{HD224465} & -0.18 & -0.12 & 0.05 &  -0.07 & -0.13 & -0.13 & -0.19 & 0.04 \\
\object{HD263175} & -0.07 & -0.06 & -0.13&  -0.07 & -0.02 & 0.02 & 0.03 & 0.23 \\
\object{BD+01 2063} & 0.00 & -0.02 & 0.07& -0.09 & -0.03 & -0.01 & -0.02 & 0.05 \\
\object{BD+12 4499} & 0.07 & -0.03 & 0.02& 0.02 & 0.01 & 0.05 & -0.07 & 0.24 \\
\hline
Hercules & & & & & & & & \\
\hline
\object{HD013403} & -0.05 & 0.10 & -0.09 & -0.03 & -0.02 & 0.16 & 0.09 & 0.15 \\
\object{HD019308} & -0.04 & -0.05 & -0.01 & -0.29 & -0.04 & -0.04 & -0.08 & -0.04 \\
\object{HD023050} & 0.09 & -0.08 & -0.04 & 0.00 & 0.08 & 0.22 & 0.26 & 0.25 \\
\object{HD030562} & -0.05 & -0.03 & 0.02 & -0.09 & 0.01 & -0.02 & -0.03 & 0.02 \\
\object{HD064606} & 0.01 & 0.16 & -0.14 & 0.11 & -0.06 & 0.13 & 0.19 & 0.40 \\
\object{HD068017} & -0.07 & 0.03 & -0.12 & -0.07 & -0.04 & 0.07 & 0.14 & 0.26 \\
\object{HD081809} & -0.08 & -0.02 & -0.15 & -0.07 & 0.07 & 0.16 & 0.04 & 0.17 \\
\object{HD107213} & 0.02 & -0.09 & 0.02 & -0.16 & -0.08 & -0.11 & 0.06 &    \\
\object{HD139323} & 0.03 & 0.04 & 0.00 & -0.10 & 0.20 & 0.14 & 0.15 & 0.10 \\
\object{HD139341} & 0.01 & 0.05 & -0.07 & -0.20 & 0.15 & 0.05 & -0.04 & 0.13 \\
\object{HD144579} & -0.16 & 0.08 & -0.25 &    & 0.08 & 0.11 & 0.15 & 0.24 \\
\object{HD159222} & -0.01 & -0.05 & -0.03 & -0.24 & 0.05 & -0.10 & -0.17 & -0.07 \\
\object{HD159909} & -0.16 & -0.11 & -0.11 & -0.24 & -0.08 & -0.09 & -0.16 & 0.03 \\
\object{HD215704} & -0.20 & 0.01 & -0.12 & 0.01 & -0.04 & -0.04 & 0.04 & 0.02 \\
\object{HD218209} & -0.04 & 0.01 & -0.01 & 0.09 & -0.02 & 0.15 & 0.13 & 0.27 \\
\object{HD221354} & -0.17 & -0.19 & -0.26 & -0.24 & -0.14 & -0.07 & -0.10 & 0.08 \\
\hline
Nonclas & & & & & & & & \\
\hline
\object{HD004628} & 0.08& 0.09& -0.04& 0.03& 0.16 & 0.13 & 0.24 & \\
\object{HD004635} & -0.13& -0.11& -0.04& -0.20& -0.11 & -0.10 & -0.13 & 0.00 \\
\object{HD010145} & -0.09& -0.11& -0.06& -0.09& 0.01 & 0.12 & 0.02 & 0.15 \\
\object{HD012051} & -0.17& -0.18& 0.10& -0.26& -0.06 & -0.13 & -0.17 & -0.07 \\
\object{HD013974} & -0.20& -0.09& -0.01& 0.00& -0.11 & 0.05 & 0.08 & 0.01 \\
\object{HD017660} & -0.08& -0.12& -0.14& -0.21& -0.20 & -0.19 & -0.09 & 0.15 \\
\object{HD020165} & -0.08& -0.10& -0.07& -0.09& -0.02 & 0.11 & -0.08 & 0.00 \\
\object{HD024206} & 0.07& 0.11& 0.03& 0.05& 0.15 & 0.17 & 0.13 & 0.07 \\
\object{HD032147} & -0.13&    & -0.04& -0.02& 0.13 & 0.02 & 0.12 & 0.06 \\
\object{HD045067} & -0.13& -0.13& 0.00& -0.25& -0.02 & -0.15 & -0.15 & -0.04 \\
\object{HD084035} & -0.05& -0.07& -0.05& -0.29& -0.05 & -0.10 & -0.12 & -0.08 \\
\object{HD086728} & -0.23& -0.26& -0.06& -0.33& -0.19 & -0.23 & -0.21 & -0.10 \\
\object{HD090875} & -0.15& -0.19& -0.01& -0.29& -0.18 & -0.15 & -0.14 & \\
\object{HD117176} & -0.09& -0.12& -0.01& -0.02& 0.01 & 0.02 & -0.01 & 0.07 \\
\object{HD117635} & 0.04& -0.01& -0.04& -0.06& 0.09 & 0.15 & 0.15 & 0.30 \\
\object{HD154931} & -0.01& -0.04& 0.01& -0.16& -0.03 & 0.01 & -0.06 & -0.01 \\
\object{HD159482} & -0.05& 0.12& -0.01& 0.06& -0.06 & 0.06 & 0.24 & 0.35 \\
\object{HD168009} & -0.19& -0.04& -0.06& -0.26& -0.08 & -0.16 & -0.14 & 0.05 \\
\object{HD173701} & -0.07& 0.02& -0.10& -0.15& -0.06 & 0.05 & -0.07 & -0.14 \\
\object{HD182736} & -0.05& 0.04& 0.05& 0.00& -0.02 & 0.05 & 0.07 &    \\
\object{HD184499} & -0.11& 0.08& -0.10& -0.02& 0.00 & 0.07 & 0.21 & 0.37 \\
\object{HD184768} & -0.08& -0.06& -0.09& -0.10& 0.00 & 0.06 & -0.04 & 0.11 \\
\object{HD186104} & -0.06& -0.13& -0.05& -0.19& -0.11 & -0.10 & 0.00 & 0.09 \\
\object{HD215065} & -0.14& -0.05& -0.16& -0.06& -0.03 & 0.12 & 0.19 &    \\
\object{HD219134} & -0.11&    & -0.03&    & 0.12 & 0.11 & -0.09 & -0.11 \\
\object{HD219396} & -0.15&    & -0.09& -0.06& 0.00 & 0.21 & 0.05 &    \\
\object{HD224930} & 0.07& 0.13& -0.09& 0.06& -0.09 & 0.02 & 0.24 &    \\
\hline
\end {longtable}


\begin{thebibliography}{}

\bibitem[1973]{al73}
Allen, C.W. 1973, Astrophysical Quantities (London: Athlone Press)

\bibitem[2004]{ar04}
Allende Prieto, C., Garcia Lopez, R.J., Lambert, D.L. \& Gustafsson, B.
1999, \apj, 527, 879

\bibitem[2009]{an09}
Andrievsky, S. M., Spite, M., Korotin, S. A. et al. 2009, \aap, 494, 1083

\bibitem[2011]{arcones11}
Arcones, A., \&  Montes, F. 2011, \apj, 731, 5

\bibitem[1996]{ba96}
Baranne, A., Queloz, D., Mayor, M. et al. 1996, \aaps, 119, 373

\bibitem[2003]{be03}
Bensby, T., Feldzing, S., \&  Lungstrem, I. 2003, \aap, 410, 527

\bibitem[2005]{be05}
Bensby, T., Feltzing, S., Lundstrom, I., \& Ilyin, I. 2005, \aap, 443, 185

\bibitem[2007]{be07}
Bensby, T., Zenn, A.R., Oey, M.S., \& Feltzing, S. 2007, \aap, 663, L13

\bibitem[2006]{bien06}
Bienaym\'{e}, O., Soubiran, C., Mishenina, T. et al. 2006, \aap, 446, 933

\bibitem[2011]{bisterzo:11}
Bisterzo, S., Gallino, R., Straniero, O., Cristallo, S., \& K\"{a}ppeler, F. 2011,
\mnras, 418, 284

\bibitem[2010]{bravo:10}
Bravo, E., Dominguez, I., Badenes, C., Piersanti, L., \&  Straniero, O. 2010, \apj, 711, 66

\bibitem[2006]{br06}
Brewer, M.M., \& Carney, B.  2006, \aap, 131, 431

\bibitem[1957]{bur57}
Burbidge, E. M., Burbidge, G. R., Fowler, W.A., \& Hoyle, F. 1957,
  Reviews of Modern Physics,  29,  547

\bibitem[1957]{cameron:57}
Cameron, A. G. W. 1957, \aj, 62, 9

\bibitem[1982]{cam82}
Cameron,  A.G.W., 1982, \aaps, 82, 123

\bibitem[1986]{ca86}
Carlsson, M. 1986, Uppsala Obs. Rep. 33

\bibitem[2010]{cas10}
Casagrande, L.,  Ramirez, I., Mel\'{e}ndez, J., Bessell, M., \&  Asplund, M.
 2010,   \aap, 512, 54

\bibitem[2011]{cas11}
Casagrande, L., Schonrich, R., Asplund, M., Cassisi, S., Ramirez, I. et al.
  2011,   \aap, 530, 138

\bibitem[2011]{ch11}
Chiappini, C., Frischknecht, U., Meynet, G. et al. 2011, \nat, 474, 666

%\bibitem[1967]{clayton:67}
%Clayton, D.D. \& Rassbach, M.E. 1967, \apj, 168, 69

\bibitem[1974]{cr74}
Crandall, D.H., Dunn, G.H., Gallagher, A. et al. 1974, \apj 191, 789

\bibitem[2006]{dasilva:06}
da Silva, L., Girardi, L., Pasquini, L. et al. 2006, \aap, 458, 609

\bibitem[2009]{dorazi:09}
D'Orazi, V., Magrini, L., Randich, S., et al. 2009, \apj, 693, 31

\bibitem[2012]{dorazi:12}
D'Orazi, V., Biazzo,K., Desidera,S., et al. 2012, \mnras, 423, 2789

\bibitem[1993]{ed93}
Edvardsson, B., Andersen, J., Gustafsson, B., et al. 1993, \aap, 275, 101

\bibitem[1958]{eg58}
Eggen, O.J.  1958, \mnras, 118, 154

\bibitem[2005]{fa05}
Famaey, B., Jorissen, A., Luri, X., et al. 2005, \aap, 430, 165

\bibitem[2009]{farouqi09}
Farouqi, K., Kratz, K.-L., Mashonkina, L. I., et al. 2009, \apj, 694, 49

\bibitem[2001]{feltzing01}
Feltzing, S., Holmberg, J., \& Hurley, J. R. 2001, \aap, .377, .911

\bibitem[2009]{fe09}
Feltzing, S., Oey, S., \& Bensby, T. Proc.IAU Symp., 2009, 254, 197.

\bibitem[1999]{fr99}
Freiburghaus, C., Rosswog S., \& Thielemann, F.-K. 1999, \apj, 525, L121

\bibitem[2012]{fricknekt12}
Frischknecht, U., Hirschi, R., \& Thielemann, F.-K. 2012, \aap, 538, 2

\bibitem[2006]{froelich06}
Frohlich, C., Martinez-Pinedo, G., Liebendorfer, M., et al. 2006, \prl,
96, 142502

\bibitem[2004]{fu04}
Fuhrmann, K. 2004,  Astronomische Nachrichten, 325, 3

\bibitem[2001]{fux01}
Fux, R. 2001, \aap, 373, 511

\bibitem[1998]{gallino:98}
Gallino, R., Arlandini, C., Busso, M., et al. 1998, \apj, 497, 388

\bibitem[1992]{ga92}
Galazutdinov, G.A. 1992, Preprint SAO RAS, n92

\bibitem[1983]{gi83}
Gilmore, G. \&  Reid, N. 1983, \mnras, 202, 1025

\bibitem[2000]{girardi00}
Girardi, L.; Bressan, A., Bertelli, G., \& Chiosi, C. 2000, \aaps, 141, 371

\bibitem[2012]{girardi12}
Girardi online http://stev.oapd.inaf.it/~lgirardi/cgi-bin/param\_1.0

\bibitem[2006]{haywood06}
Haywood, M. 2006, \mnras,  371,  1760

\bibitem[2000]{hillebrandt:00}
Hillebrandt, W., Niemeyer, J. C., \& Reinecke, M. 2000, \araa, 38, 191

\bibitem[1996]{hoffman96}
Hoffman, R. D., Woosley, S. E., Fuller, G. M., \& Meyer, B. S. 1996, \apj, 460, 478

\bibitem[1979]{ho79}
Hofsaess, D. 1979, ADNDT 24, 285

\bibitem[2009]{holm09}
Holmberg, J., Nordstrom, B., \& Andersen, J. 2009,  \aap, 501, 941

\bibitem[2006]{ho06}
Honda, S., Aoki, W., Ishimaru, Y., Wanajo, S., \& Ryan, S.G. 2006, \apj, 643, 1180

\bibitem[2005]{jor05}
Jorgensen, B. R.\& Lindegren, L. 2005, \aap,  436, 127

\bibitem[1989]{kap89}
K\"{a}ppeler, F., Beer, H., Wisshak, K. 1989, Reports on Progress in Physics,  52,  945

\bibitem[1998]{ka98}
Katz, D., Soubiran, C., Cayrel, R. et al., 1998, \aap, 338, 151

\bibitem[2011]{kl11}
Klochkova, V., Mishenina, T., Korotin, S., et al. 2011, \apss, 335, 141

\bibitem[2012]{korobkin:12}
Korobkin, O., Rosswog, S., Arcones, A., Winteler, C. 2012, \mnras, 426, 1940

\bibitem[1999]{ko99}
Korotin, S.A., Andrievsky, S.M., \& Luck, R.E. 1999, \aap 351, 168

\bibitem[2011]{ko11}
Korotin, S., Mishenina, T., Gorbaneva, T., \& Soubiran, C.
2011, \mnras, 415, 2093

\bibitem[2004]{ko04}
Kovtyukh, V.V., Soubiran, C., \& Belik, S.I. 2004, \aap, 427, 923

\bibitem[1993]{ku93}
Kurucz R.L. 1993, CD ROM n13

\bibitem[2011]{kusakabe:11}
Kusakabe, M., Iwamoto, N., \& Nomoto, K. 2011, \apj, 726, 25

\bibitem[2007]{le07}
van Leeuwen F., 2007, ASSL, 350

\bibitem[2012]{maiorca:12}
Maiorca, E., Randich, S., Busso, M., Magrini, L., \& Palmerini, S.E. 2011, \apj,
736, 120

\bibitem[2005]{ma05}
Marsakov, V.A., Borkova, T.V. 2005, Astronomy Letters, 31, 515

\bibitem[2000]{ma00}
Mashonkina, L., Gehren, T. 2000, \aap, 364, 249

\bibitem[2000]{mas00}
Mashonkina, L.I. 2000, Astr. Rep. 44, 558

\bibitem[2001]{ma01}
Mashonkina, L.I., Gehren, T. 2001, \aap, 376, 232

\bibitem[2004]{ma04}
Mashonkina, L.I., Kamaeva, L.A., Samotoev, V.A., \& Sakhibullin, N.A. 2004, Astr.
Rep., 48, 185

\bibitem[2011]{ma11}
Mashonkina, L.I., Gehren, T., Shi, J.-R., Korn, A.J., Grupp, F. 2011, \aap, 528,
A87

\bibitem[2006]{matteucci:06}
Matteucci, F., Panagia, N., Pipino, A., Mannucci, F., Recchi, S., Della Valle,
M. 2006, \mnras, 372, 265

\bibitem[2012]{Minchev:12}
Minchev, I., Chiappini, C., Martig, M. 2012, arXiv1208.1506

\bibitem[2000]{mishenina:00}
Mishenina, T. V., Korotin, S. A., Klochkova, V. G., Panchuk, V. E. 2000, \aap,
353, 978

\bibitem[2001]{mi01}
Mishenina, T.V., Kovtyukh, V.V.  2001, \aap, 370,  951

\bibitem[2004]{mi04}
Mishenina, T.V., Soubiran, C., Kovtyukh, V.V., Korotin, S.A. 2004, \aap, 418,
551

\bibitem[2006]{mi06}
 Mishenina, T. V., Bienaym\'{e}, O., Gorbaneva, T. I. et al. 2006, \aap , 456, 1109

\bibitem[2008]{mi08}
Mishenina, T.V.,  Soubiran C., Kovtyukh V. V., et al., 2008, \aap, 489, 923

\bibitem[2012]{mowlavi12}
Mowlavi, N., Eggenberger, P., Meynet, G., et al. 2012, \aap, 541, 41

\bibitem[2009]{neves:09}
Neves, V., Santos, N.C., Sousa, S.G., Correia, A.C.M., \& Israelian, G.  2009, \aap, 497, 563


\bibitem[2006]{nikmish06}
 Nikityuk, T., \& Mishenina, T. 2006, \aap, 456, 969

\bibitem[2006]{nishimura:06}
Nishimura, S., Kotake, K., Hashimoto, M., et al. 2006, \apj, 642, 410

\bibitem[2008]{ni08}
Nissen P.E., \& Schuster W.J. 2008, Proc. IAU Symposium, 254, 103

\bibitem[2011]{pakh11}
Pakhomov, Yu.V., Antipova, \&  L.I., Boyarchuk, A.A. 2011, Astr. Rep. 55, 256

\bibitem[2005]{pe05}
del Peloso, E.F., da Silva, L., \& Porto de Mello, G.F. 2005, \aap, 434, 285

\bibitem[2010]{pignatari:10}
Pignatari, M., Gallino, R., Heil, M., et al. 2010, \apj, 710, 1557

\bibitem[2008]{pignatari08}
Pignatari, M., Gallino, R., Meynet, G., et al. 2008, \apj, 687, 95

\bibitem[2000]{pr00}
Prochaska, J.X., Naumov, S.O., Carney, B.W., McWilliam, A., \& Wolfe, A.M. 2000,
\aj, 120, 2513

\bibitem[2008]{qi08}
Qian, Y.-Z., Wasserburg, G.J.  2008, \apj, 687, 272

\bibitem[2002]{rauscher:02}
Rauscher, T., Heger, A., Hoffman, R. D., Woosley, S. E. 2002, \apj, 576, 323

\bibitem[2006]{re06}
Reddy,  B.E., Lambert, D.L., \& Allende Prieto, C. 2006, \mnras, 367, 1329

\bibitem[1978]{rut78}
Rutten R.J. 1978, SoPh 56, 237

\bibitem[1998]{sc98}
Schoening, T., Butler, K., 1998, \aaps, 128, 581

\bibitem[2009]{se09}
Serminato A., Gallino R., Travaglio C., Bisterzo S., Straniero O. 2009,
\pasa, 26, 153

\bibitem[1981]{so81}
Sobelman I.I., Vainshtein L.A., Yukov E., 1981, Excitation of Atoms
and Broadening of Spectral Lines, Springer Ser. in Chem. Phys.,
Berlin, Springer

\bibitem[2005]{so05}
Soubiran, \& C., Girard, F. 2005, \aap, 438, 139

\bibitem[2008]{soub08}
Soubiran, C., Bienaym\'{e}, O., Mishenina, T. V.,\&  Kovtyukh, V. V. 2008, \aap, 480, 91

\bibitem[2008]{sneden08}
Sneden, C., 2008, Bull. AAS, 40, 264

\bibitem[2006]{stein06}
Steinmetz, M., Zwitter, T., Siebert, A. et al. 2006, \aj, 132, 1645

\bibitem[2003]{st03}
Straniero, O., Dominguez, I., Imbriani, G., \& Piersanti, L. 2003, \apj, 583,  878

\bibitem[2008]{su08}
Surman, R., McLaughlin, G. C., Ruffert, M., Janka, H.-Th., \& Hix, W. R. \apj, 679,
L117

\bibitem[2007]{the:07}
The, L-S, El Eid, M. F., \& Meyer, B. S.  2007, \apj, 655, 1058

\bibitem[1986]{thielemann:86}
Thielemann, F.-K., Nomoto, K., \& Yokoi, K. 1986, \aap, 158, 17

\bibitem[2004]{thielemann:04}
Thielemann, F.-K., Brachwitz, F., Hoflich, P., Martinez-Pinedo, G., \& Nomoto, K.
2004, \nar, 48, 605

\bibitem[2011]{thielemann:11}
Thielemann, F.-K., Arcones, A., Kappeli, R., et al.
2011, Progress in Particle and Nuclear Physics, 66, 346

\bibitem[1995]{timmes95}
Timmes, F. X., Woosley, S. E., \& Weaver, T. A. 1995, \apjs, 98, 617

\bibitem[2003]{timmes:03}
Timmes, F.X., Brown, E.F., \& Truran, J.W. 2003, \apj, 590, 83

\bibitem[2009]{tolstoy09}
Tolstoy, E., Hill, V., Tosi, M. 2009, \araa, 47, 371

\bibitem[2004]{tr04}
Travaglio, C., Gallino, R., Arnone, E., Cowan, J., Jordan, F., \& Sneden, C. 2004,
\apj, 601, 864

\bibitem[2005]{travaglio:05}
Travaglio, C., Hillebrandt, W., \& Reinecke, M. 2005, \aap, 443, 1007


\bibitem[2011]{travaglio:11}
Travaglio, C., Ropke, F. K., Gallino, R., \& Hillebrandt, W. 2011, \apj, 739, 93

\bibitem[2002]{truran02}
Truran, J. W., Cowan, J. J., Pilachowski, C. A., \& Sneden, Ch.
2002, \pasp, 114, 1293

\bibitem[1996]{ts96}
Tsymbal, V.V. 1996, ASP Conf. Ser. 108, 198

\bibitem[1962]{re62}
van Regemorter, H., 1962, \apj, 136, 906

\bibitem[2012]{winteler:12}
Winteler, C., Kappeli, R., Perego, A., et al. 2012, \apj, 750, 22

\bibitem[1994]{wo94}
Woosley, S.E., Wilson, J.R., Mathews, G.J., Hoffman, R.D., \& Meyer, B.S. 1994, \apj, 433, 229

\end{thebibliography}
\end{document}